\documentclass[fleqn,usenatbib]{mnras}

\usepackage[T1]{fontenc}

\DeclareRobustCommand{\VAN}[3]{#2}
\let\VANthebibliography\thebibliography
\def\thebibliography{\DeclareRobustCommand{\VAN}[3]{##3}\VANthebibliography}

\usepackage{siunitx}
\usepackage{supertabular}
\usepackage{caption}

\usepackage{graphicx}
\usepackage{amsmath}
\usepackage{amssymb}
\usepackage{afterpage}
\usepackage {ulem}

\title[AstraLux Search for Stellar Companions of Exoplanet Host Stars]{Search for Stellar Companions of Exoplanet Host Stars with AstraLux/CAHA 2.2\,m}

\author[S. Schlagenhauf et al.]{Saskia Schlagenhauf$^{1,2,3}$\thanks{E-mail: saskia.schlagenhauf@armagh.ac.uk}, Markus Mugrauer$^{1}$, Christian Ginski$^{4}$, Sven Buder$^{5,6}$,
\newauthor Matilde Fern\'andez$^{7}$, Richard Bischoff$^{1}$\\
$^{1}$Astrophysikalisches Institut und Universit\"{a}ts-Sternwarte Jena, Jena, Germany\\
$^{2}$Armagh Observatory and Planetarium, Armagh, UK\\
$^{3}$Queen's University Belfast, UK\\
$^{4}$University of Galway, Galway, Ireland\\
$^{5}$Research School of Astronomy $\&$ Astrophysics, Australian National University, Canberra, Australian Capital Territory, Australia\\
$^{6}$ARC Center of Excellence for All Sky Astrophysics in 3 Dimensions (ASTRO 3D), Australia\\
$^{7}$Instituto de Astrof\'isica de Andaluc\'ia CSIC, Glorieta de la Astronomia, Granada, Spain\\}

\date{Accepted 2024 February 14. Received 2024 February 14; in original form 2023 September 27}

\pubyear{2024}

\begin{document}
\label{firstpage}
\pagerange{\pageref{firstpage}--\pageref{lastpage}}

\maketitle

\begin{abstract}
Stellar multiplicity is a key aspect of exoplanet diversity, as the presence of more than one star in a planetary system can have both devastating and positive effects on its formation and evolution. In this paper, we present the results of a Lucky Imaging survey of 212 exoplanet host stars performed with AstraLux at CAHA 2.2\,m. The survey includes data from seven observing epochs between August 2015 and September 2020, and data for individual targets from four earlier observing epochs. The targets of this survey are nearby, bright, solar-like stars with high proper motions. In total, we detected 46 co-moving companions of 43 exoplanet host stars. Accordingly, this survey shows that the minimum multiplicity rate of exoplanet host stars is $20 \pm 3\,\%$. In total, 33 binary and ten hierarchical triple star systems with exoplanets have been identified. All companions were found to have a common proper motion with the observed exoplanet host stars, and with our astrometry we even find evidence of orbital motion for 28 companions. For all targets, we determined the detection limit and explore the detection space for possible additional companions of these stars. Based on the reached detection limit, additional co-moving companions beyond the detected ones can be excluded around all observed exoplanet host stars. The increasing number of exoplanets discovered in multiple stellar systems suggests that the formation of planets in such systems is by no means rare, but common. Therefore, our study highlights the need to consider stellar multiplicity in future studies of exoplanet habitability.
\end{abstract}


\section{Introduction}

Stellar companions in an extrasolar planetary system can dramatically alter the dynamics of the system. A famous example of this is the Kozai mechanism. If a planet in an S-type orbit around its host star is sufficiently inclined to the binary orbital plane, its eccentricity and hence its periapsis will oscillate with the Kozai period, which in combination with tidal friction will then shrink the orbit of the planet \cite[Kozai migration, see e.g.][]{Wu2007}. Furthermore, a stellar companion in a binary system limits the parameter space for long-term stable orbits of planets around their host stars \cite[][]{Holman1999}, and even distant stellar companions can significantly influence the evolution of a planet's orbit \cite[][]{Kaib2013}. However, planet formation is expected to be different in multiple star systems compared to single star systems because stellar companions change the properties of protoplanetary disks around the host stars in which planet formation takes place \cite[][]{Marzari2000, Mayer2005, Thebault2015}. In particular, stellar multiplicity has an effect on each stage of planet formation. This shows that multiple star systems are extreme environments for planet formation and evolution and can therefore help to understand the process of planet formation and the dynamics of planetary systems themselves, which also significantly affect the habitability of these worlds. As multiple star systems are very common in the Milky Way, this has important implications for the overall abundance of planets and their properties in our Galaxy.

Prior to this study, there have already been two multiplicity surveys of exoplanet host stars carried out at the Astrophysical Institute and University Observatory (AIU) Jena using the Lucky Imaging camera AstraLux \citep{Hormuth2008} at the 2.2\,m telescope of the Centro Astron\'omico Hispano en Andaluc\'ia (CAHA) in Spain \cite[see][]{Ginski2012, Ginski2016}. In the survey presented here, we will evaluate some of the stellar companion-candidates of exoplanet host stars detected in previous surveys, but also assess new targets, mainly located in the northern sky. For objects further south, there is a complementary survey carried out with the extreme adaptive optics imager SPHERE \citep{Beuzit2019} at ESO's Very Large Telescope (VLT) in Chile \citep[for first results see, ][]{Ginski2021}.

In addition, several other studies focus on the search for (sub)stellar companions of exoplanet host stars using data from the ESA-Gaia mission, see for example \cite{Mugrauer2019, Michel2021,Fontanive2021, Michel2024}, among others. A similar survey using ESA-Gaia data is being carried out among (Community) TESS\footnote{Transiting Exoplanet Survey Satellite \citep{Ricker2015}.} objects of interest that are either confirmed transiting exoplanet host stars or promising candidates \citep[see][]{Mugrauer2020, Mugrauer2021, Mugrauer2022b, Mugrauer2023}. These surveys are all designed to find stellar companions at larger angular separations from their host stars than the AstraLux and SPHERE surveys. Beside these imaging surveys, there are also literature studies that compile the multiplicity status of exoplanet host stars, such as that of \cite{Cuntz2022}, which explicitly focuses on exoplanet systems consisting of three or more stars.

The observations with AstraLux are described in the next section. This is followed by the properties of the exoplanet host stars observed with AstraLux in this survey. The image processing and data analysis methods are described in the fourth section of this paper. The detected stellar companions are discussed thereafter. Finally, we present the detection limit for each exoplanet host star, which is used to characterize the detection space for potential companions of these stars.

\section{Observations with AstraLux}

This study analyzes data from seven AstraLux observing epochs between August 2015 and September 2020, each consisting of three nights, as well as data for some additional targets from four previous observing epochs. In total, 314 observations of 212 exoplanet host stars have been made during this survey.

As a default setting for the Lucky Imaging observations with AstraLux, 50000 individual frames are taken in the i$'$-band of the Sloan Digital Sky Survey photometric system \citep{Fukugita1996}. In general, a detector integration time (DIT) of 30\,ms is used, which is sufficiently short to capture the speckle pattern in the seeing disk of the observed exoplanet host stars and to avoid saturation or operation in the non-linear regime of the AstraLux detector. This gives a total execution time of 25\,min for each observation. For the Lucky Imaging data reduction we select the best 10\,\% of all frames (5000), using the Strehl ratio to assess the image quality, resulting in a total integration time (TIT) of 2.5\,min per observation. The chosen frame selection criterion in the Lucky Imaging data processing is a compromise between the achievable image quality and the total execution time required to reach sufficient sensitivity.

\begin{figure}
\includegraphics[width=1\columnwidth]{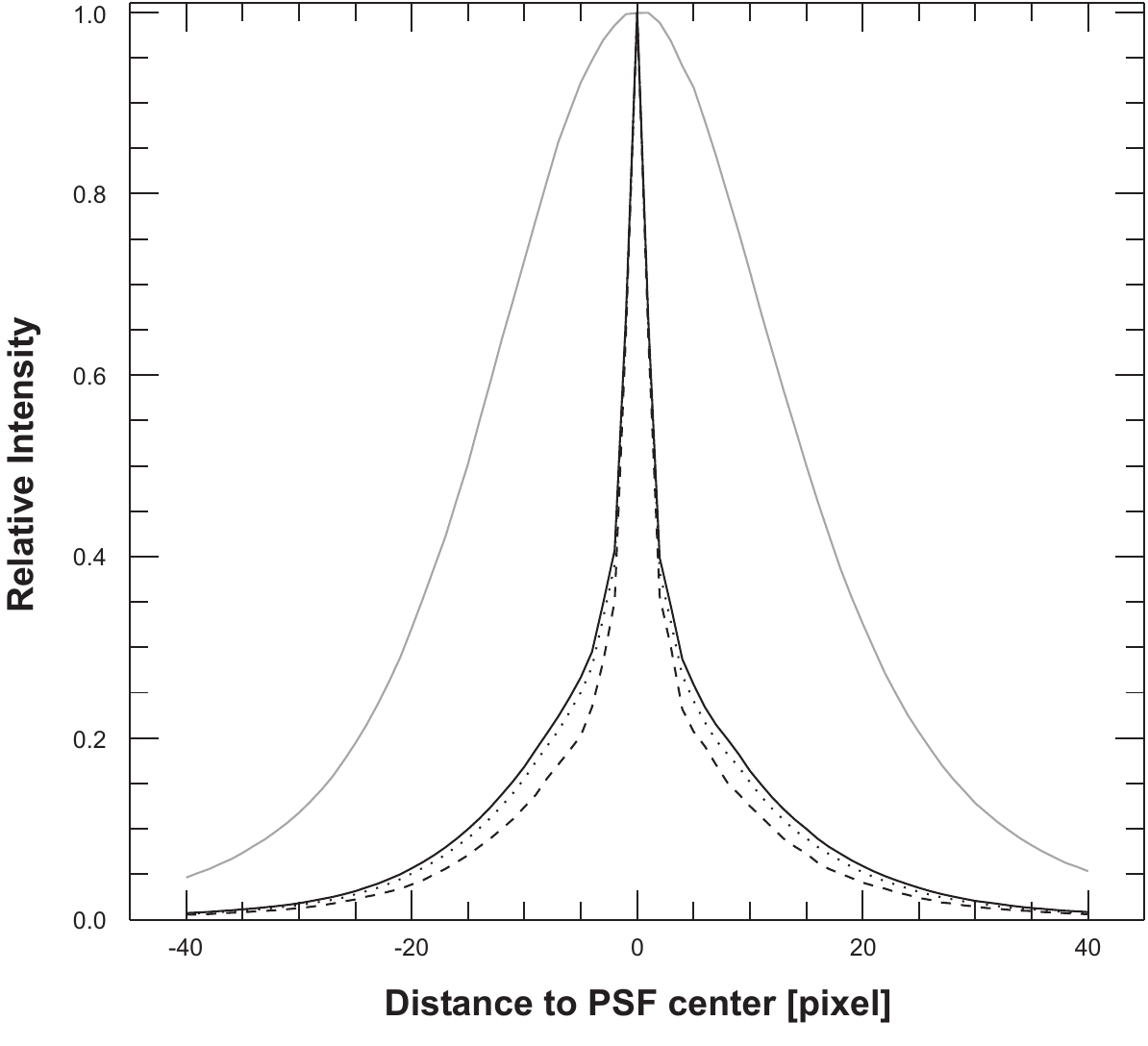}
\caption{The radial profile of the point spread function (PSF) of a star observed with AstraLux in the i$'$-band, reduced with and without our Lucky Imaging data processing pipeline. The grey line is the profile of the seeing limited PSF of the star. The black solid, dotted and dashed lines are the PSF profiles of the star for Lucky Imaging frame selection rates of 10, 5 and 1\,\% used, respectively.} \label{psf}
\end{figure}

In Figure\,\ref{psf} we show the radial profile of a point spread function (PSF) of a star observed with AstraLux in the i$'$-band as a function of the selected frame selection rate of the Lucky Imaging data processing applied. A significant improvement in angular resolution is obtained by using the Lucky Imaging data reduction procedure (grey versus black solid PSF profile). In contrast, reducing the frame selection rate from 10 to 1\,\% (black solid versus black dashed and black dotted PSF profiles) only slightly improves the image resolution, while the sensitivity for companion detection in the background limited region around the bright exoplanet host star is significantly reduced due to the shorter total integration time used. The electron multiplication (EM) gain is always chosen at the start of each observation so that the peak value of the PSF of the exoplanet host stars does not exceed 7000\,ADU. For bright targets, shorter DITs and/or the non-EM readout mode of AstraLux has to be used to avoid saturation of the instrument's detector. For the faintest targets, longer DITs and higher EM gain settings are required to meet the integration criterion described above. The details of all observations are summarized in Table\,\ref{tab_oblsog}. For each AstraLux observation, the equatorial coordinates of the targets, the observation epoch, the total number of frames, the number of frames selected for Lucky Imaging data processing, the detector integration time (DIT), the total integration time (TIT) of the reduced image, and the EM gain used are listed.

For the reduction and processing of the obtained Lucky Imaging data we use our Lucky Imaging pipeline software, which is based on  \textsc{ESO-Midas} \citep{Warmels1992} and \textsc{ESO-Ecipse} \citep{Devillard1997} software packages. The first step in the Lucky Imaging data processing is to characterize the quality of all images taken during the observation of a target by measuring the Strehl ratio of the exoplanet host star in all frames. Then only the best 10\,\% of all images, i.e. the frames where the targets have the highest Strehl ratios, are selected for dark subtraction and flat field correction. The dark frames are always taken with the same camera settings as the science frames, but with the camera shutter closed to avoid illumination of the AstraLux detector during the dark frame acquisition. For the flatfield correction, we use both dome-flats and sky-flats taken during morning or evening twilight, whenever possible, on each observing night. The reduced images are then shifted so that in each image the brightest pixel in the speckle pattern of the target is aligned. The shifted images are finally combined to produce the fully reduced AstraLux image of the observed exoplanet host star.

All AstraLux images are astrometrically calibrated with astrometric data of the third data release \citep[Gaia DR3 from hereon][]{GaiaCollaboration2023} of the ESA Gaia mission \citep{GaiaCollaboration2016}. As astrometric calibrators, we use the central region of the globular cluster M15 (centre $\text{RA(J2000)} = 21^\text{h}29^\text{m}58^\text{s}$ and $\text{Dec(J2000)} = +12^\circ09'56''$) as well as the wide binaries HIP\,59585, HIP\,65205, HIP\,67099 and HIP\,80953.

For the astrometric calibration of the Astralux detector with M15 we use a total of 100 stars located in the AstraLux field of view around the center of the cluster, which have the most accurate astrometric positions listed in the Gaia DR3 with an average position uncertainty of 0.23\,mas. The selected astrometric reference stars have G-band magnitudes between 12.5\,mag and 15.5\,mag and are therefore all well detected in the AstraLux images of M15. The binaries complement the astrometric calibration and also ensure the stability of the instrument during the individual nights of each observing epoch.

The astrometric calibration of the instrument for the observing epochs is summarized in Table\,\ref{table_calib}. In April 2017 a different calibration has to be used for the third night (19 April 2017) than for the first two nights (17 \& 18 April 2017) due to instrument maintenance. The average pixel scale (PS) of the AstraLux detector for all observation epochs is 47.030\,mas/pixel. With the AstraLux detector size of 512\,pixel $\times$ 512\,pixel, this yields the field of view of the instrument of about 24\,arcsec $\times$ 24\,arcsec. The detector position angle\footnote{The detector position angle is the angle to be added to a measured position angle in an image to obtain the true position angle on the sky.} (DPA) is on average 1.33\,$^\circ$. We find that the absolute calibration of the instrument DPA is stable within $1.1\,^\circ$ across all our observing epochs. With dedicated calibrators we achieve a higher astrometric accuracy for individual epochs of $0.04\,^\circ$ on average.

\begin{table}
\centering
\caption{The astrometric calibration of AstraLux. The pixel scale (PS) of the instrument as well as the detector position angle (DPA) are listed for all observing epochs.}
\begin{tabular}{lcc}
\hline
Epoch               & PS [mas/pixel]     & DPA [$^\circ$]\\
\hline
Jan 2009            & $47.230 \pm 0.022$ & $-0.56 \pm 0.03$\\
Jan 2011            & $47.318 \pm 0.022$ & $-1.74 \pm 0.07$\\
Jul 2011            & $47.303 \pm 0.026$ & $-1.66 \pm 0.05$\\
Aug 2014            & $46.844 \pm 0.030$ & $-1.87 \pm 0.02$\\
Aug 2015            & $46.868 \pm 0.015$ & $-1.62 \pm 0.05$\\
Apr 2016            & $46.911 \pm 0.020$ & $-2.63 \pm 0.03$\\
Oct 2016            & $46.873 \pm 0.038$ & $-1.94 \pm 0.06$\\
Apr 2017 (17 \& 18) & $46.930 \pm 0.035$ & $-1.05 \pm 0.04$\\
Apr 2017 (19)       & $46.890 \pm 0.055$ & $+0.79 \pm 0.04$\\
Oct 2017            & $46.871 \pm 0.035$ & $-0.28 \pm 0.02$\\
May 2018            & $46.773 \pm 0.070$ & $-0.23 \pm 0.02$\\
Sep 2020            & $47.546 \pm 0.016$ & $-3.15 \pm 0.02$\\
\hline
\end{tabular} \label{table_calib}
\end{table}

\section{Target Properties}

The individual properties of the exoplanet host stars observed in this survey are compiled from various catalogues\footnote{Online available in the VizieR database:\newline \url{https://vizier.cds.unistra.fr/viz-bin/VizieR}} and the Extrasolar Planets Encyclopaedia (EPE)\footnote{Online available at: \url{http://exoplanet.eu/}} or are derived with these parameters. The parallax $\varpi$ of the targets and their proper motion in right ascension $\mu_{\rm RA}$ and declination $\mu_{\rm Dec}$ are taken from the Gaia DR3, or, if not available, from the second data release of the ESA Gaia mission \citep[Gaia DR2 from here on, ][]{GaiaCollaboration2018}, or from the new reduction of the astrometric data of the Hipparcos mission \cite[see][]{vanLeeuwen2007}. The effective temperature ($T_{\rm eff}$), surface gravity ($log(g)$) and V-band extinction estimate ($A_{\rm V}$) of the targets are from the Starhorse and Starhorse 2 catalogues \cite[SHC from hereon][]{Anders2019,Anders2022} where available. The V-band extinction estimate $A_{\rm V}$ of the targets is used to derive the extinction estimate in the i$'$-band ($A_{\rm i'}$) using the relation $A_{\rm i'} = (0.61^{+0.03}_{-0.00}) \cdot A_{\rm V}$ from \cite{Fiorucci2003}. Most of the apparent i$'$-band magnitudes of the targets are taken from the catalogues \mbox{UCAC 4} \citep{Zacharias2013}, \mbox{URAT 1} \citep{Zacharias2015}, \mbox{APASS DR 9} \citep{Henden2015}, and \mbox{APASS DR 10} \citep{Henden2018}. For some exoplanet host stars only magnitudes in the Johnson-Cousins photometric standard system are available. In this case, the apparent i$'$-band magnitude of the stars is derived using the target magnitudes and the empirical color-transformations from \cite{Jordi2006}. The absolute i$'$-band magnitude of the stars is calculated from their apparent magnitude in this band, their Gaia DR3 parallax and their i$'$-band extinction estimates. The mass of all targets, as well as their age is taken from the EPE, if available.

In Table\,\ref{table_properties} we list for all the target properties their median and extreme values, as well as the 16\,\% and 84\,\% percentiles of their distribution. The histogram and distribution of each property are shown in Figure\,\ref{fig_Histograms} and are listed for all targets in the Appendix \ref{app_catalogues}.

\begin{table}
\centering
\caption{The distance, proper motion ($\mu$), apparent magnitude \& extinction estimate ($A_{\rm i'}$) in the i$'$-band, mass, and age of all exoplanet host stars observed with AstraLux in this survey. We list the minimum (Min), maximum (Max) and median (Mdn) values, as well as the 16\,\% and 84\,\% percentiles of the distribution of the individual target properties.}
\begin{tabular}{lccccc}
\hline
                          & Min  & 16\,\% & Mdn  & 84\,\% & Max\\	
\hline
Distance [pc]             & 3    & 29     & 81   & 292    & 1473\\
$\mu$ [mas/yr]            & 4    & 20     & 85   & 276    & 4812\\
i$'$ [mag]                & 14.6 & 10.0   & 7.9  & 5.9    & 1.0\\
$A_{\rm i'}$ [mag]        & 0.00 & 0.00   & 0.10 & 0.22   & 0.50\\
Mass [M$_\odot$]          & 0.1  & 0.8    & 1.1  & 1.6    & 4.5\\
Age [Gyr]                 & 0.08 & 2      & 4    & 7      & 13\\
\hline
\end{tabular} \label{table_properties}
\end{table}

\begin{figure}
\includegraphics[width=1\columnwidth]{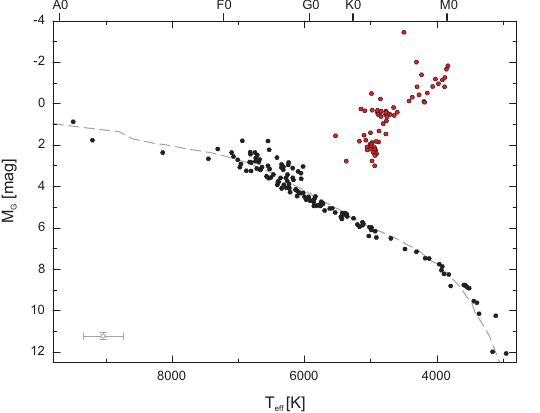}
\caption{$T_{\rm eff}$-$M_{\rm G}$-diagram of all exoplanet host stars observed in this survey and listed in the SHC, together with the main sequence from \protect\cite{Pecaut2013}, shown as gray dashed line. Red dots indicate a surface gravity of $\log(g\,[cm/s^2])<3.7$, black dots indicate $\log(g\,[cm/s^2])>3.7$.} \label{fig_HRD}
\end{figure}

\begin{figure*}
\includegraphics[width=0.329\textwidth]{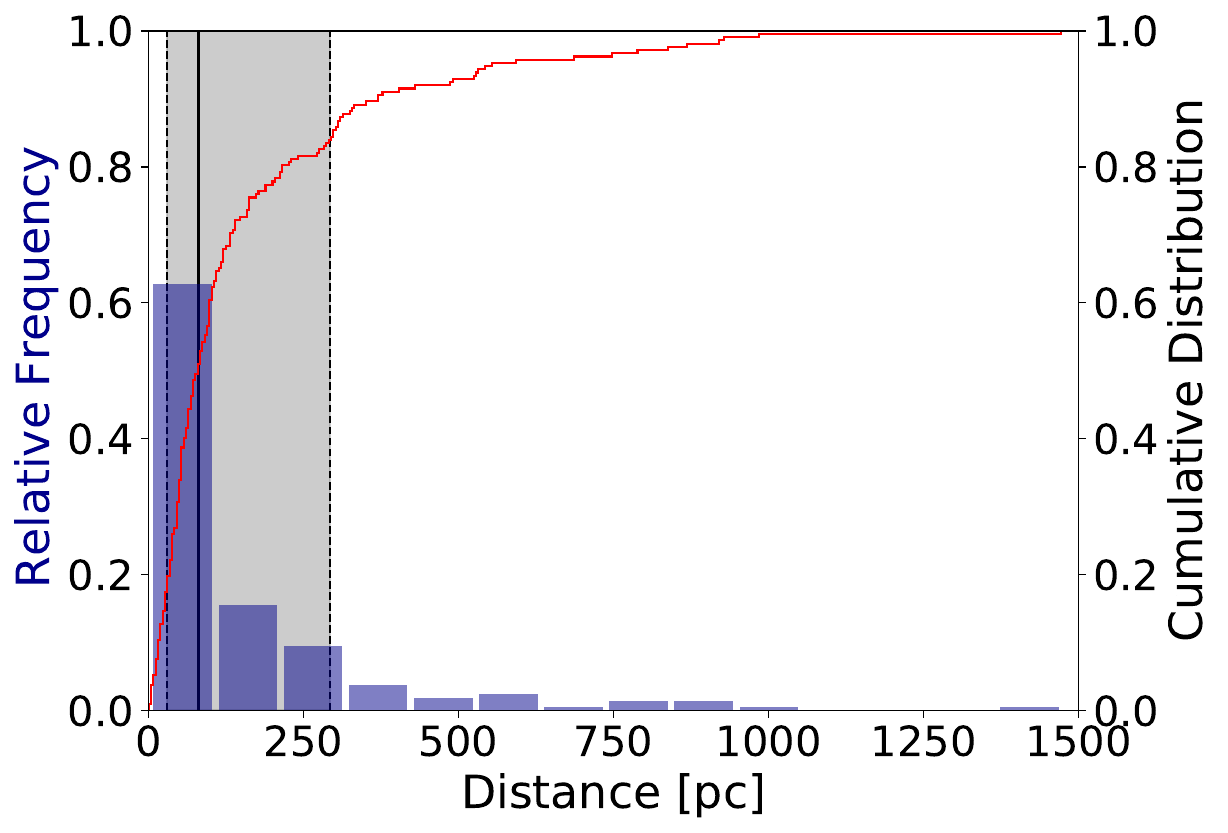}
\includegraphics[width=0.329\textwidth]{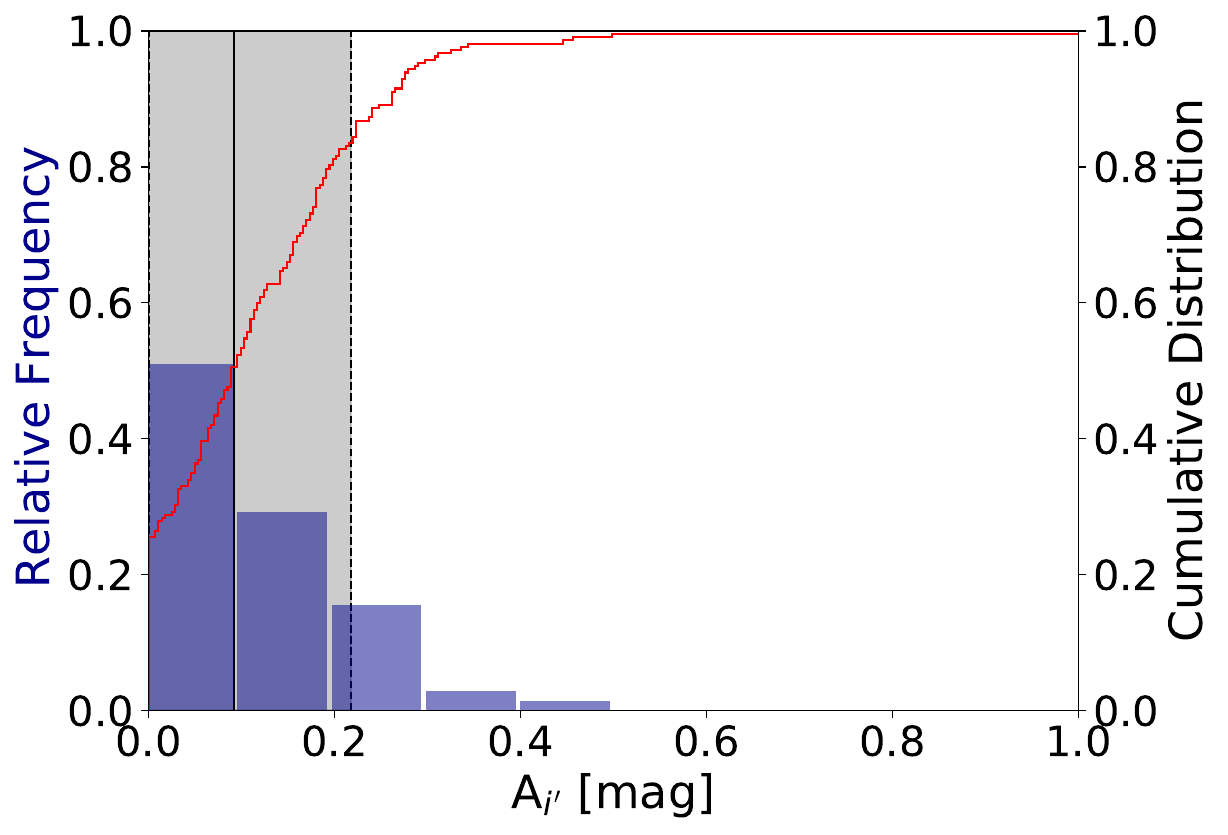}
\includegraphics[width=0.329\textwidth]{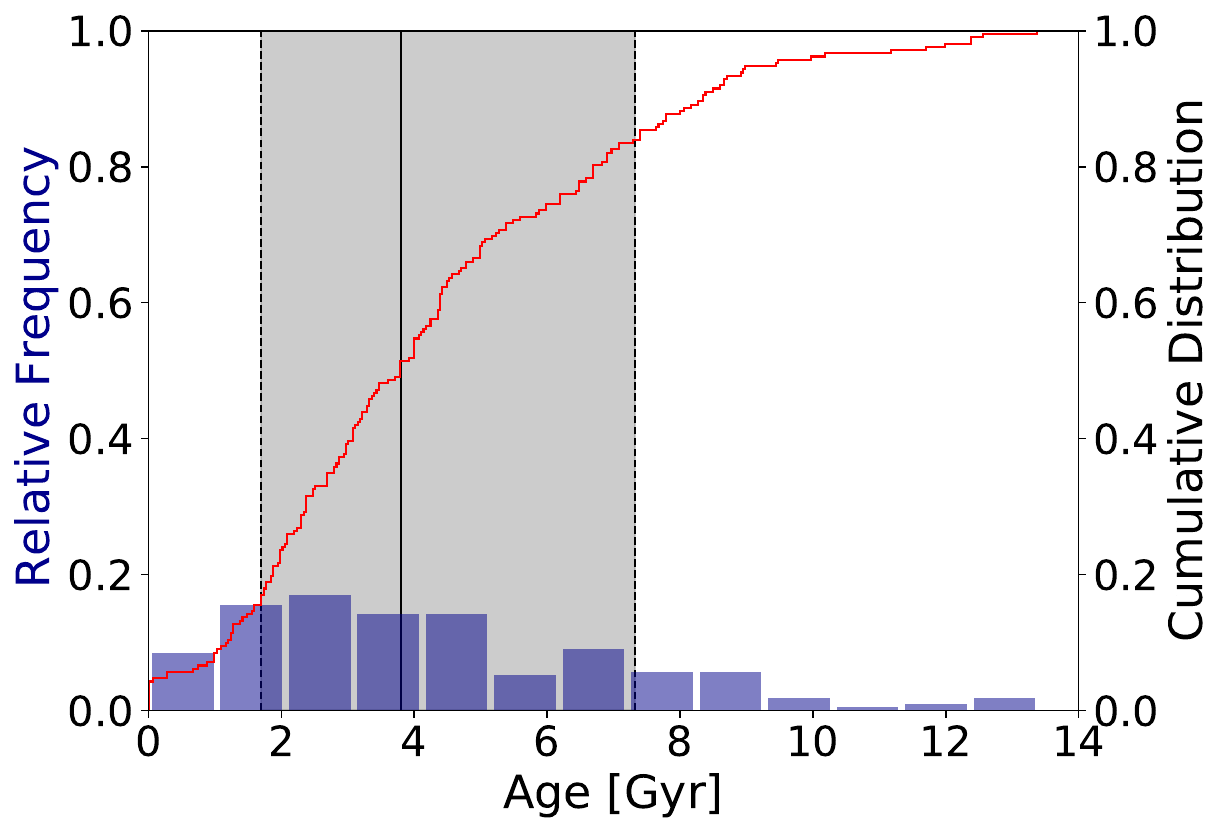}
\includegraphics[width=0.329\textwidth]{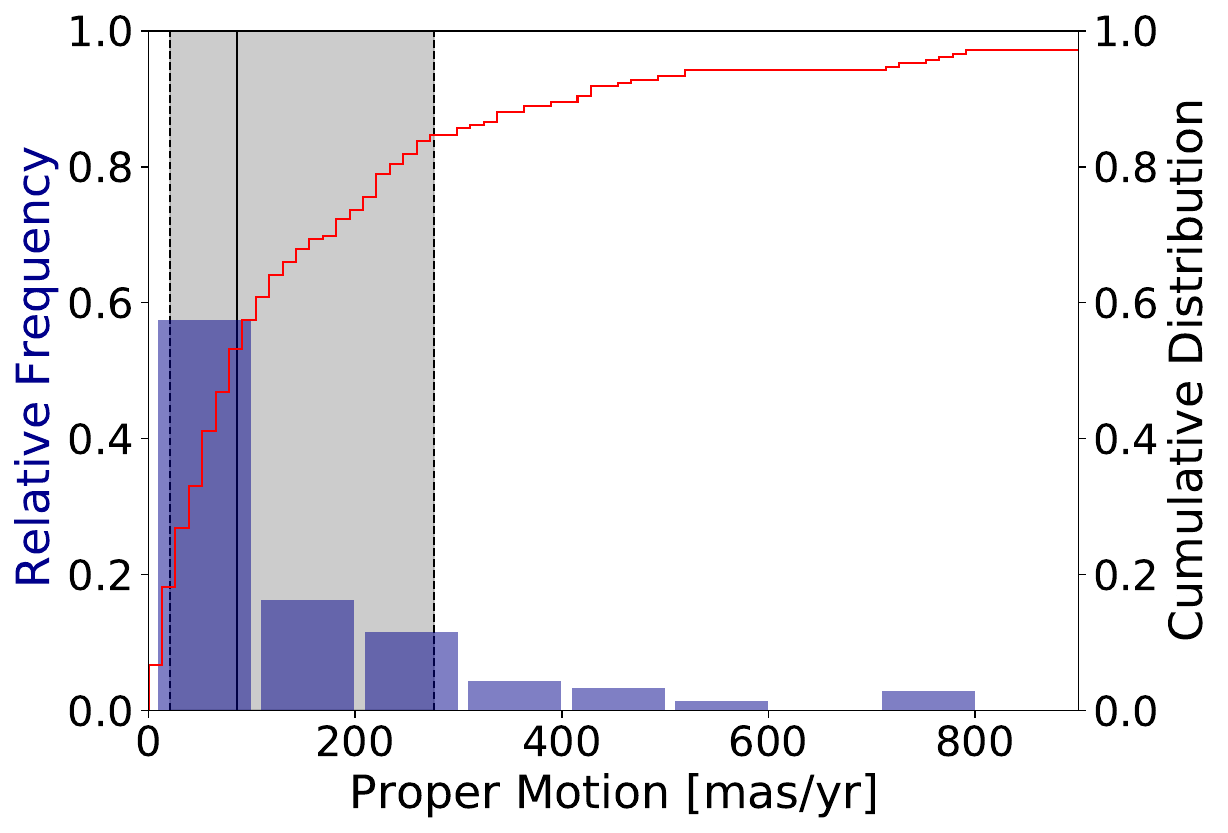}
\includegraphics[width=0.329\textwidth]{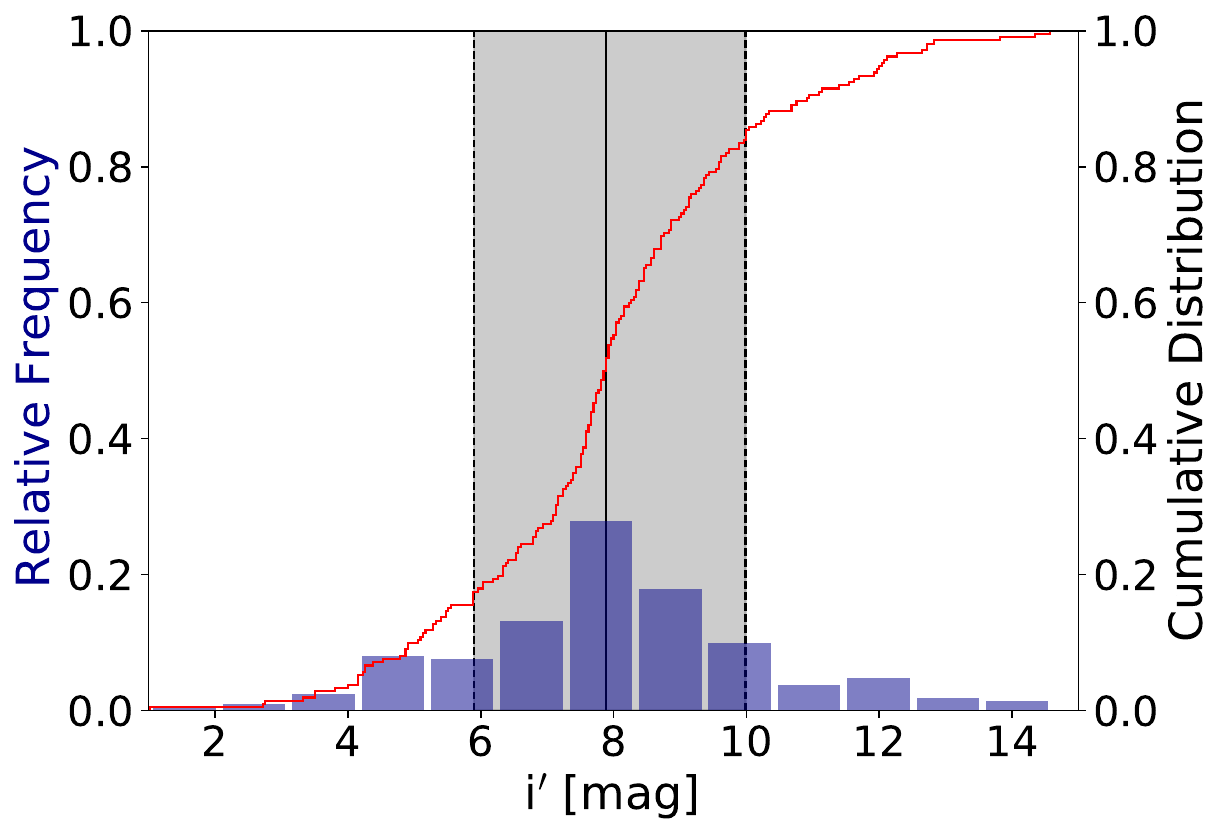}
\includegraphics[width=0.329\textwidth]{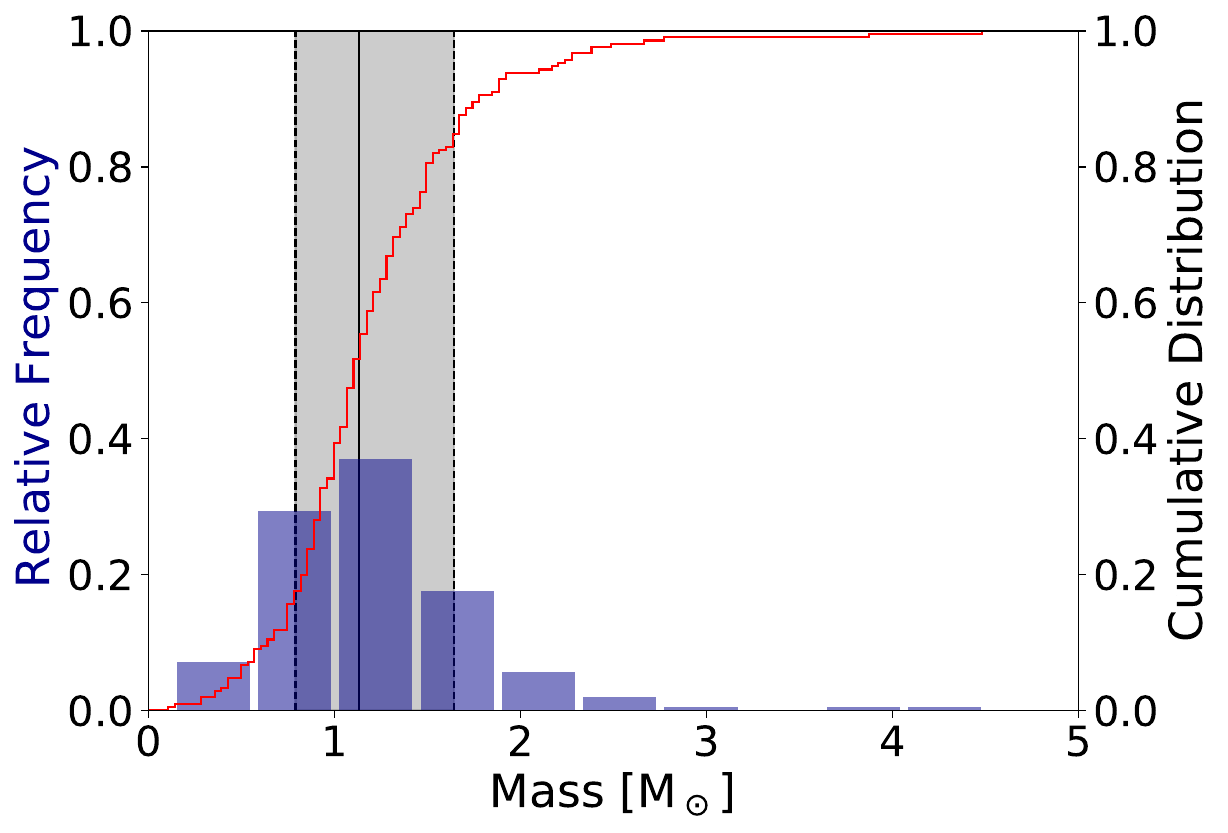}
\caption{Histograms and cumulative distribution functions of the target properties observed in this survey. The solid vertical lines show the median values and the dashed lines show the 16\,\% and 84\,\% percentiles of the distribution of the individual target properties. Note that 3\,\% of all targets exceed a proper motion of 900\,mas/yr.} \label{fig_Histograms}
\end{figure*}

In order to characterize the evolutionary stage of the targets of this survey, we present them in a $T_{\rm eff}$-$M_{\rm G}$-diagram, which is shown in Figure\,\ref{fig_HRD}. For exoplanet host stars listed in the SHC (199 stars), their absolute G-band magnitude ($M_{\rm G}$) is plotted against their effective temperature ($T_{\rm eff}$). The surface gravity ($\log(g)$) of the stars is shown with different color. For orientation, the main sequence from \cite{Pecaut2013} is plotted as a grey dashed line. The majority of all targets (67\,\% of all) are main sequence stars, the others that are above the main sequence and have a lower surface gravity ($\log(g\,[cm/s^2])<3.7$) are evolved stars or even giants already.

The exoplanet host stars observed in this survey are old ($\rm{Mdn}(\rm{age})=4$\,Gyr), predominantly solar-like stars ($\rm{Mdn}(\rm{mass})=1.1\,\rm M_\odot$) located in the solar neighbourhood at a mean distance of 81\,pc as derived from their Gaia DR3 parallax. On the one hand, the proximity of the targets explains their high proper motion ($\rm{Mdn}(\mu)=85$\,mas/yr), which facilitates the detection of co-moving companions and their separation from unrelated and slow-moving background sources. On the other hand, it explains the high apparent brightness of the exoplanet host stars ($\rm{Mdn(i')}=7.9$\,mag), which makes them ideal targets for Lucky Imaging observations, as well as their low i$'$-band extinction, which does not exceed 0.5\,mag. Therefore, for all targets not listed in the SHC, we assume $A_{\rm i'}=0$\,mag. For those targets for which no age information is available (9 stars), we adopt the median age of the target sample of 4\,Gyr.

\section{Image Processing and Data Analysis}

All additional point sources, that are detected with AstraLux around the much brighter exoplanet host stars, are considered here as companion-candidates to these stars. In order to accurately measure the astrometry and photometry of the companion-candidates, we subtracted the PSF of the exoplanet host stars from all fully reduced AstraLux images using the roll subtraction technique \cite[described in ][]{Ginski2014, Mugrauer2022a}. The PSF subtraction and the astrometric and photometric measurements was performed with \textsc{ESO-Midas}.

The astrometry and photometry of the exoplanet host star were measured in the fully reduced AstraLux images, those of the detected companion-candidates in the PSF subtracted images. The position of all sources was determined by fitting a two-dimensional Gaussian function to their PSF. With the astrometric calibration of the AstraLux detector we derived for all detected companion-candidates their relative astrometry (angular separation and position angle) to the exoplanet host stars. All astrometric measurements are listed in Table\,\ref{tab_astr1}, \ref{tab_astr2}, \ref{tab_astr3} and \ref{tab_astr4}. On median, the angular separation measurements have an uncertainty of 11\,mas, those of the position angle of 0.14\,$^\circ$.

The differential photometry between the exoplanet host stars and the detected companion-candidates was determined using aperture photometry. We used an aperture diameter set to twice the full width half maximum (FWHM) of the PSF of the exoplanet host stars, which gives the best photometric accuracy. The background flux was measured in a circular annulus around the aperture with inner and outer annulus diameters of typically 5.6 and 6.1\,arcsec, wide enough to avoid background contamination by the detected sources. The obtained differential magnitude of the detected companion-candidates and the given apparent i$'$-band magnitude of the targets yields the apparent i$'$-band magnitude of the candidates, which is summarized in Table\,\ref{tab_astr3} and \ref{tab_photo1}. On median, the apparent photometry of the detected companion-candidates is determined at the 0.09\,mag level.

\section{Detected Stellar Companions}\label{sec_comp}

\begin{figure}
\includegraphics[width=1\columnwidth]{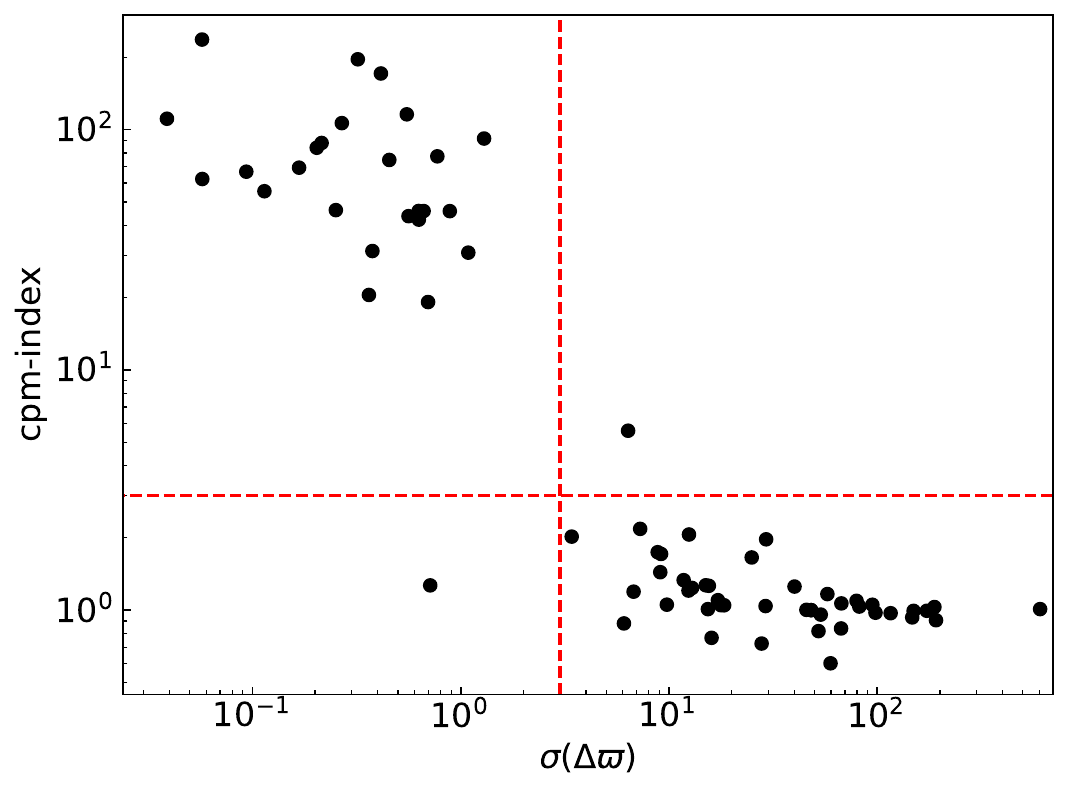}
\caption{Common proper motion index (cpm-index) over the significance of the parallax difference for all detected companion-candidates also listed in the Gaia DR3. The dashed red lines indicate the $\sigma(\Delta\varpi)=3$ and $\text{cpm-index}=3$ thresholds for companionship. Objects in the upper left corner of the plot are classified as equidistant co-moving companions of the exoplanet host stars.} \label{fig_Gaia_comoving}
\end{figure}

For each companion-candidate detected around the exoplanet host stars within the AstraLux field of view we performed the astrometric test for companionship. Several of these candidates are already listed with parallaxes and proper motions in the Gaia DR3. In this case, we use the Gaia astrometry of the detected companion-candidates to test for equidistance and common proper motion with the exoplanet host stars. As defined in \cite{Mugrauer2019}, the equidistance criterion is met if the parallaxes of the candidates and targets are within 3\,$\sigma$. This takes into account the uncertainty of the parallax as well as the astrometric excess noise of the individual objects. Accordingly, the degree of common proper motion of the detected candidates is characterized by the common proper motion (cpm) index:
\begin{equation}
\text{cpm-index} = \left| \boldsymbol{\mu}_{\text{PH}} + \boldsymbol{\mu}_{\text{CC}} \right|/\left| \boldsymbol{\mu}_{\text{PH}} - \boldsymbol{\mu}_{\text{CC}} \right|
\end{equation}
with the proper motion of the exoplanet host stars $\boldsymbol{\mu}_{\text{PH}}$ and the proper motion of the detected companion-candidates $\boldsymbol{\mu}_{\text{CC}}$. Detected candidates are considered to be co-moving companions of the observed exoplanet host stars if $\text{cpm-index} \geq 3$. In addition, we estimated the escape velocity $\mu_\text{esc}$ of all detected co-moving and equidistant companions and compared it with their differential proper motion $\mu_\text{rel}$ to the exoplanet host stars. All results are listed in Table\,\ref{tab_astr1} and \ref{tab_astr3}, and are visualized in Figure\,\ref{fig_Gaia_comoving}. We plot the cpm-index against the significance of the parallax difference. The red dotted lines indicate the threshold of companionship described above. All candidates appearing in the upper left-hand corner of the plot are equidistant co-moving companions of the exoplanet host stars. In total, we have applied this test to 71 companion-candidates and detected 25 co-moving companions that also have Gaia DR3 astrometry confirming their companionship with the observed exoplanet host stars. The other 46 companion-candidates turned out to be background stars.

For 20 of these companions, their proper motion differs significantly from that of the associated exoplanet host star ($\sigma(\mu_\text{rel})>3$), indicating the detection of orbital motion in these stellar systems. It is indicated in Table\,\ref{tab_astr1}, in which systems orbital motion is significantly detected. In the case of the companion HD\,132563\,A(SB), the differential proper motion of the companion even exceeds the estimated escape velocity. This is because the escape velocity is estimated for binary systems composed of the exoplanet host star and the detected companion. In contrast, HD\,132563 is a hierarchical triple system, with the A component being a close binary system itself. In this context, it is worth noting that none of the companions also detected by Gaia have a non-single star solution in the Gaia DR3, i.e. there is no indication of additional close companions of these stars based on their Gaia DR3 astrometry.

We then derived the masses of the detected co-moving companions from their apparent i$'$-band photometry, measured as described above in our AstraLux images. The absolute i$'$-band magnitudes of the companions were derived from their apparent i$'$-band magnitude, the i$'$-band extinction of the associated exoplanet host star, and its parallax, which gives the distance modulus. The mass of the companions is then determined from their absolute i$'$-band magnitude using the stellar evolutionary models of low-mass stars from \cite{Baraffe2015}. Since the evolutionary models provide the relation between the absolute I-band magnitude in the Johnson-Cousins photometric standard system and the stellar mass, we transform them to the i$'$-band magnitude using the color-transformation relations from \cite{Jordi2006}. The isochrone closest to the age of the exoplanet host star is always used, and the mass-magnitude relation is completed by linear interpolation. The masses of the companions derived with the 4\,Gyr isochrone differ by only 0.1\,\% on median from those determined with the correct isochrone, regardless of the true age of the star. This justifies the above approximation to assume an age of 4\,Gyr for all stars for which no age is available. The mass estimate of all detected companions can be found in Table\,\ref{tab_photo1}. The masses of the companions are determined with a precision of 0.02\,$\rm M_\odot$, in median.

For detected companion-candidates whose parallax and proper motion are not listed in the Gaia data releases, their companionship can be tested by comparing their relative AstraLux astrometry from different observing epochs with the co-moving and background hypothesis. For the background hypothesis, we assume that a distant background star has $\varpi=0$\,mas and $\mu=0$\,mas/yr, and thus the relative motion is caused only by the proper and parallactic motion of the exoplanet host star. Thus, for an unrelated background object, its angular separation and position angle to the target should change with time, while for a perfectly co-moving companion both should remain constant. However, due to orbital motion, both the angular separation and the position angle of a co-moving companion can vary over time, limited by the escape velocity of the stellar system. This means that a deviation from the strict co-moving hypothesis is possible due to orbital motion, which is estimated here by assuming a circular orbit with a radius equal to the projected separation of the companion. While the maximum change in position angle is derived by assuming a face-on orbit, the maximum change in separation is estimated by assuming an edge-on orbit.

\begin{figure*}
\includegraphics[width=0.49\textwidth]{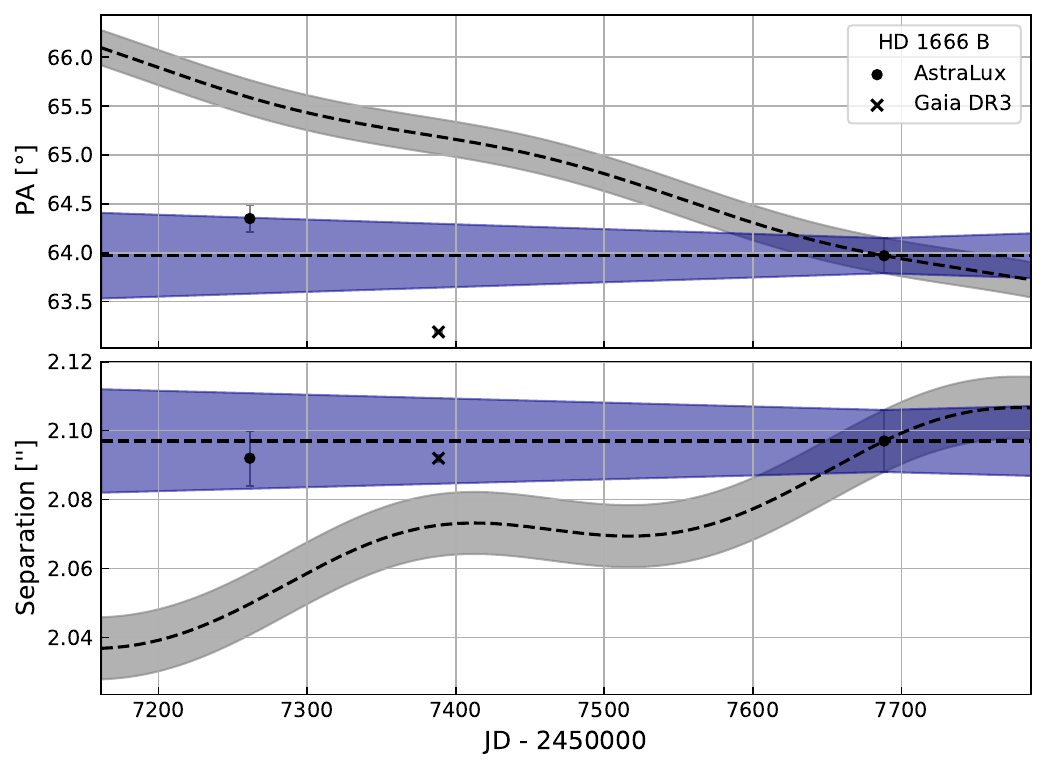}
\includegraphics[width=0.49\textwidth]{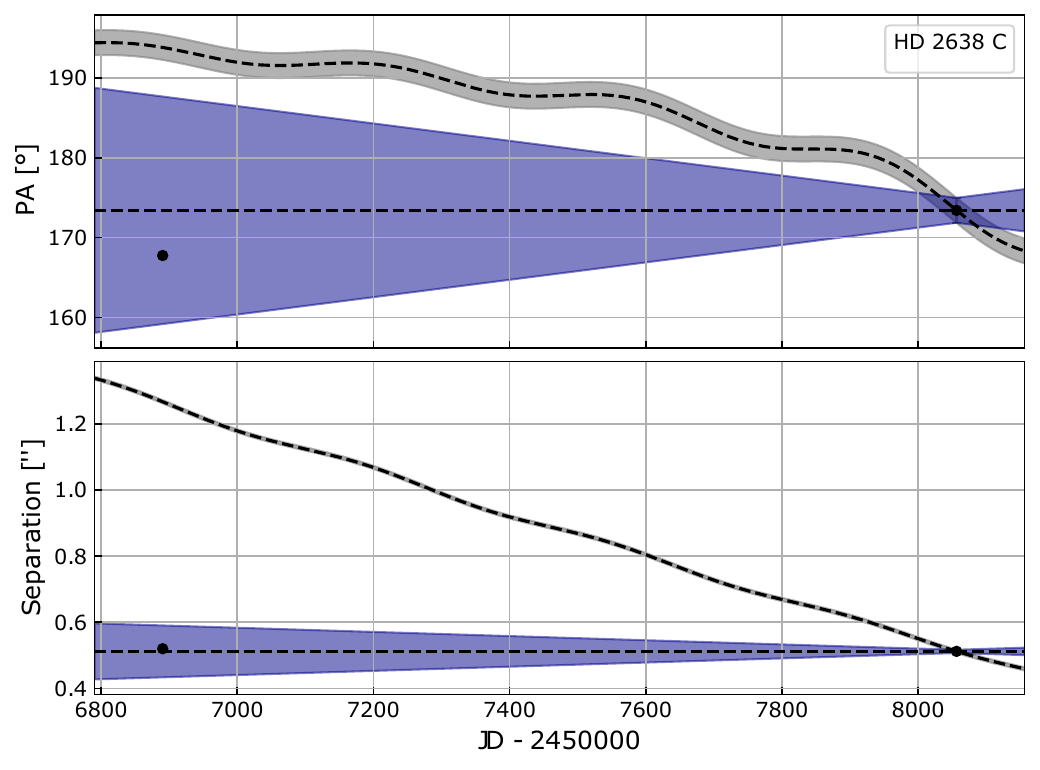}
\includegraphics[width=0.49\textwidth]{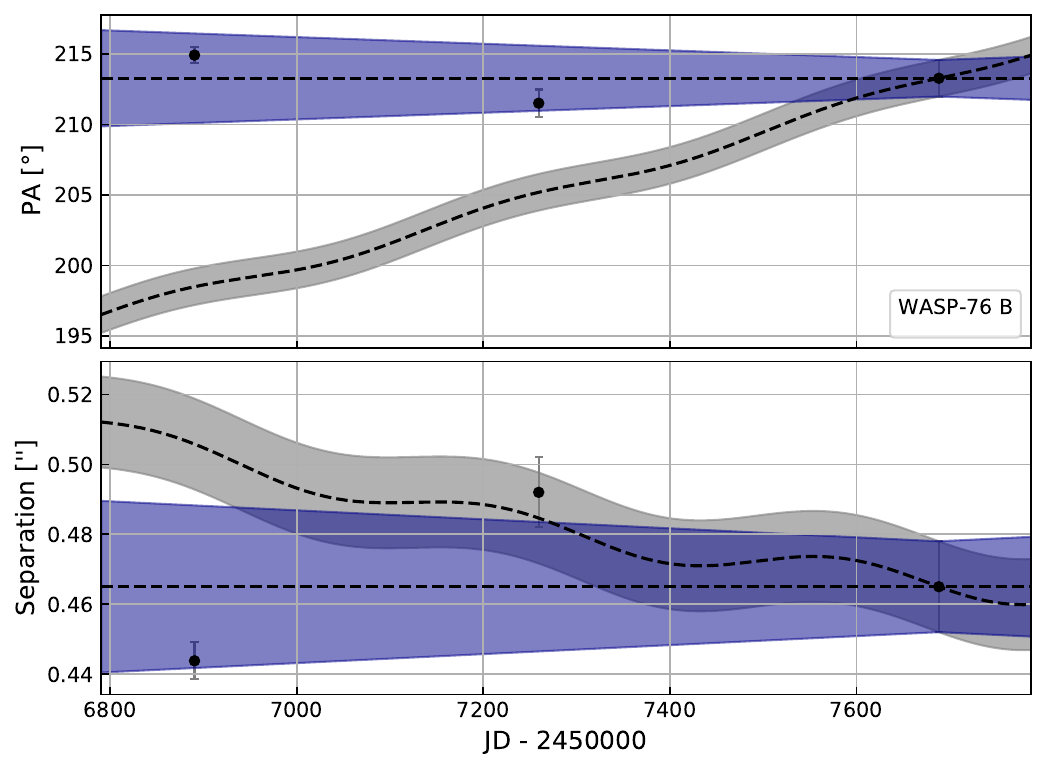}
\includegraphics[width=0.49\textwidth]{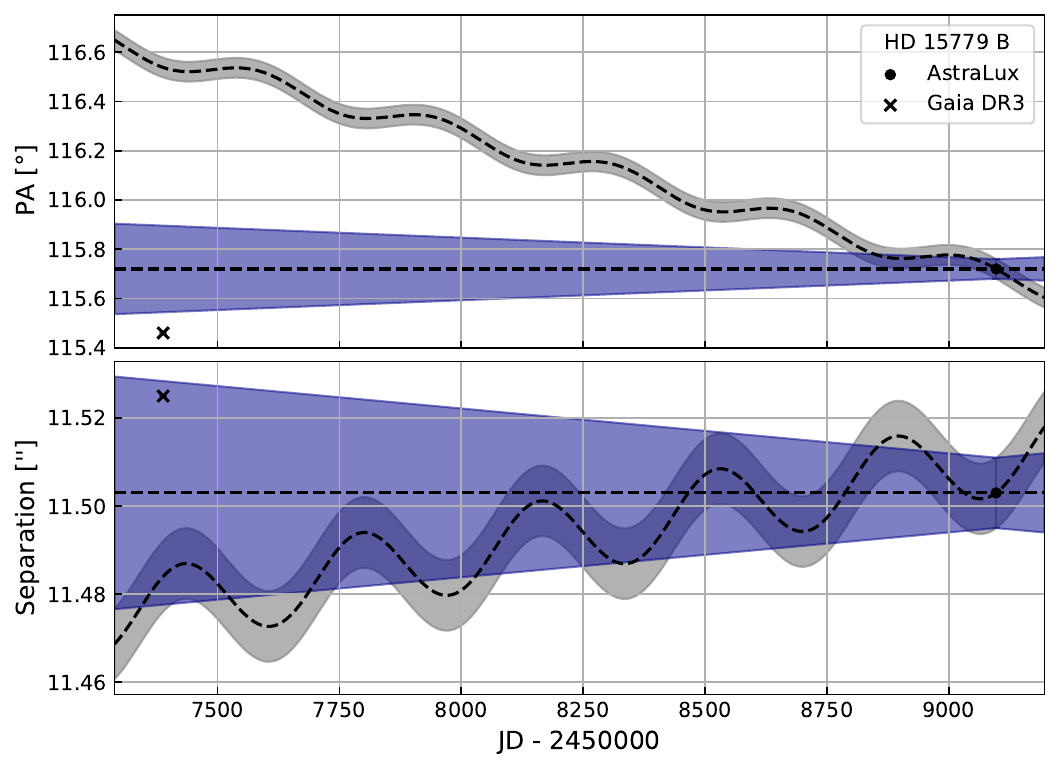}
\includegraphics[width=0.49\textwidth]{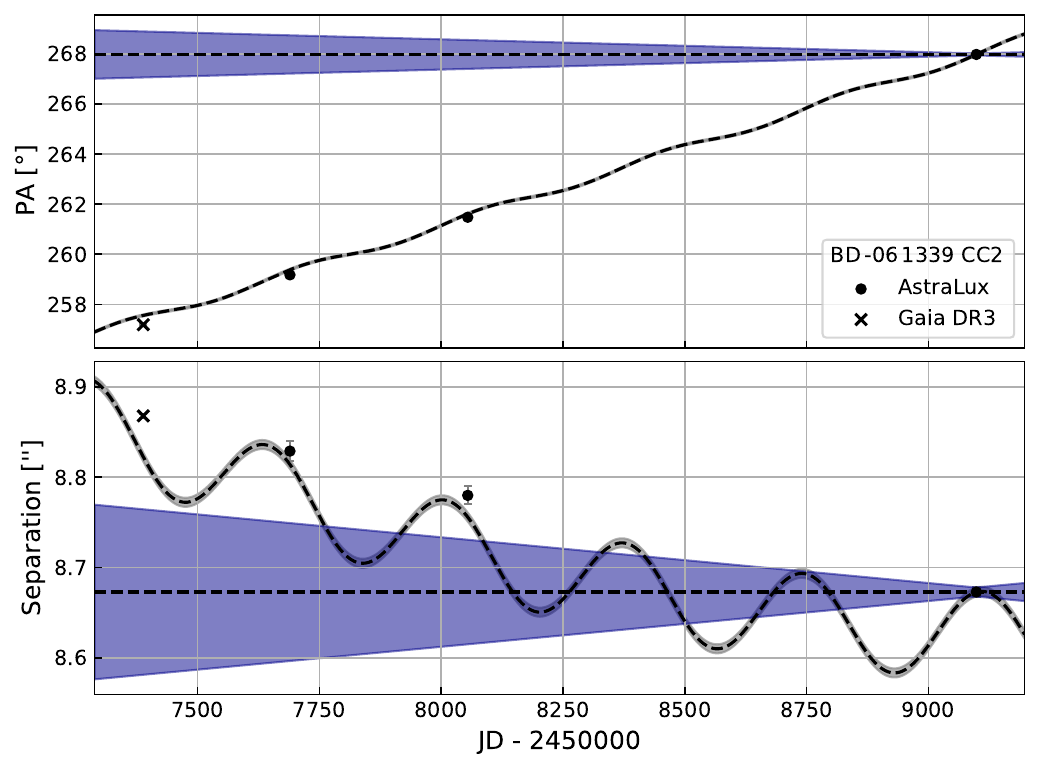}
\includegraphics[width=0.49\textwidth]{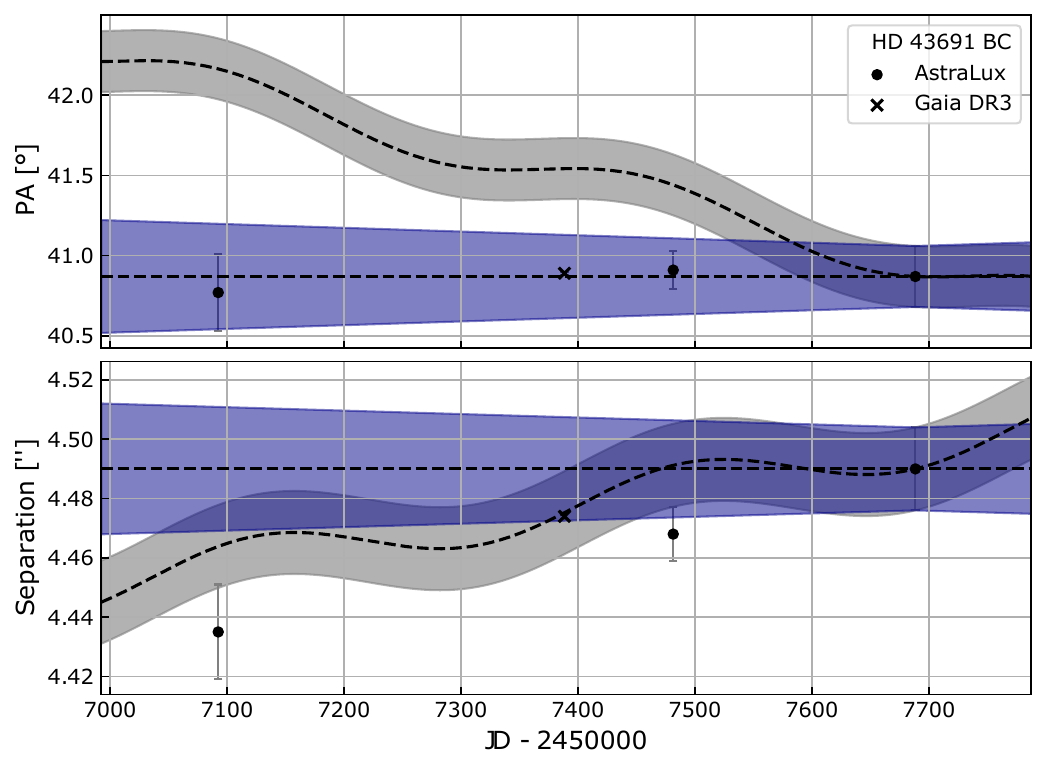}
\caption{Angular separation and position angle (PA) of companion-candidates over time. The grey area represents the expected motion of a background object and the blue area that of a co-moving companion on a gravitationally bound orbit.} \label{fig_comoving}
\end{figure*}

The reference epoch is always the most recent AstraLux observation epoch and the relative motion of a detected companion-candidate is calculated backwards. The astrometric measurements of all 19 detected companion-candidates are shown in Figure\,\ref{fig_comoving}, together with the background and the co-moving hypothesis. The dashed line in the blue area represents the strict co-moving hypothesis, where the blue area is the estimate of possible orbital motion. The dashed line in the grey region indicates the expected relative motion of a background star. Where available, the relative astrometry of the companions, derived from Gaia DR3 astrometry, is included in the plots shown in Figure\,\ref{fig_comoving}. In the case of HD\,87646 two positions from the Washington Double Star Catalogue \cite[WDS,][]{Mason2001} are included, because the epoch difference of the AstraLux observations of this exoplanet host star is too close to test the common proper motion of the detected companion. In Table\,\ref{tab_astr2} and \ref{tab_astr4} show the significance level at which the background and strict co-moving hypotheses can be rejected. For all 17 detected co-moving companions the background hypothesis can be rejected by more than 3\,$\sigma$. In addition, the strict co-moving hypothesis can be significantly rejected for eight of these companions, indicating the detection of orbital motion for these companions. Out of the 19 companion-candidates, two were identified as background stars.

\begin{figure*}
\includegraphics[width=0.49\textwidth]{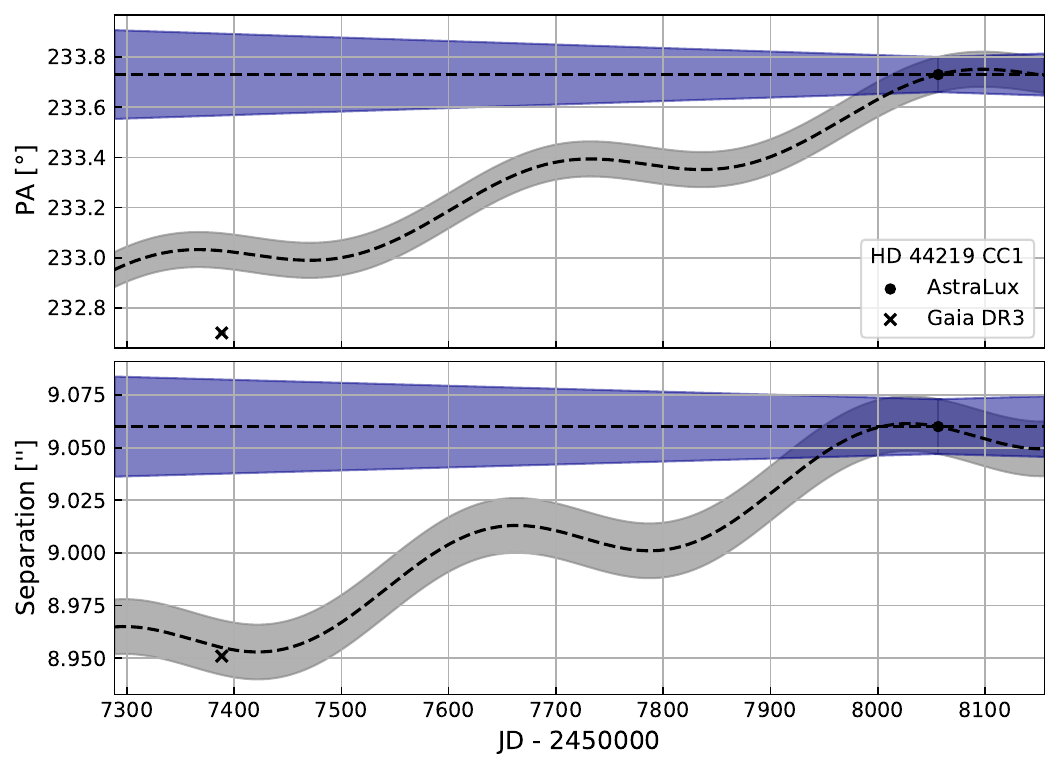}
\includegraphics[width=0.49\textwidth]{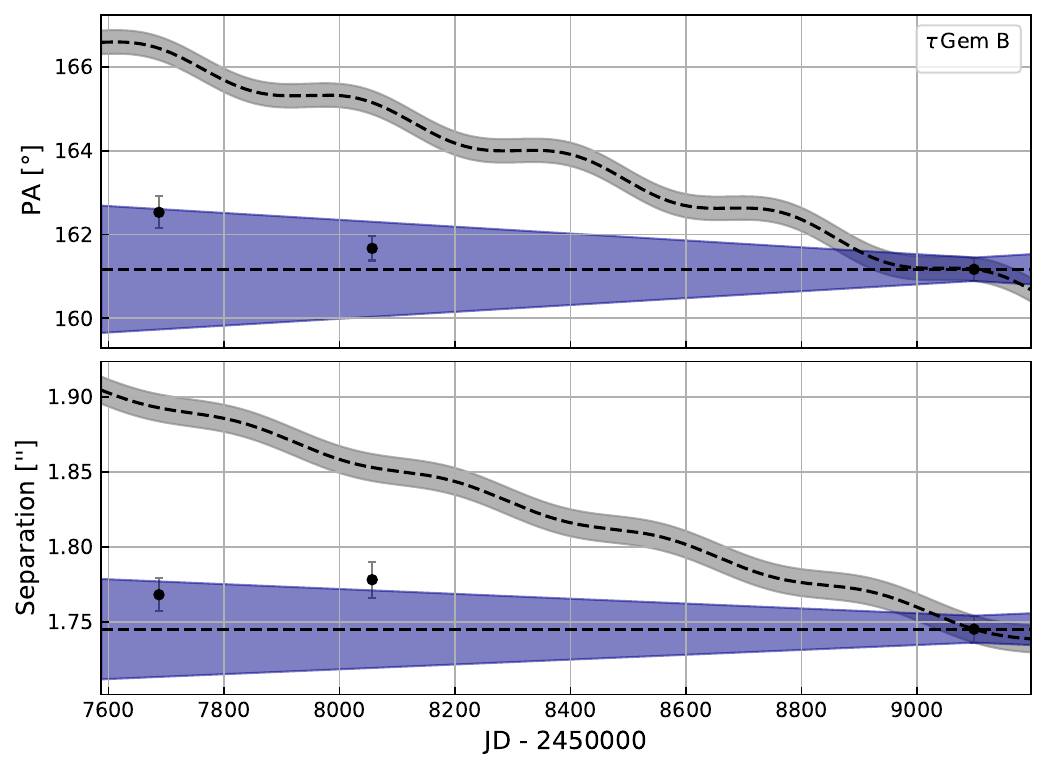}
\includegraphics[width=0.49\textwidth]{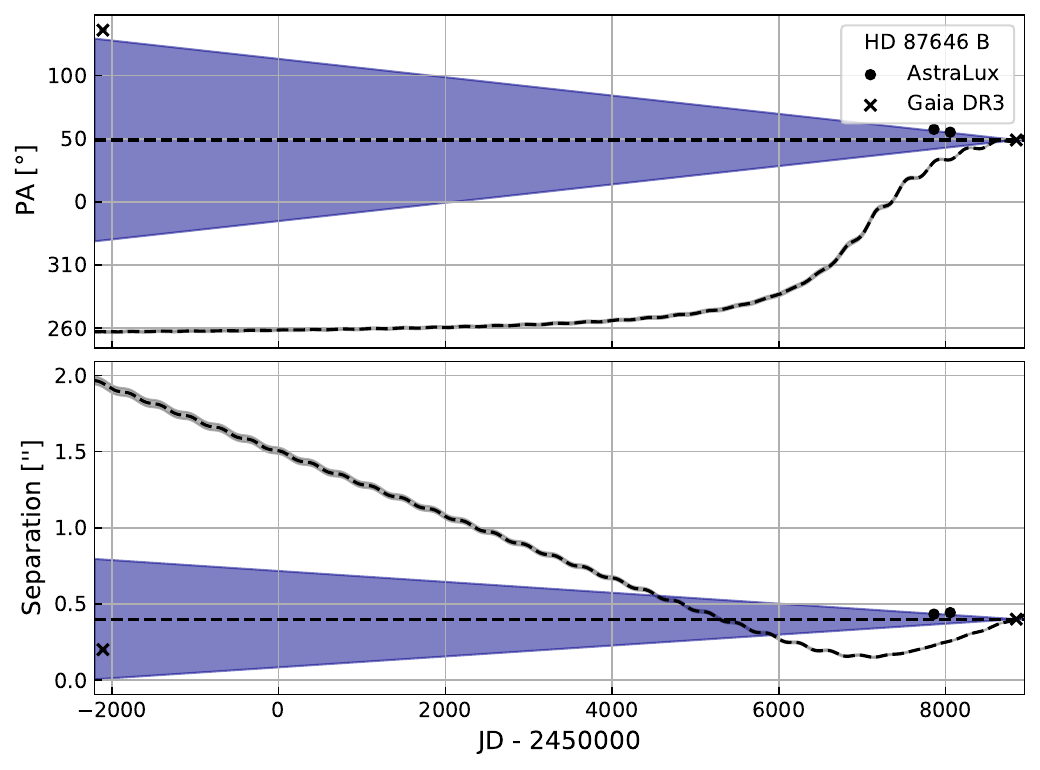}
\includegraphics[width=0.49\textwidth]{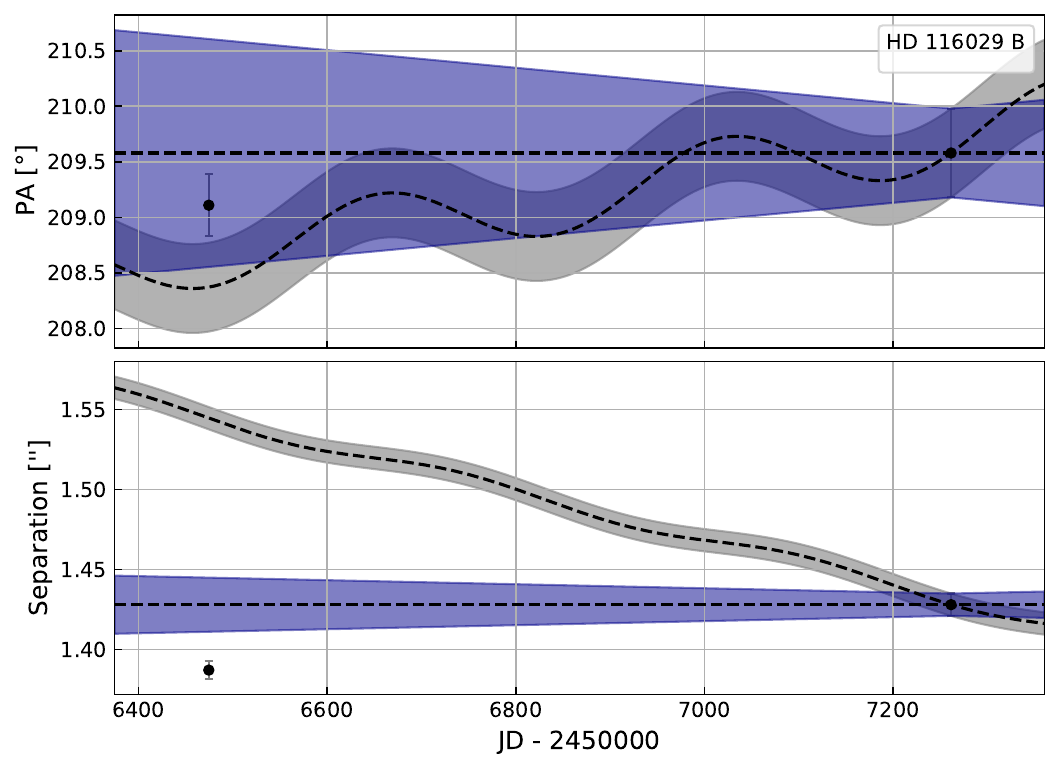}
\includegraphics[width=0.49\textwidth]{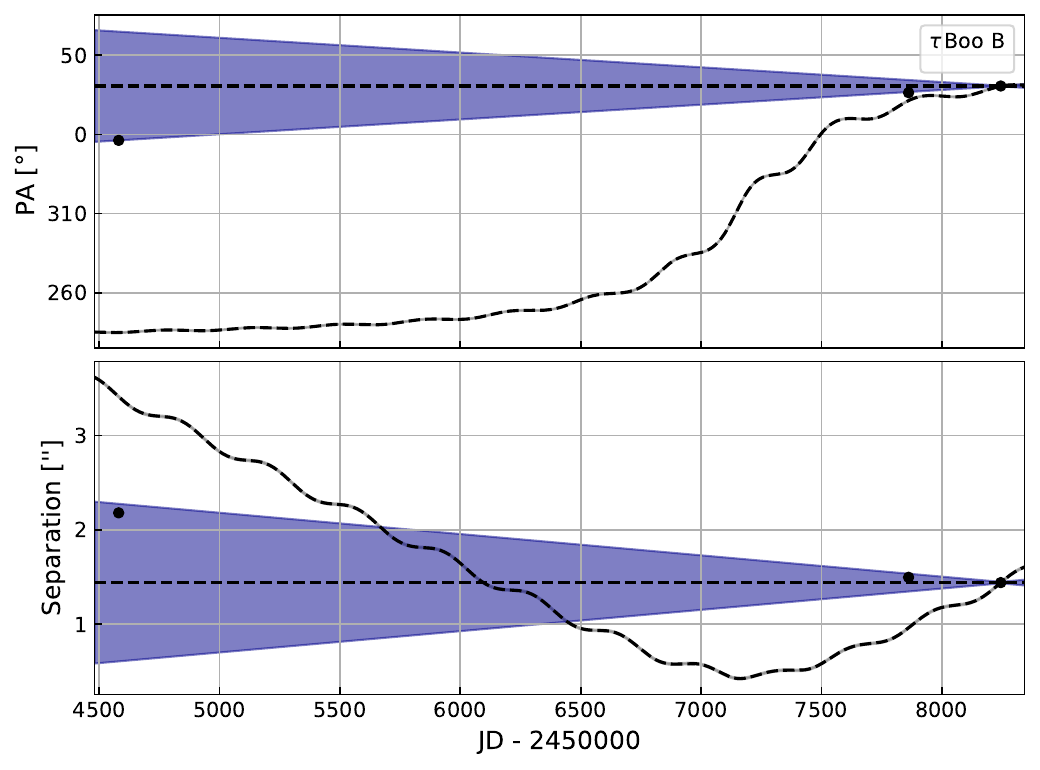}
\includegraphics[width=0.49\textwidth]{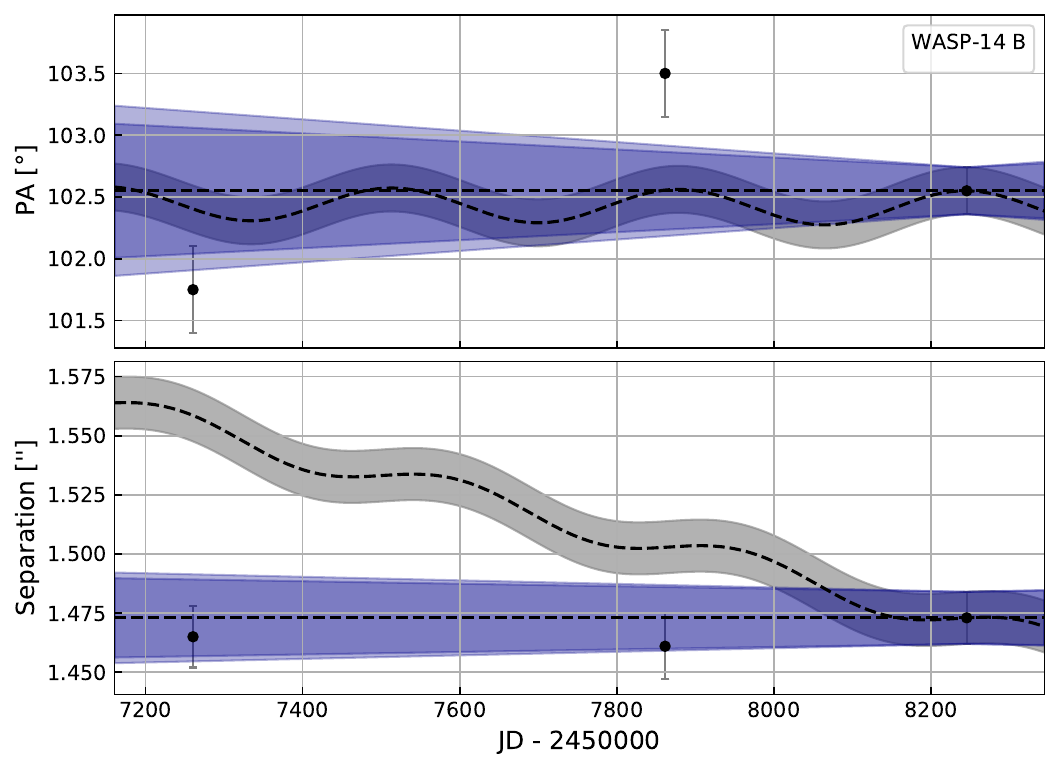}

\textbf{Figure\,\ref{fig_comoving}.} continued.
\end{figure*}
\begin{figure*}
\includegraphics[width=0.49\textwidth]{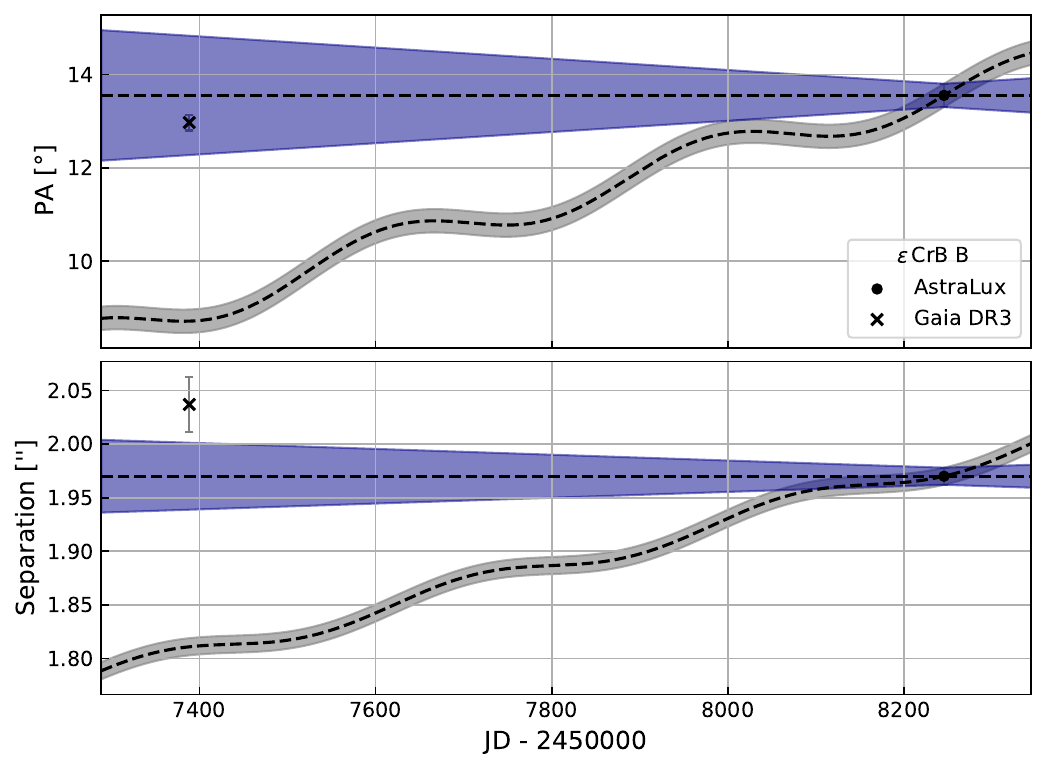}
\includegraphics[width=0.49\textwidth]{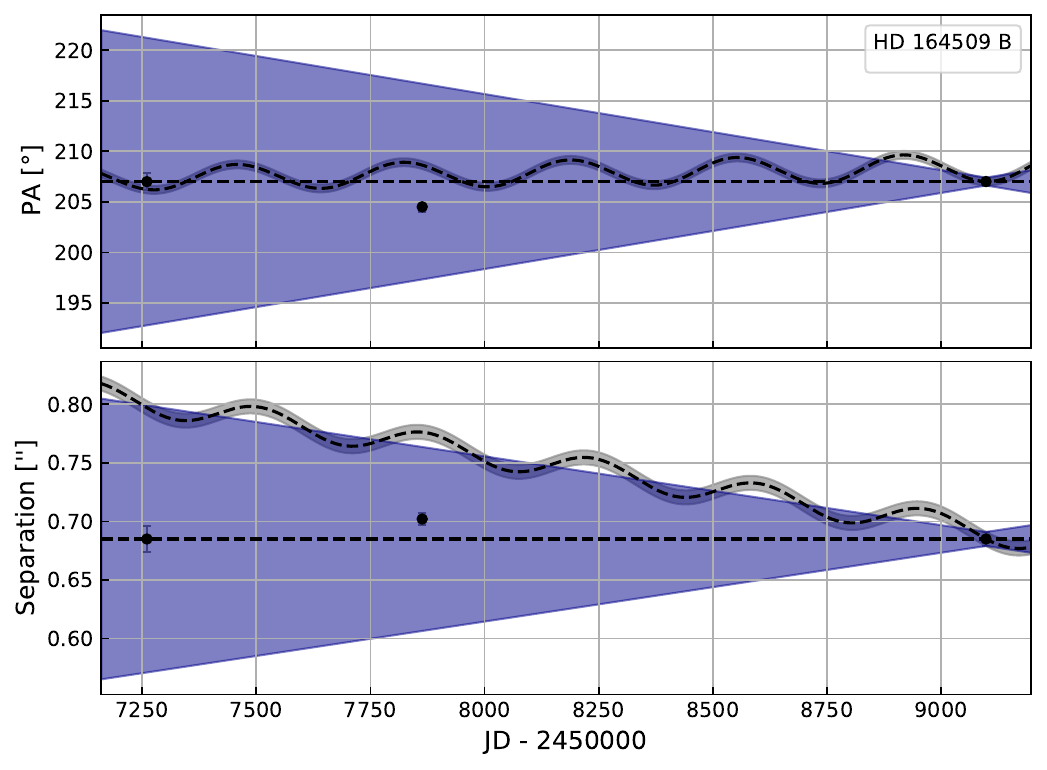}
\includegraphics[width=0.49\textwidth]{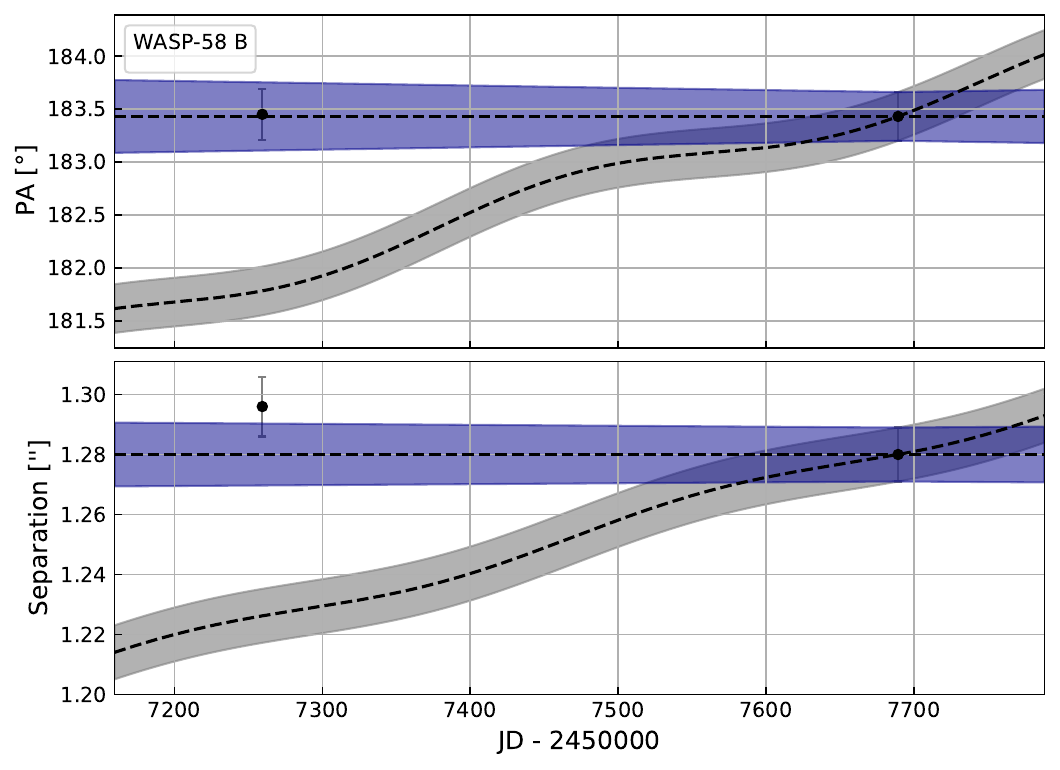}
\includegraphics[width=0.49\textwidth]{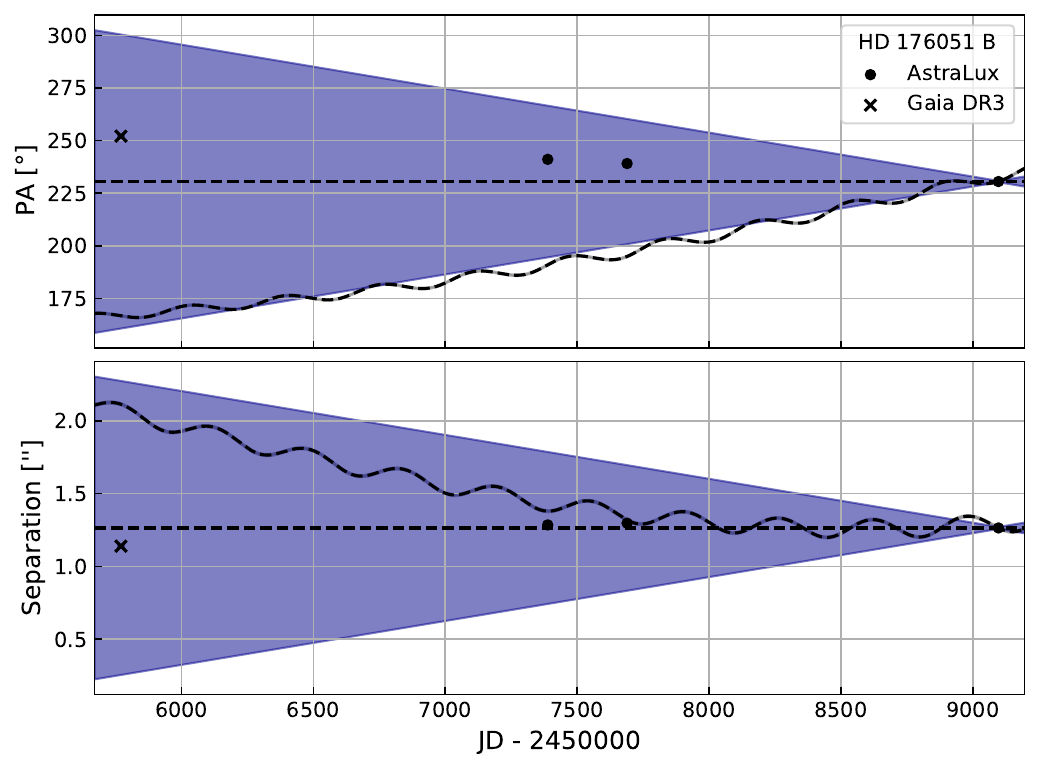}
\includegraphics[width=0.49\textwidth]{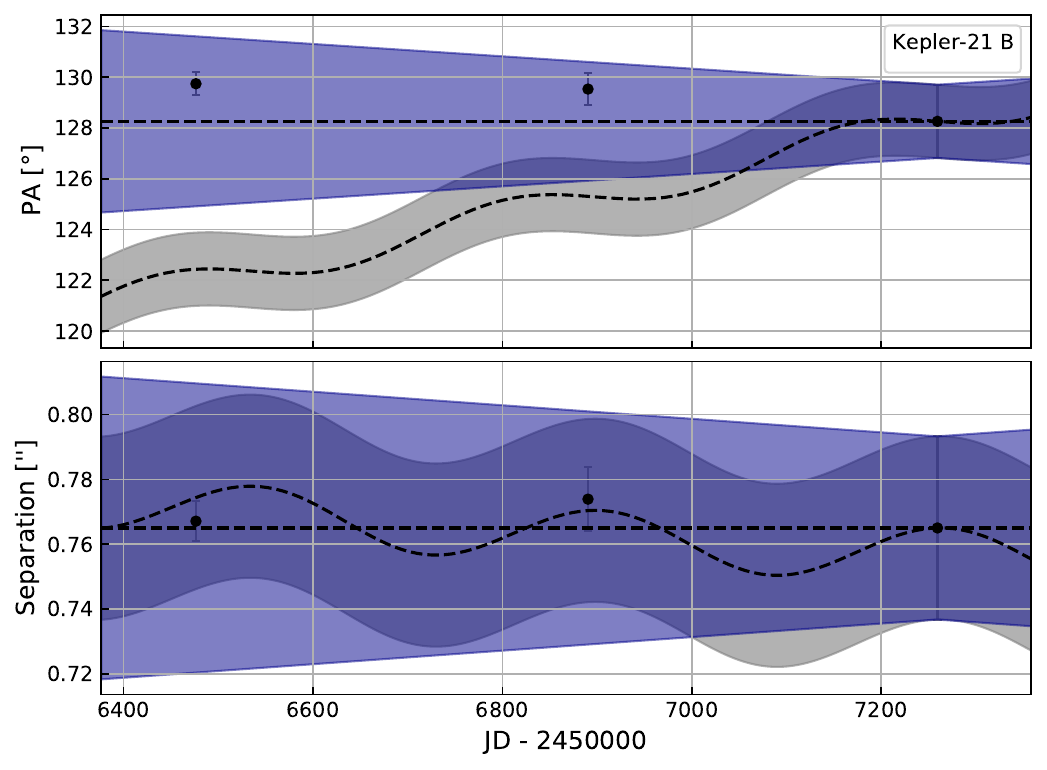}
\includegraphics[width=0.49\textwidth]{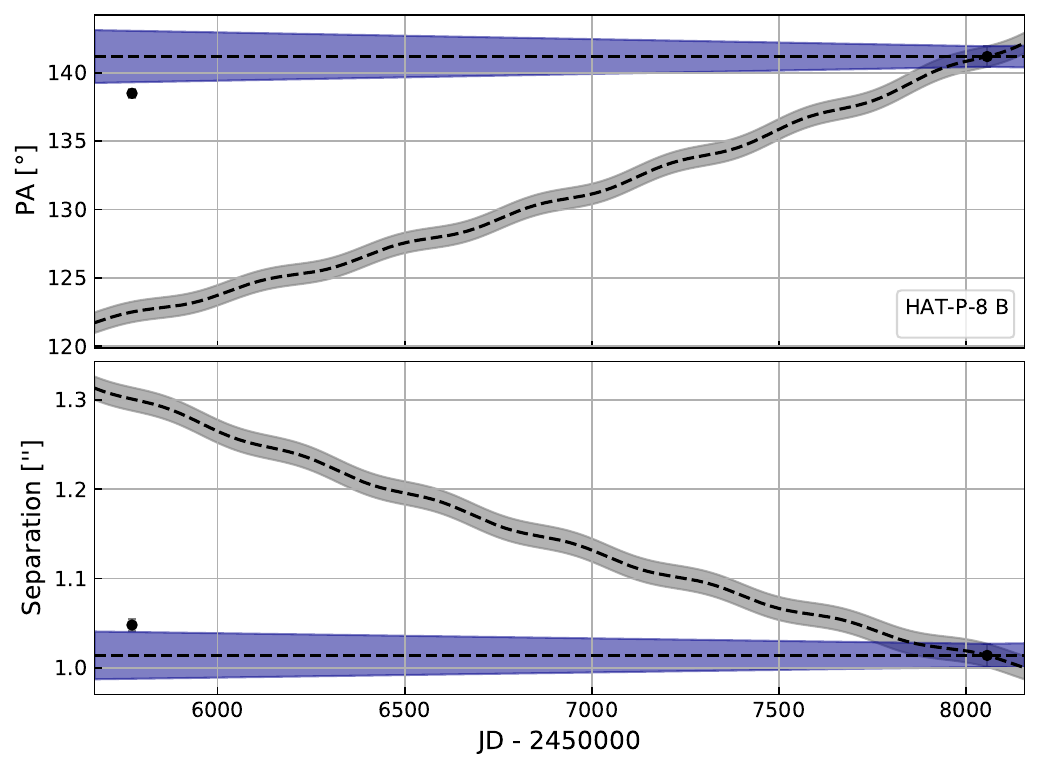}

\textbf{Figure\,\ref{fig_comoving}.} continued.
\end{figure*}
\begin{figure}
\includegraphics[width=0.49\textwidth]{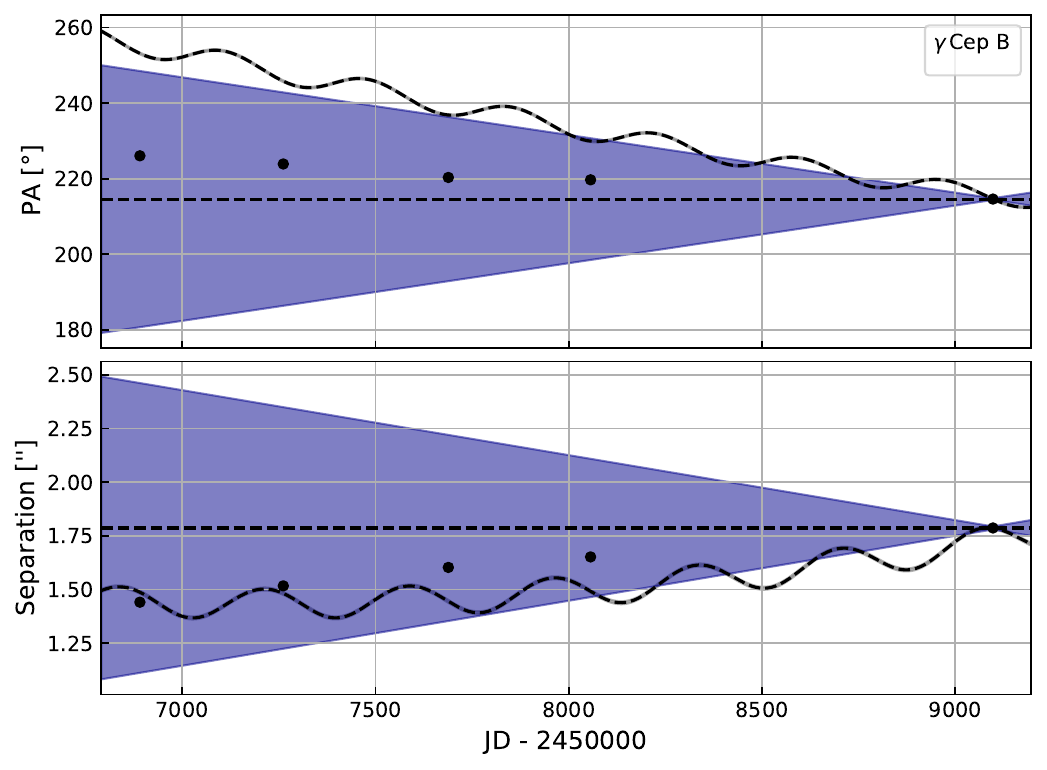}

\textbf{Figure\,\ref{fig_comoving}.} continued.
\end{figure}

Additionally, there are four companion-candidates, which have no parallaxes and proper motions in the Gaia DR3 and there is only data from one observation epoch available. Nevertheless, these candidates (HAT-P-35\,C, $\gamma\,$Leo\,B, HAT-P-57\,B, and HAT-P-57\,C) can be classified as companions of the associated exoplanet host stars, as their companionship has already been confirmed in other surveys \citep{Woellert2015,Romanenko2014,Bohn2020}. Their AstraLux astrometry is listed in Table\,\ref{tab_astr_extra}.

\begin{figure}
\includegraphics[width=1\columnwidth]{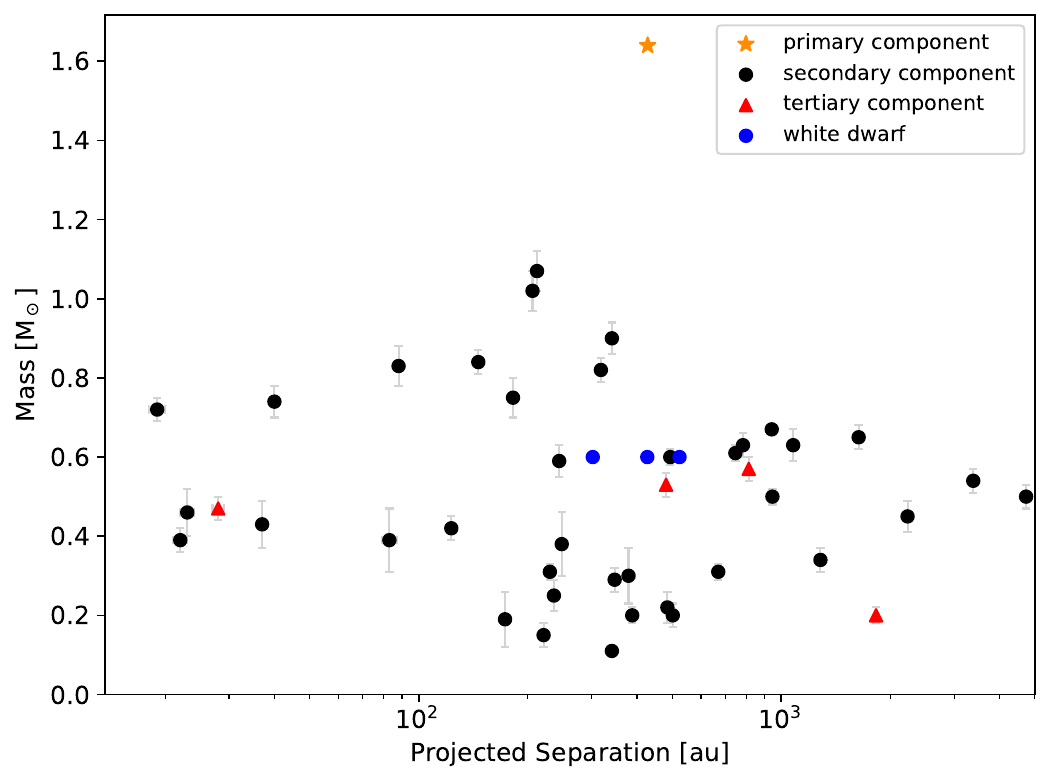}
\caption{Mass over projected separation of all companions of exoplanet host stars detected with AstraLux in this survey. The symbols indicate whether the companion is the primary (star), secondary (circle) or tertiary (triangle) component of the stellar system. White dwarfs are shown as blue symbols.} \label{fig_comps_plot}
\end{figure}

\begin{figure*}
\includegraphics[width=0.32\textwidth]{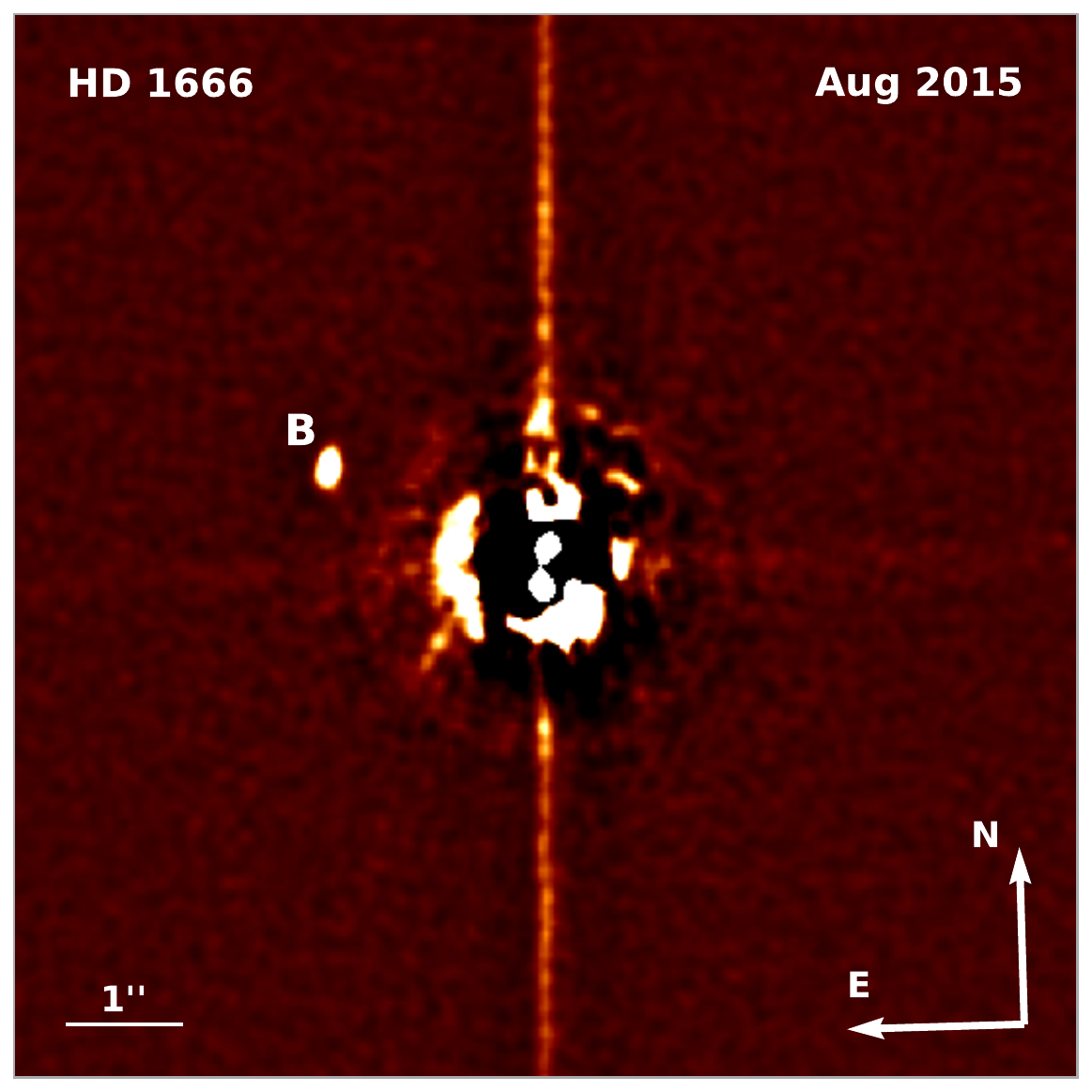}
\includegraphics[width=0.32\textwidth]{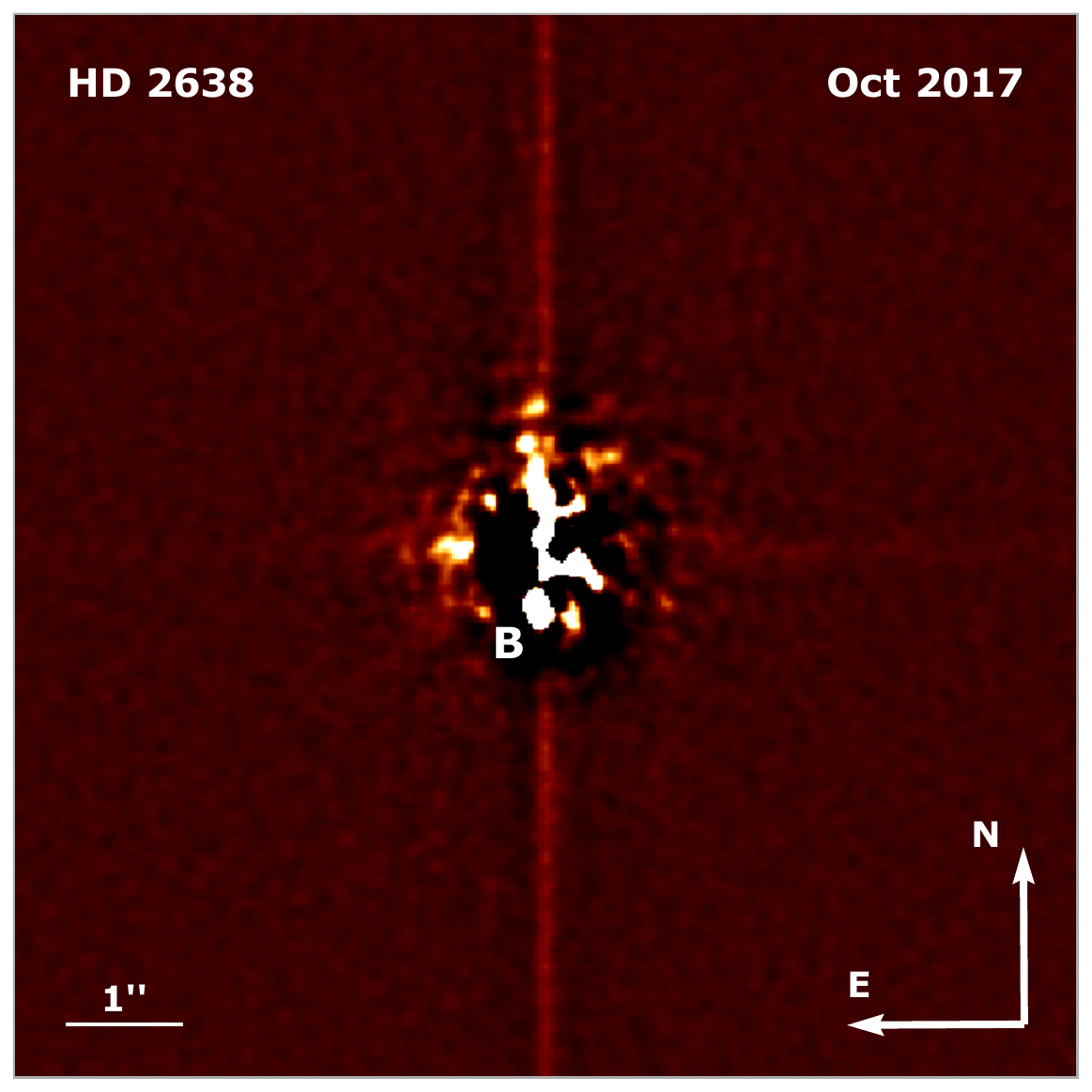}
\includegraphics[width=0.32\textwidth]{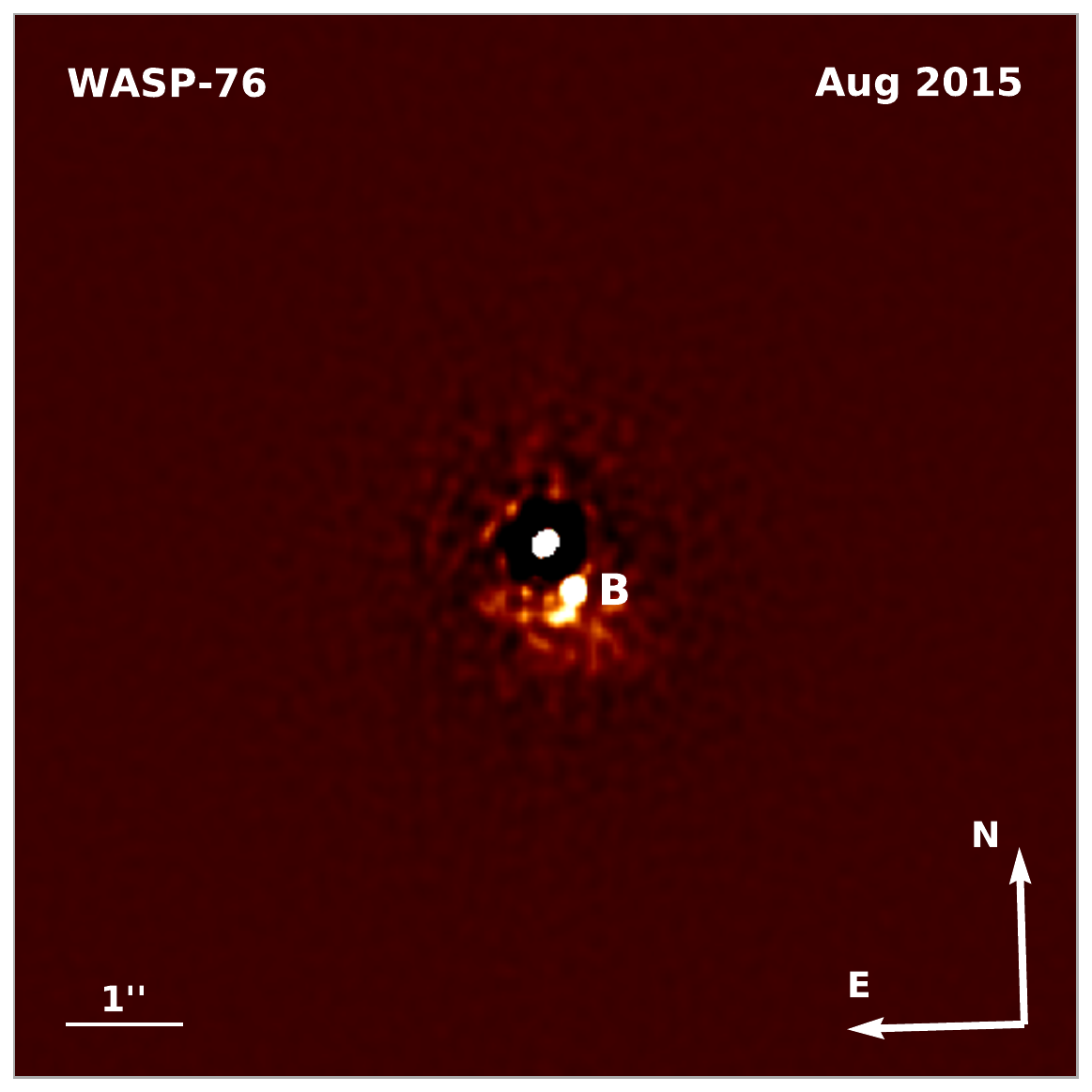}
\includegraphics[width=0.32\textwidth]{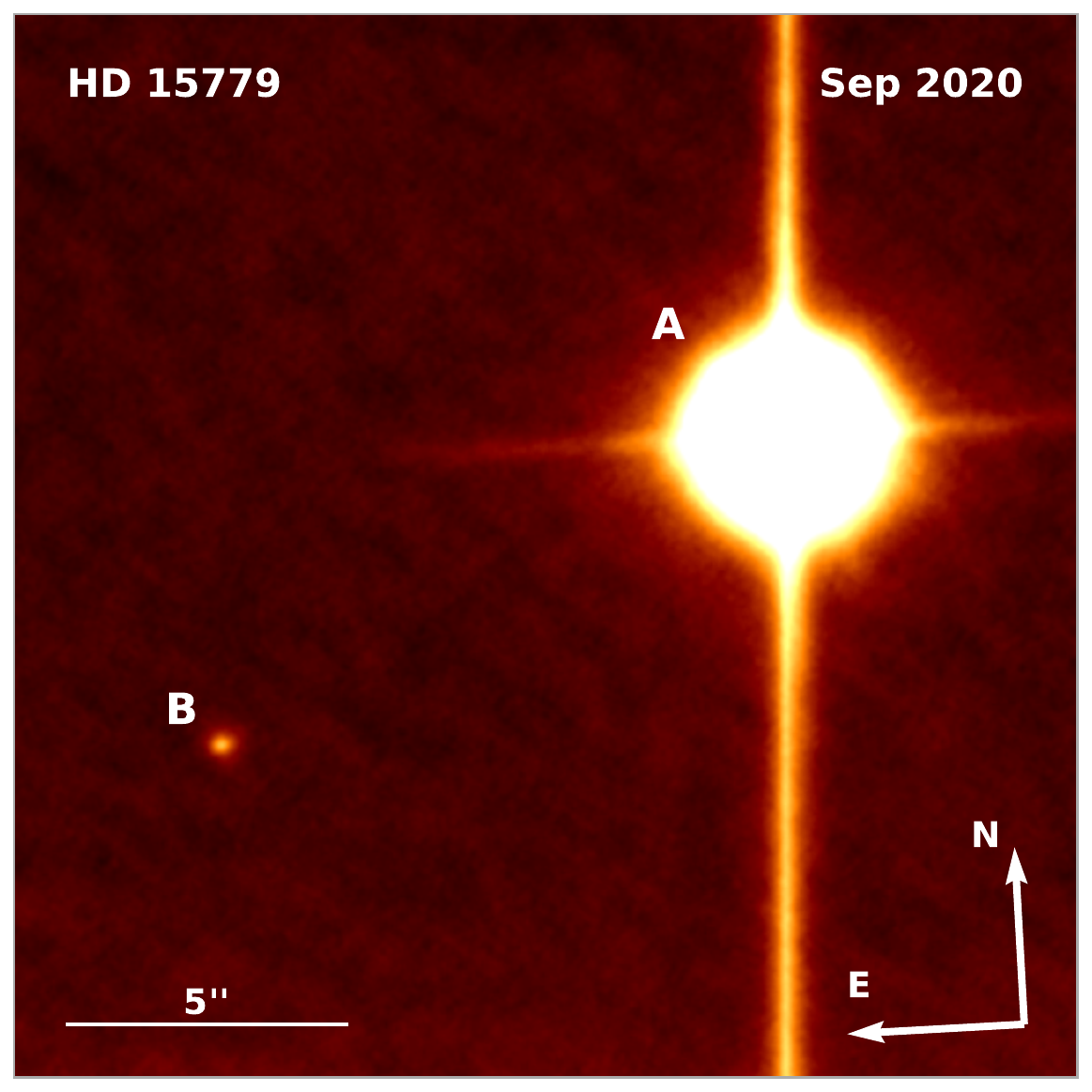}
\includegraphics[width=0.32\textwidth]{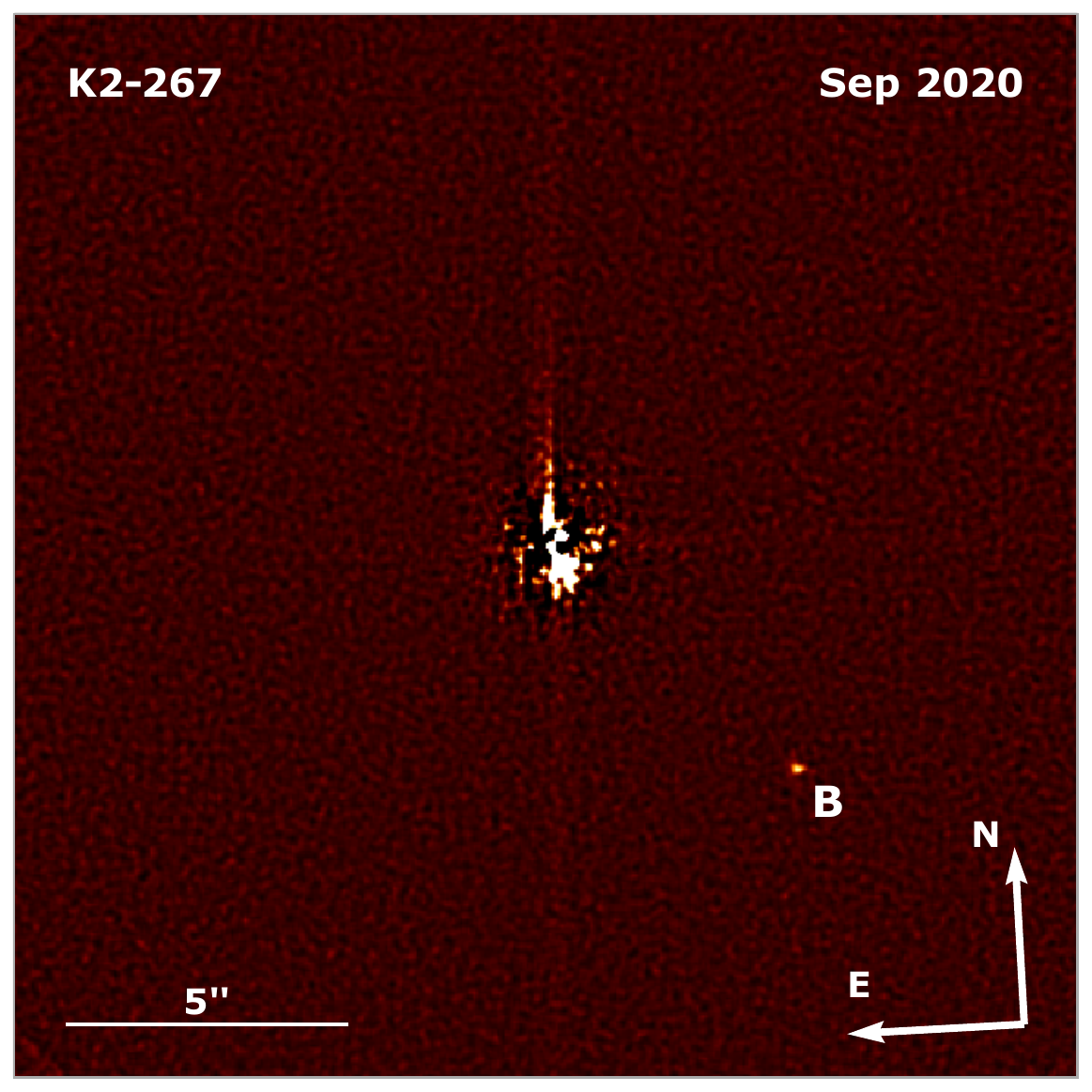}
\includegraphics[width=0.32\textwidth]{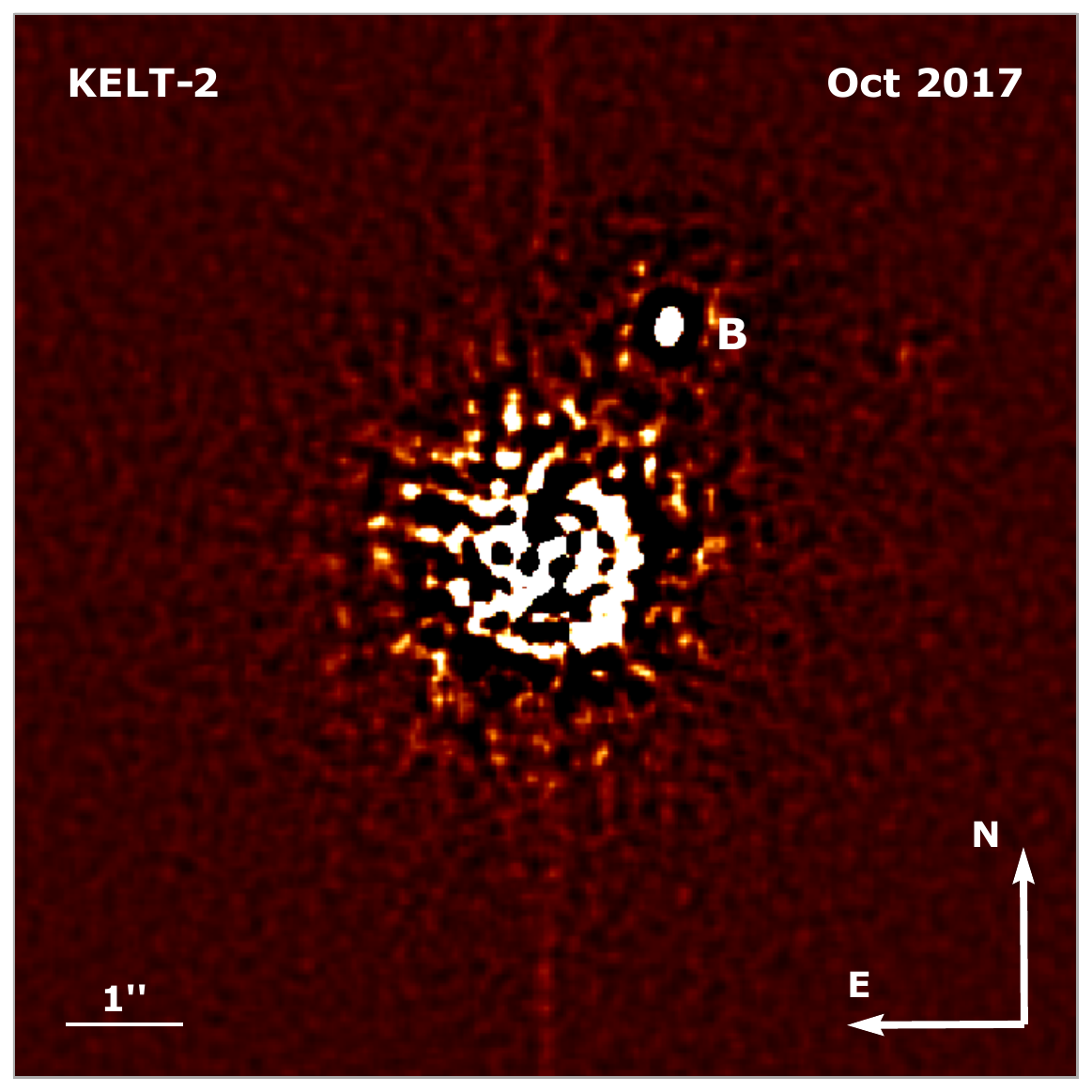}
\includegraphics[width=0.32\textwidth]{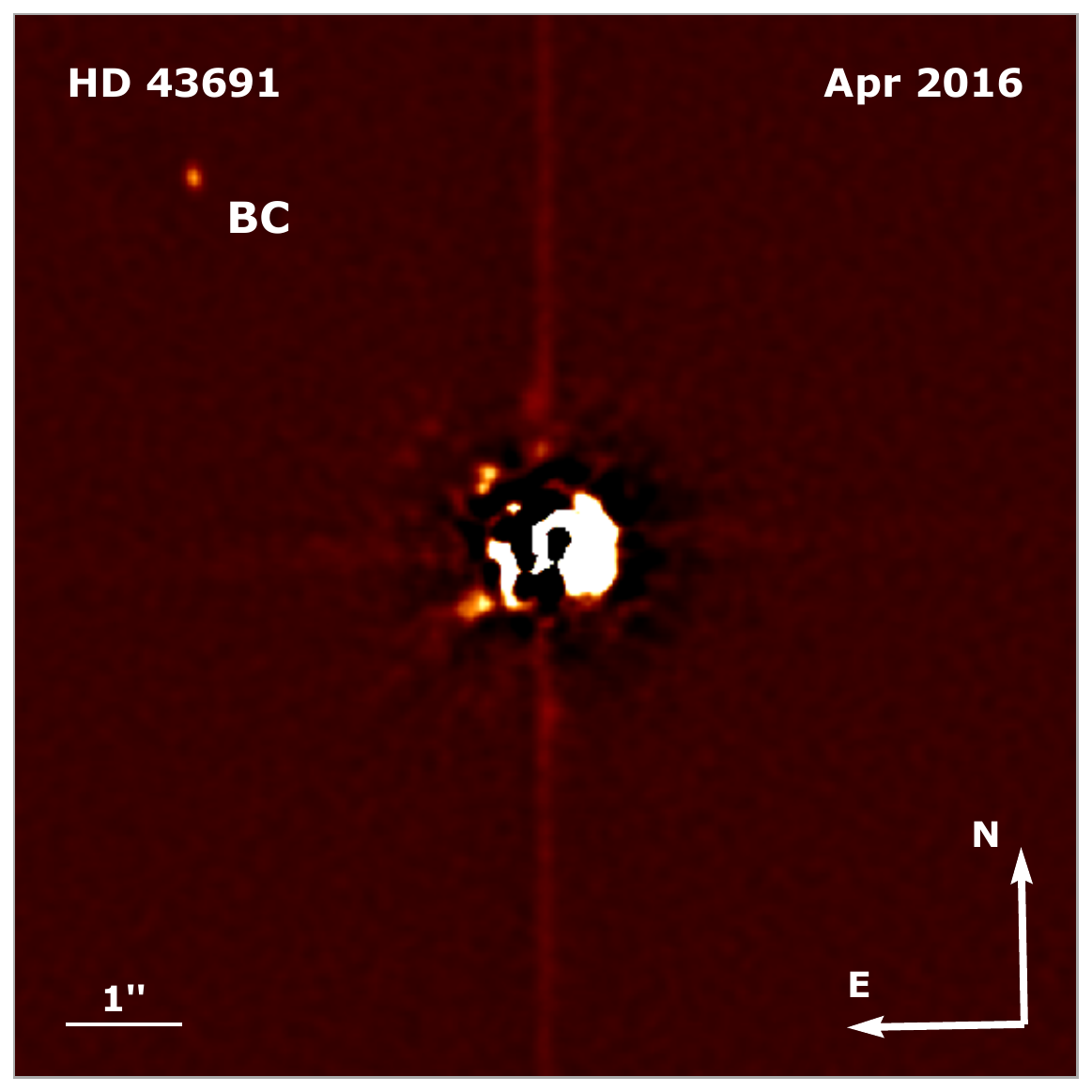}
\includegraphics[width=0.32\textwidth]{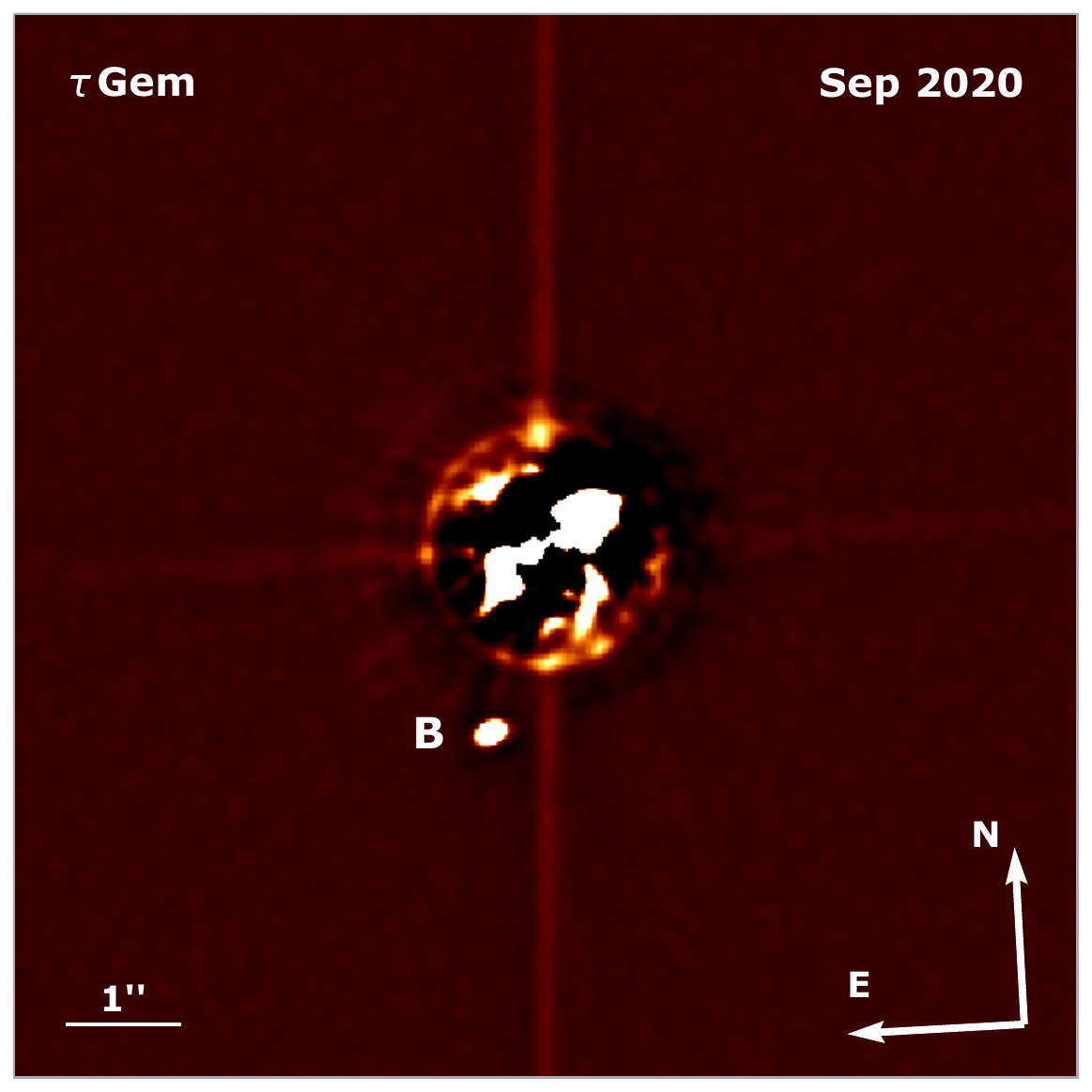}
\includegraphics[width=0.32\textwidth]{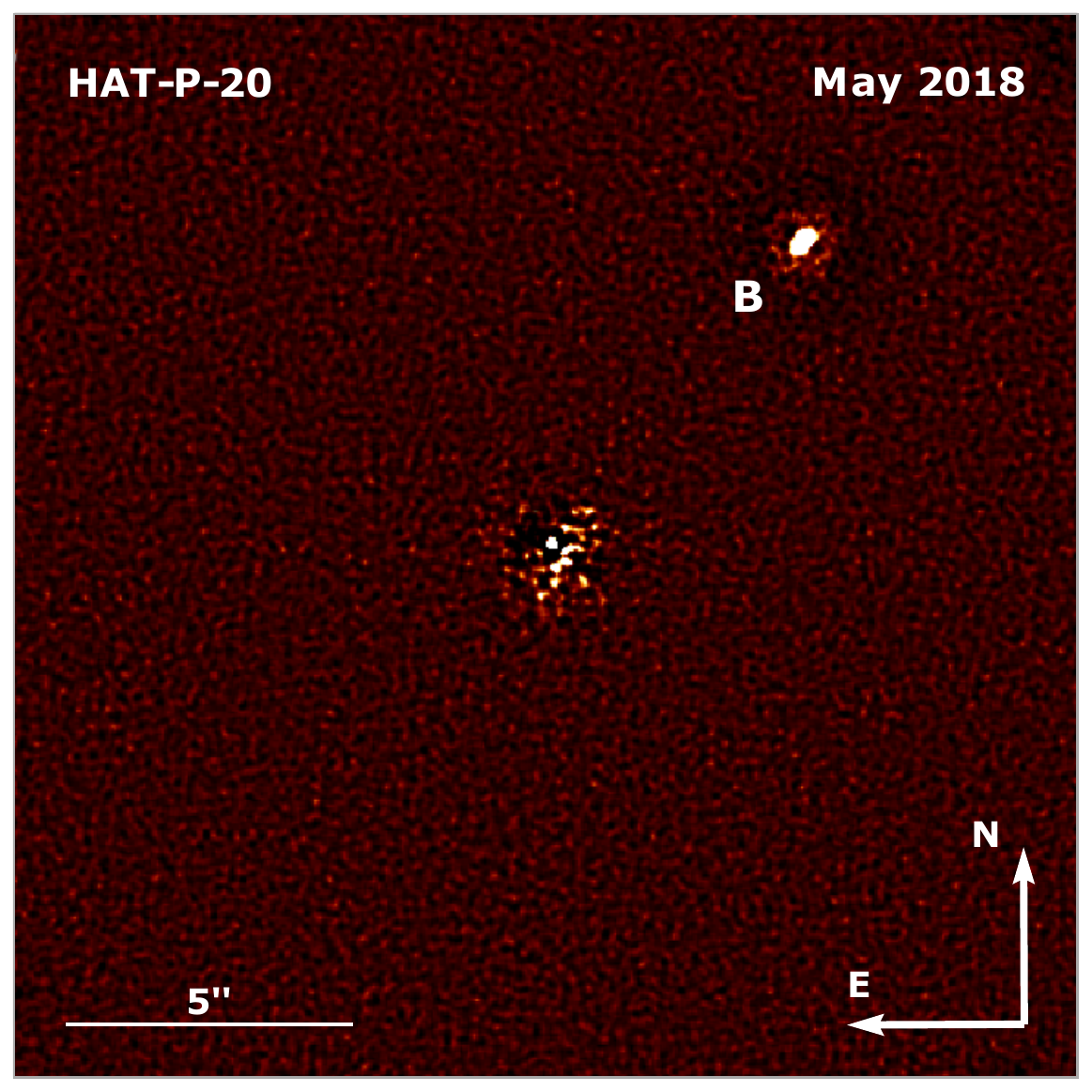}
\includegraphics[width=0.32\textwidth]{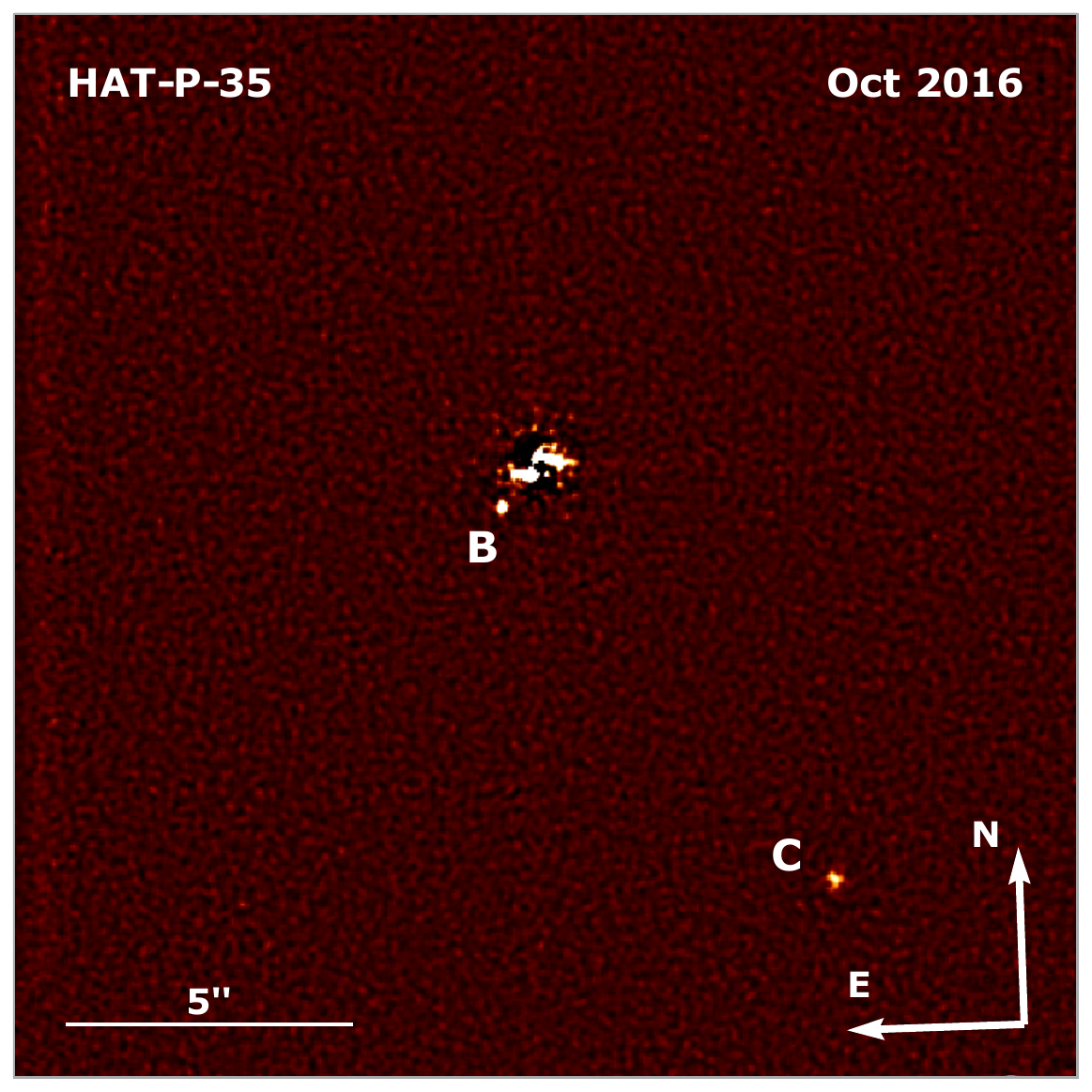}
\includegraphics[width=0.32\textwidth]{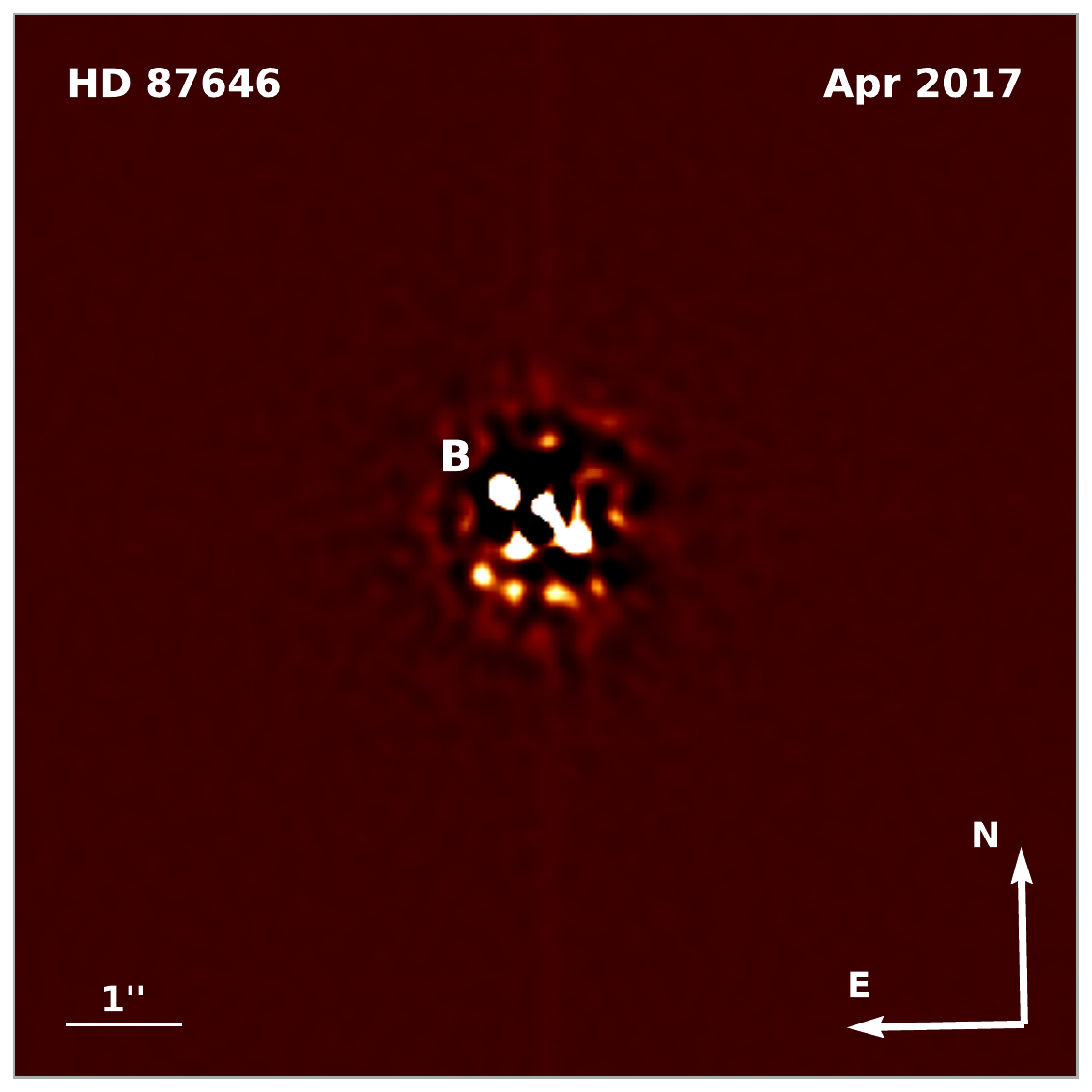}
\includegraphics[width=0.32\textwidth]{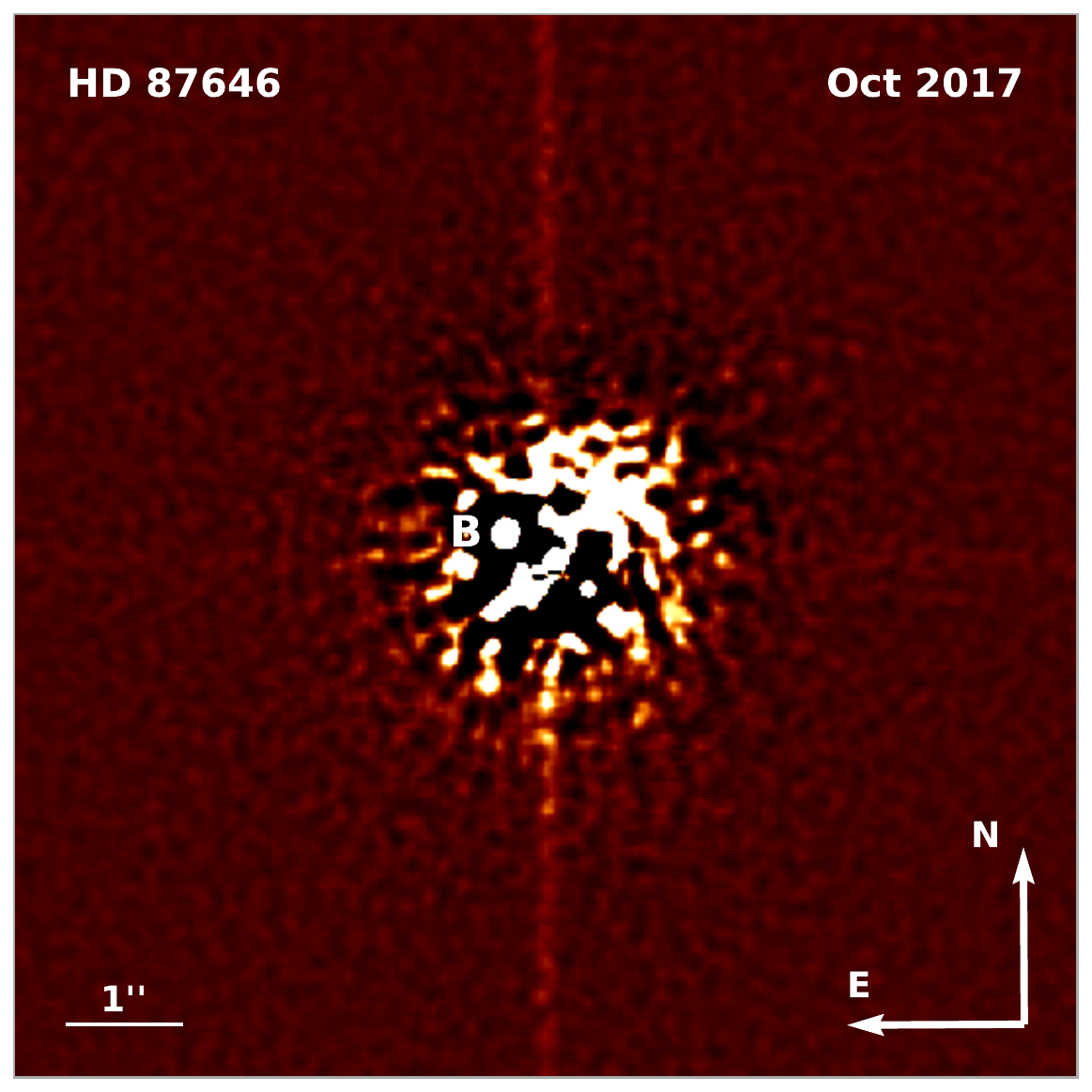}
\caption{Co-moving companions of exoplanet host stars (typically located in the center of all images unless otherwise indicated) detected with AstraLux in this survey. Except for the images of HD\,15779, $\gamma\,$Leo, WASP-85, HD\,132563, and HD\,176551, the PSF of the exoplanet host star is subtracted and the images are spatially high-pass filtered.} \label{fig_comps}
\end{figure*}

All detected co-moving companions are shown in Figure\,\ref{fig_comps}, with the PSF of the exoplanet host star subtracted in most images, which are also spatially high-pass filtered. The angular and projected separations as well as the mass of all detected co-moving companions are listed in Table\,\ref{table_comps}. The companions have angular separations between 0.4 and 11.5\,arcsec from the exoplanet host stars, corresponding to projected separations between 19\,au and 4731\,au (median projected separation is 344\,au). The companions have masses between 0.11\,$\rm M_\odot$ and 1.64\,$\rm M_\odot$ (median mass of 0.52\,$\rm M_\odot$). Figure\,\ref{fig_comps_plot} shows a mass-separation plot for all detected companions. In summary, 46 companions were found around 43 exoplanet host stars. As 212 targets were examined in total in this survey, this gives a minimum multiplicity rate of $20 \pm 3\,\%$. This rate is consistent with the previously found minimum multiplicity rates of exoplanet host stars \citep[$16\pm2$\,\%, ][]{Michel2021} or potential exoplanet host stars \citep[$20.1 \pm 0.9$\,\%, ][]{Mugrauer2022b}. Of the 46 companions detected, one is the primary component, 41 are the secondary component, and 4 are the tertiary component of their systems. Of the 43 detected systems, 33 are binary systems and 10 are hierarchical triple star systems. In the case of the exoplanet host stars HAT-P-35, WASP-14 and HAT-P-57, two co-moving companions are detected with AstraLux in this survey, while the co-moving companions of HD\,34691, KELT-4, HD\,142245, HD\,185269 and HAT-P-8 were previously resolved as close binaries by high-contrast adaptive optics imaging observations \cite[see][respectively]{Ngo2017,Eastman2016,Mugrauer2015,Ilic2022,Bechter2014}.

\subsection{Notes On Individual Companions}

\textbf{HD\,109271\,B, HAT-P-18\,B, and HIP\,116454\,B:} These companions are all known white dwarfs as reported in \cite{Ginski2021}, \cite{Mugrauer2021} and \cite{Mugrauer2019}. Thus, their mass is significantly underestimated (0.10\,$\rm M_\odot$ to 0.16\,$\rm M_\odot$) by the photometric approach described above, assuming that the detected companions are main sequence stars. We therefore adopt a mass of 0.60\,$\rm M_\odot$ for all white dwarf companions as given in the literature.\newline

\textbf{$\gamma\,$Leo\,B:} The photometric mass estimate of the companion $\gamma\,$Leo\,B gives a companion mass of about 1.45\,$\rm M_\odot$, which would be greater than the mass of the system's primary, the exoplanet host star $\gamma\,$Leo\,A, which has a mass of $1.23 \pm 0.21\,\rm M_\odot$ \cite[see][]{Han2010}. This is because the companion is not a main sequence star but is in fact a G7 giant with a mass of $0.75 \pm 0.05\,\rm M_\odot$, \cite{daSilva2015}. This mass estimate of the companion is used here for further analysis.\newline

\textbf{HD\,132563\,A}: This companion is itself a spectroscopic binary. The masses of the components HD\,132563\,Aa and HD\,132563\,Ab are about 1.08\,$\rm M_\odot$ and 0.56\,$\rm M_\odot$ respectively, as determined by \cite{Desidera2011}. We use the combined mass of 1.64\,$\rm M_\odot$ for the primary component of this hierarchical triple star system.\newline

\textbf{HD\,2638\,C}: The exoplanet host star HD\,2638 is actually the secondary of a very wide hierarchical triple with the primary component HD\,2638\,A (alias HD\,2567) at $\rho=839.39522 \pm 0.00003$\,arcsec and $\text{PA}=231.080145 \pm 0.000002^\circ$, listed in the Gaia DR3 at epoch 2016.0, yielding a projected separation of about 45900\,au. The Gaia DR3 parallaxes of HD\,2638 and HD\,2567 do not significantly differ from each other. Furthermore, this stellar pair exhibits a high degree of common proper motion (cpm-index = 133), and both stars share the same radial-velocity ($\langle\text{RV}\rangle = +9.73$\,km/s with $\Delta\text{RV} = 0.14 \pm 0.22$\,km/s) as expected for the components of a gravitationally bound stellar system. Therefore, the detected companion HD\,2638\,C can be classified as the tertiary component of the system.

\begin{table}

\caption{Projected separation (sep) and mass of all detected companions. For five companions we list their mass estimate from the literature.}

\begin{tabular}{lccc}
\hline
companion           & sep [au]       & mass [$\rm M_\odot$] & reference\\
\hline
HD\,1666\,B         & $248 \pm 2$    & $0.38 \pm 0.08$\\
HD\,2638\,C         & ~\,$28 \pm 1$  & $0.47 \pm 0.03$\\
WASP-76\,B          & ~\,$88 \pm 2$  & $0.83 \pm 0.05$\\
HD\,15779\,B        & $945 \pm 1$    & $0.50 \pm 0.02$\\
K2-267\,B           & $2229 \pm 5$~\,& $0.45 \pm 0.04$\\
KELT-2\,B           & $318 \pm 2$    & $0.82 \pm 0.03$\\
HD\,43691\,BC       & $388 \pm 2$    & $0.20 \pm 0.02$\\
$\tau\,$Gem\,B      & $212 \pm 2$    & $1.07 \pm 0.05$\\
HAT-P-20\,B         & $494 \pm 1$    & $0.60 \pm 0.02$\\
HAT-P-35\,B         & $4731 \pm 9$~\,& $0.50 \pm 0.03$\\
HAT-P-35\,C         & $481 \pm 7$    & $0.53 \pm 0.03$\\
HD\,87646\,B        & ~\,$40 \pm 1$  & $0.74 \pm 0.04$\\
$\gamma\,$Leo\,B    & $182 \pm 1$    & $0.75 \pm 0.05$      & (1)\\
HAT-P-22\,B         & $747 \pm 2$    & $0.61 \pm 0.02$\\
KELT-4\,BC          & $341 \pm 2$    & $0.90 \pm 0.04$\\
HD\,96167\,B        & $502 \pm 3$    & $0.20 \pm 0.03$\\
WASP-85\,B          & $206 \pm 2$    & $1.02 \pm 0.05$\\
HD\,103774\,B       & $347 \pm 2$    & $0.29 \pm 0.03$\\
11\,Com\,B          & $941 \pm 2$    & $0.67 \pm 0.01$\\
HD\,109271\,B (WD)  & $302 \pm 1$    & $\sim0.60$           & (2)\\
HD\,116029\,B       & $173 \pm 1$    & $0.19 \pm 0.07$\\
$\tau\,$Boo\,B      & ~\,$23 \pm 1$  & $0.46 \pm 0.06$\\
KELT-18\,B          & $1078 \pm 3$~\,& $0.63 \pm 0.04$\\
WASP-14\,B          & $236 \pm 3$    & $0.25 \pm 0.04$\\
WASP-14\,C          & $1825 \pm 4$~\,& $0.20 \pm 0.02$\\
Qatar\,6\,B         & $485 \pm 2$    & $0.22 \pm 0.04$\\
HD\,132563\,A (SB)  & $428 \pm 2$    & $\sim1.64$           & (3)\\
HD\,142245\,BC      & $244 \pm 1$    & $0.59 \pm 0.04$\\
$\epsilon\,$CrB\,B  & $146 \pm 1$    & $0.84 \pm 0.03$\\
HAT-P-18\,B (WD)    & $427 \pm 3$    & $\sim0.60$           & (4)\\
HAT-P-67\,B         & $3383 \pm 5$~\,& $0.54 \pm 0.03$\\
HD\,164509\,B       & ~\,$37 \pm 1$  & $0.43 \pm 0.06$\\
WASP-58\,B          & $379 \pm 3$    & $0.30 \pm 0.07$\\
HAT-P-57\,B         & $784 \pm 3$    & $0.63 \pm 0.03$\\
HAT-P-57\,C         & $813 \pm 3$    & $0.57 \pm 0.03$\\
WASP-153\,B         & $1634 \pm 7$~\,& $0.65 \pm 0.03$\\
HD 176051\,B        & ~\,$19 \pm 1$  & $0.72 \pm 0.03$\\
Kepler-21\,B        & ~\,$83 \pm 4$  & $0.39 \pm 0.08$\\
HAT-P-7\,B          & $1282 \pm 4$~\,& $0.34 \pm 0.03$\\
HD\,185269\,BC      & $230 \pm 1$    & $0.31 \pm 0.02$\\
HD\,197037\,B       & $123 \pm 1$    & $0.42 \pm 0.03$\\
HD\,214823\,B       & $670 \pm 2$    & $0.31 \pm 0.02$\\
HAT-P-8\,BC         & $221 \pm 3$    & $0.15 \pm 0.03$\\
HD\,220842\,B       & $341 \pm 1$    & $0.11 \pm 0.01$\\
HIP\,116454\,B (WD) & $524 \pm 1$    & $\sim0.60$           & (5)\\
$\gamma$\,Cep\,B    & ~\,$22 \pm 1$  & $0.39 \pm 0.03$\\
\hline
\end{tabular} \label{table_comps}
\\
\\
(1) \cite{daSilva2015}; (2) \cite{Ginski2021}; (3) \cite{Desidera2011}; (4) \cite{Mugrauer2021}; (5) \cite{Mugrauer2019}
\end{table}

\begin{figure*}
\includegraphics[width=0.32\textwidth]{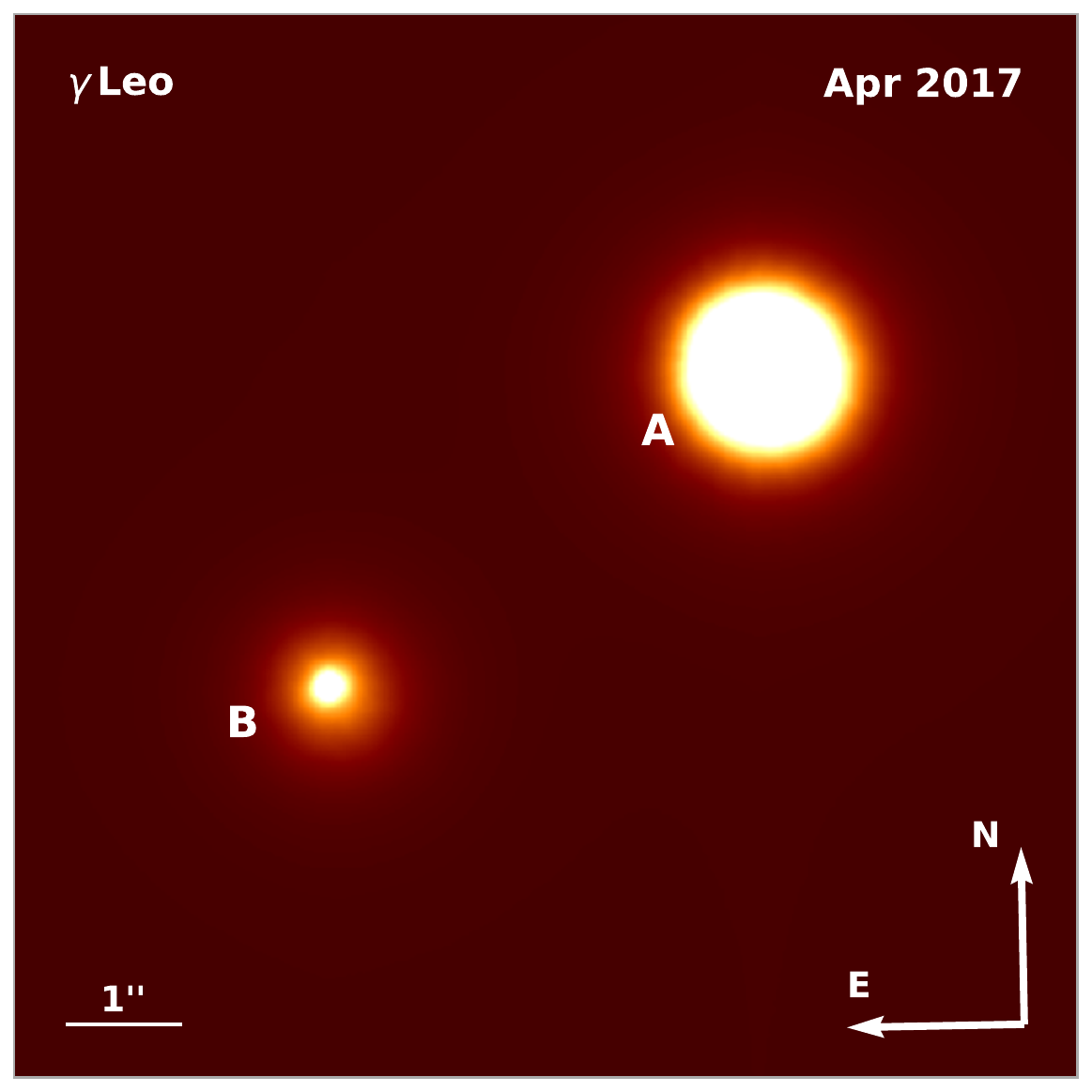}
\includegraphics[width=0.32\textwidth]{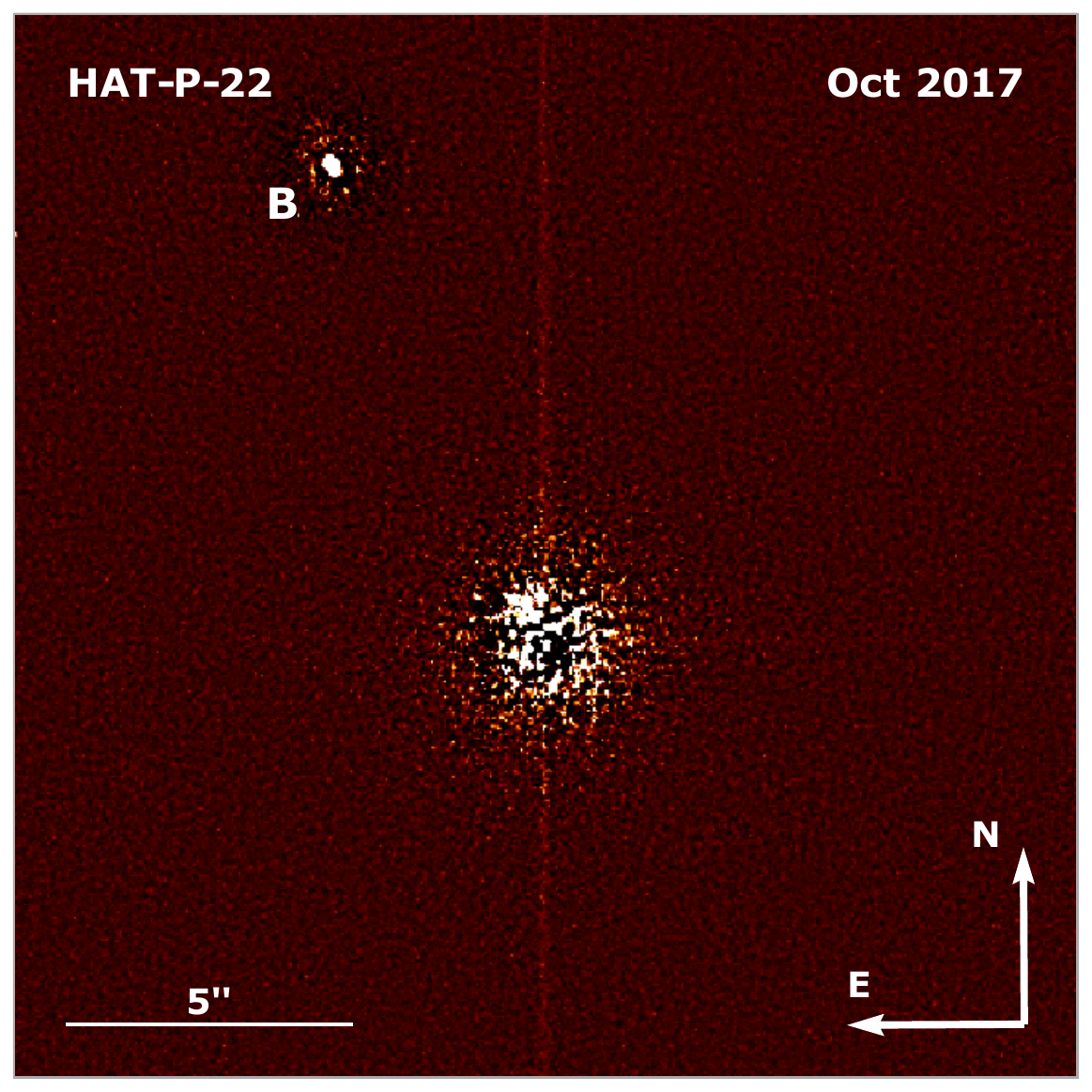}
\includegraphics[width=0.32\textwidth]{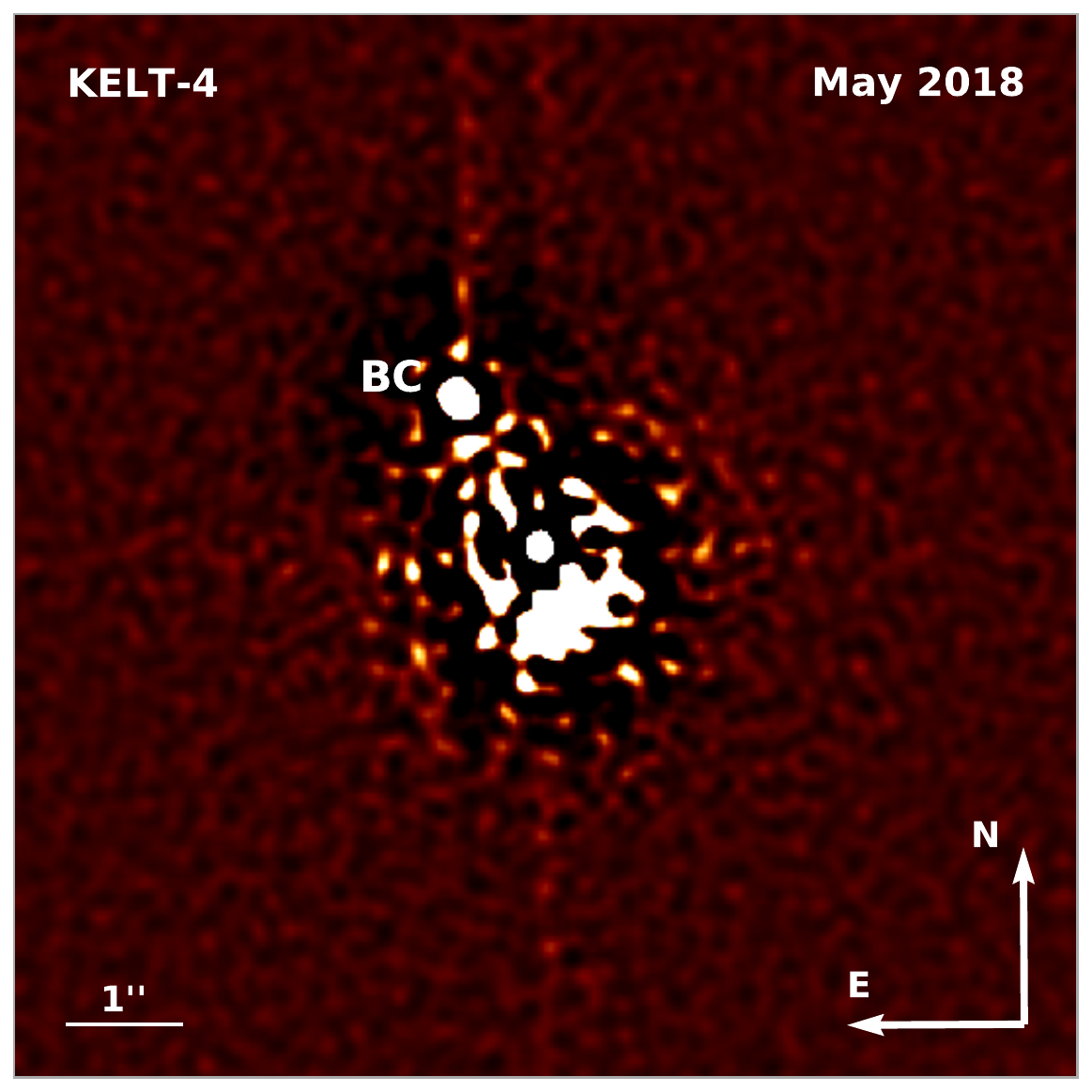}
\includegraphics[width=0.32\textwidth]{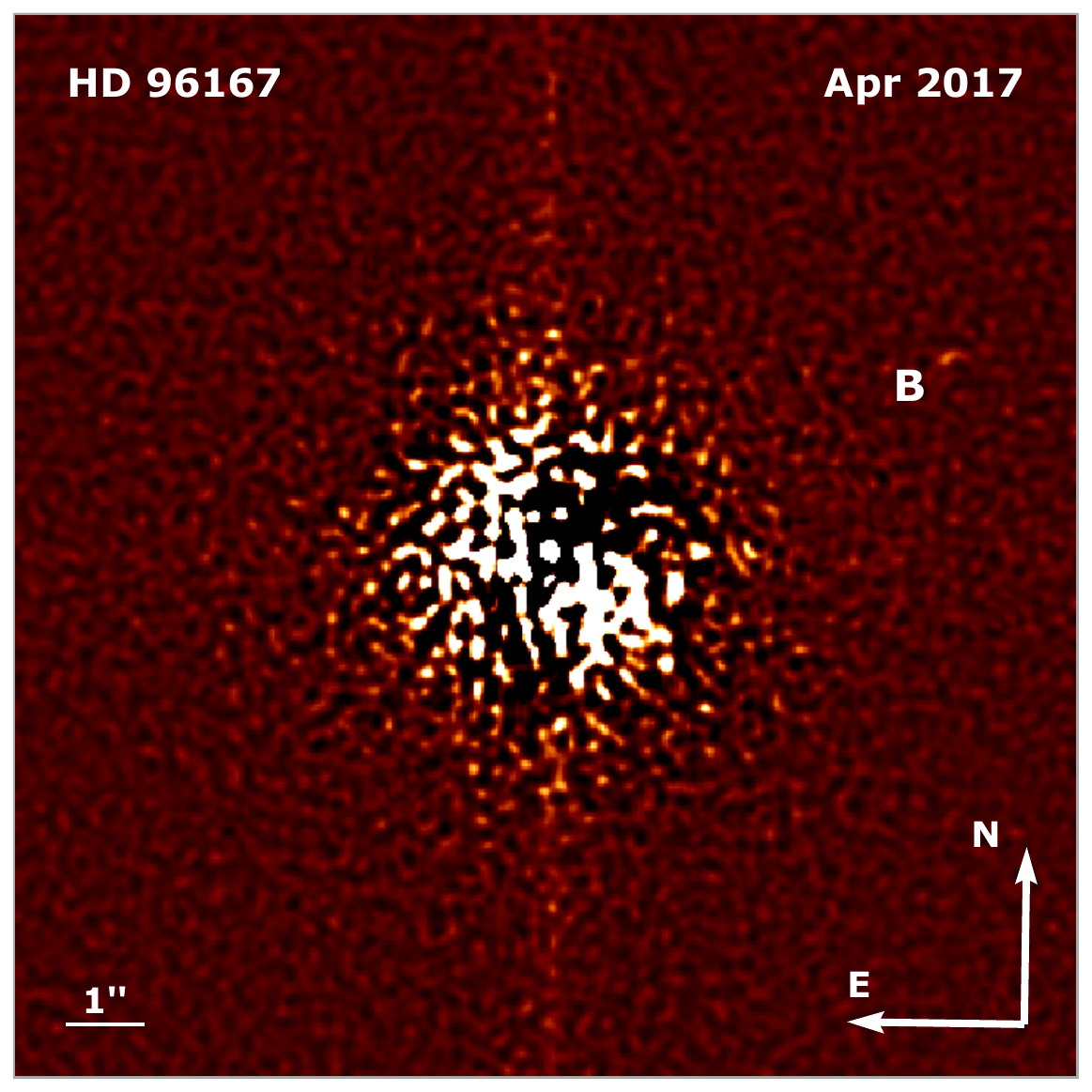}
\includegraphics[width=0.32\textwidth]{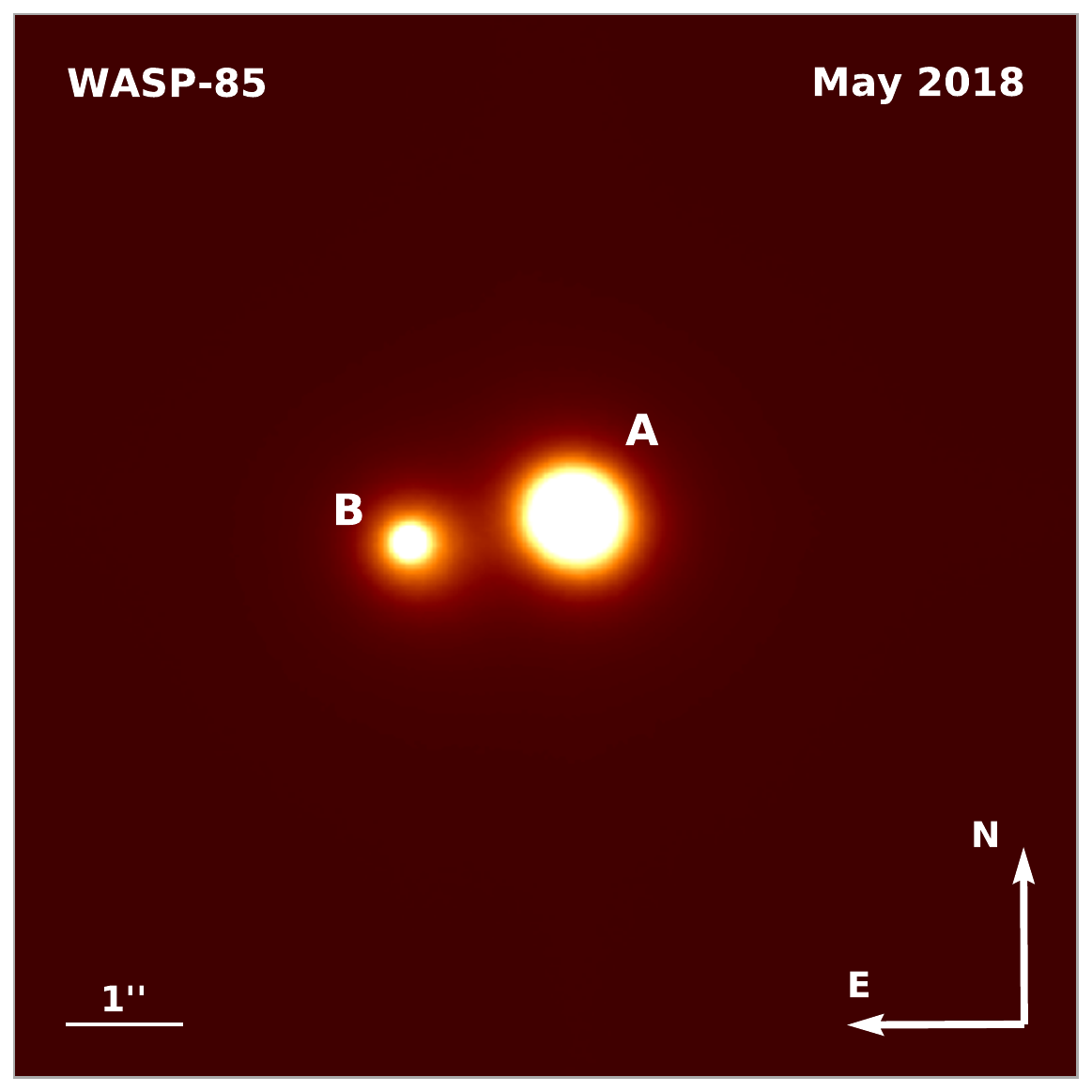}
\includegraphics[width=0.32\textwidth]{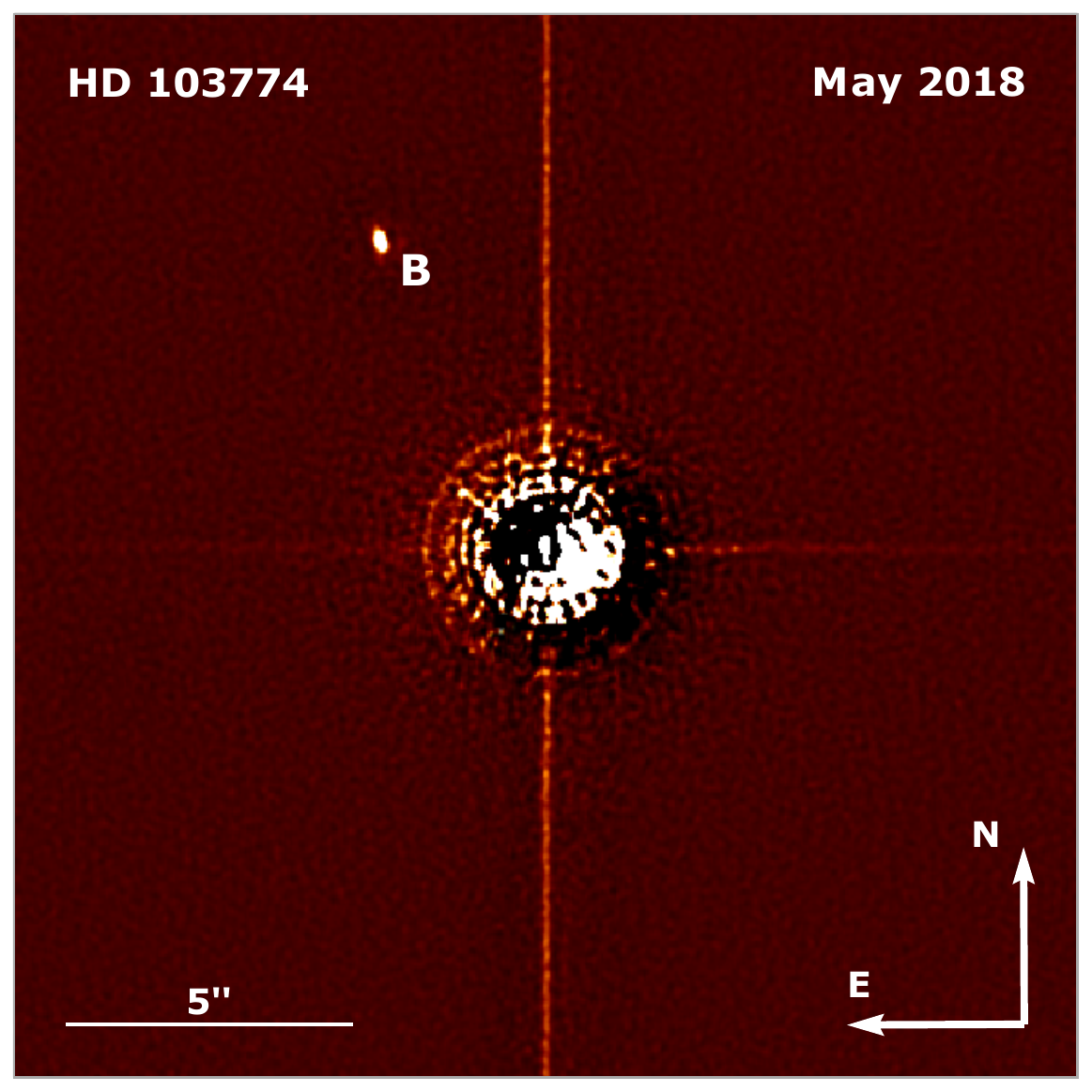}
\includegraphics[width=0.32\textwidth]{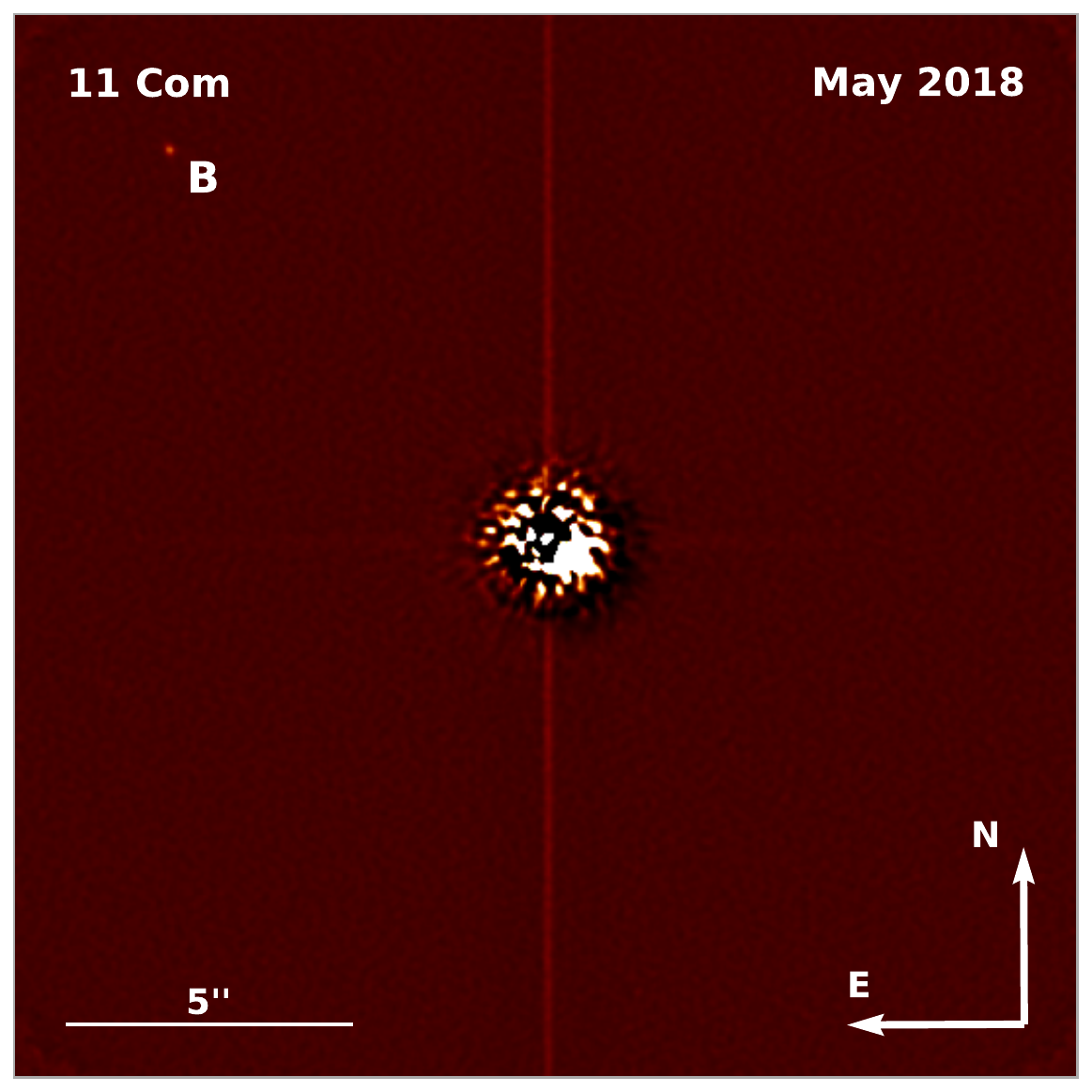}
\includegraphics[width=0.32\textwidth]{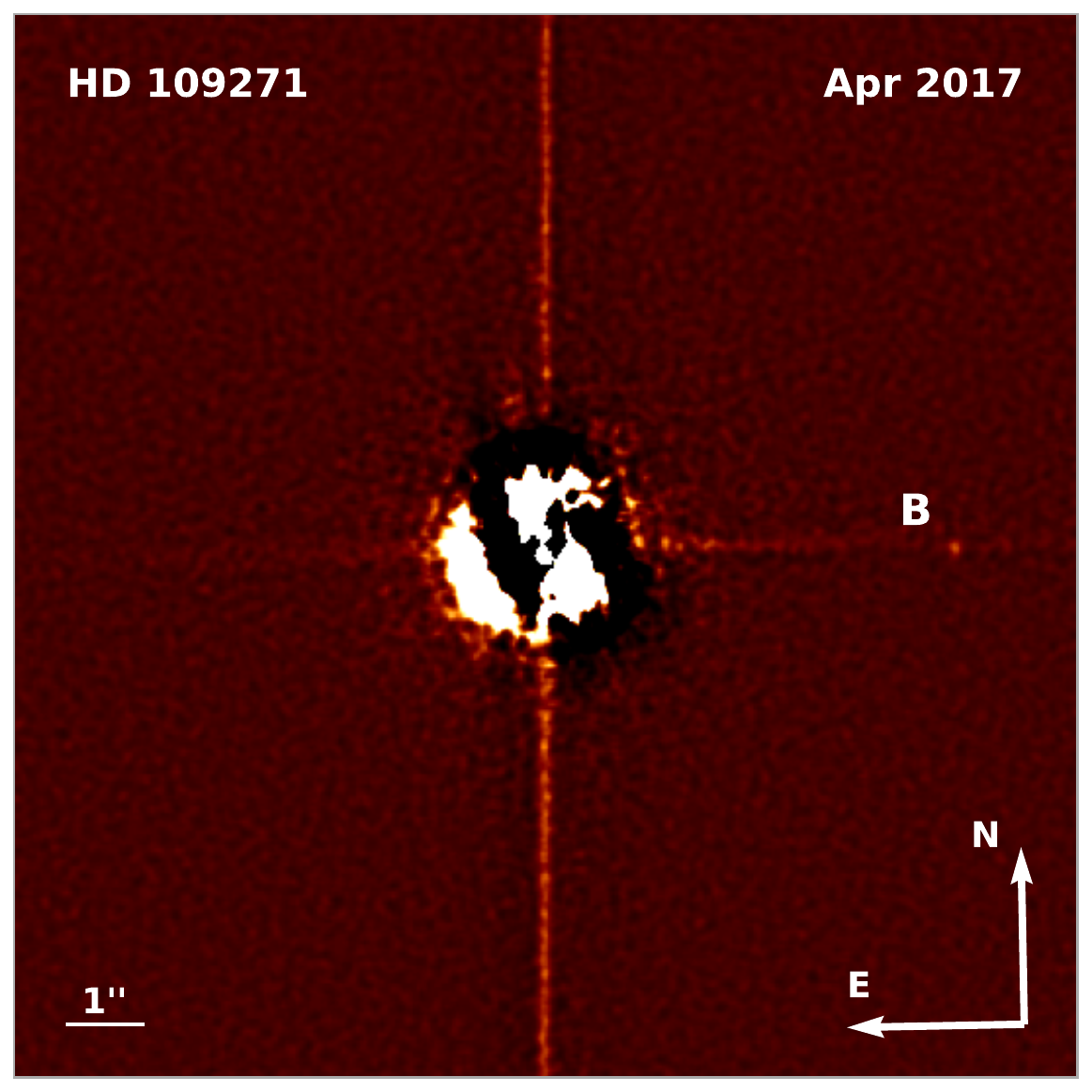}
\includegraphics[width=0.32\textwidth]{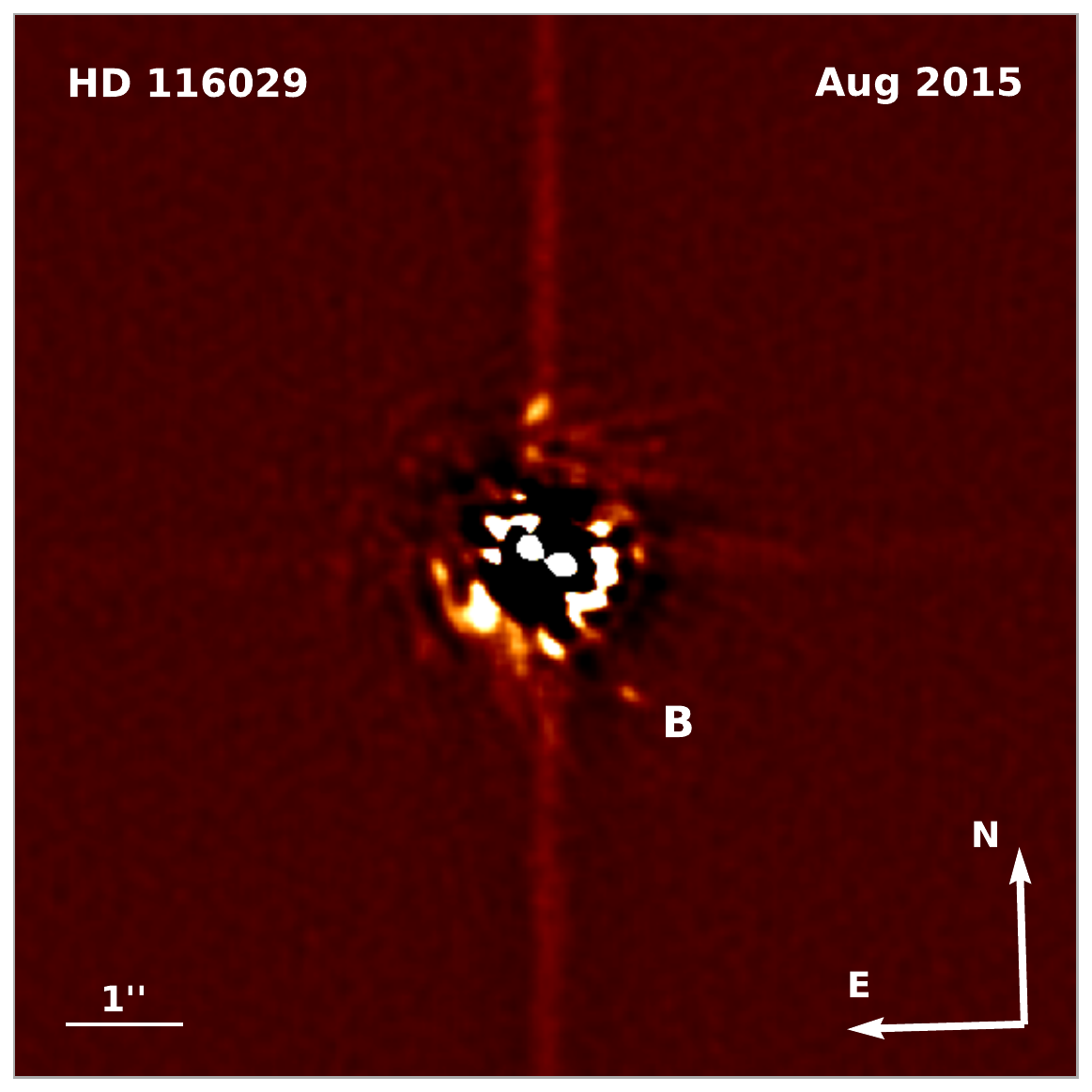}
\includegraphics[width=0.32\textwidth]{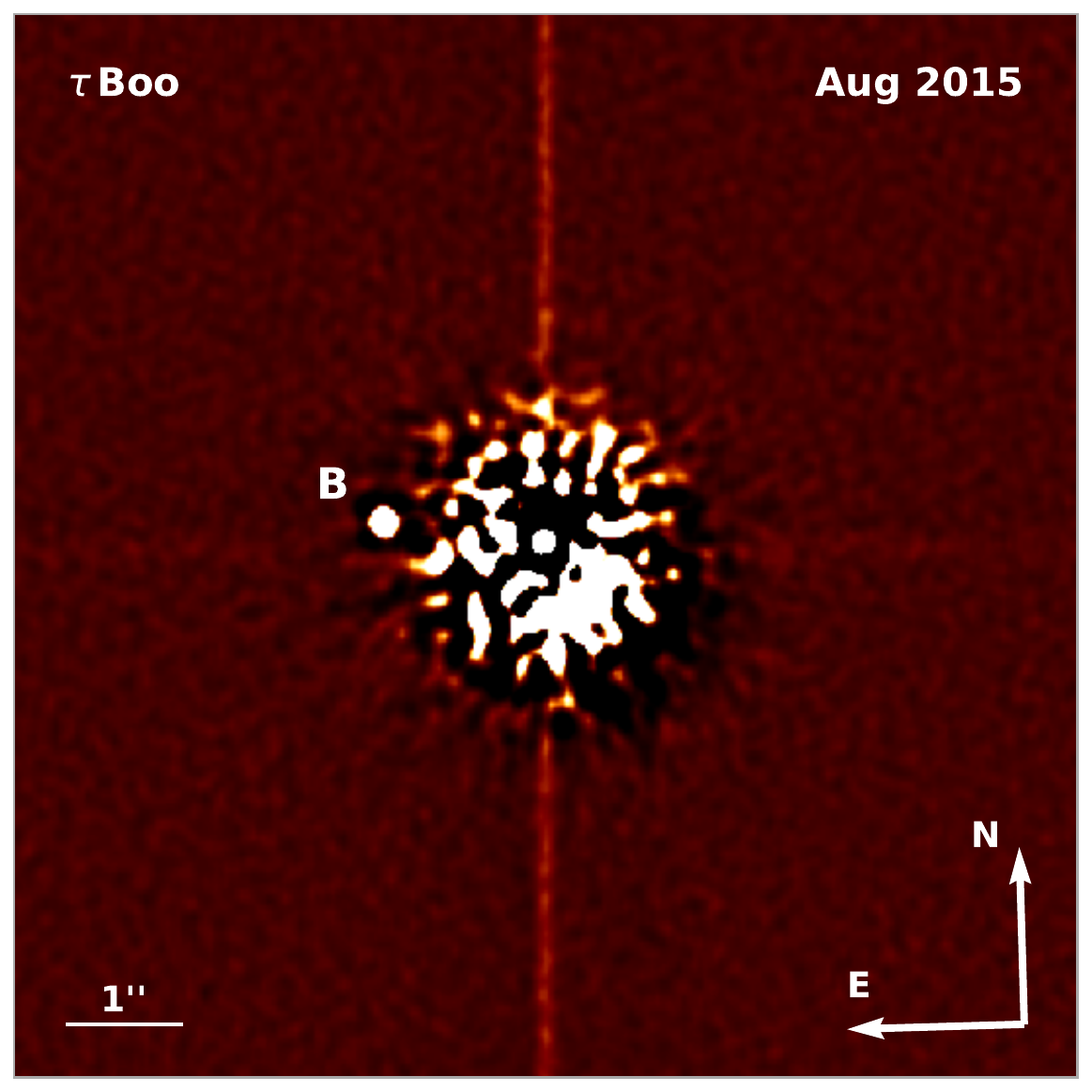}
\includegraphics[width=0.32\textwidth]{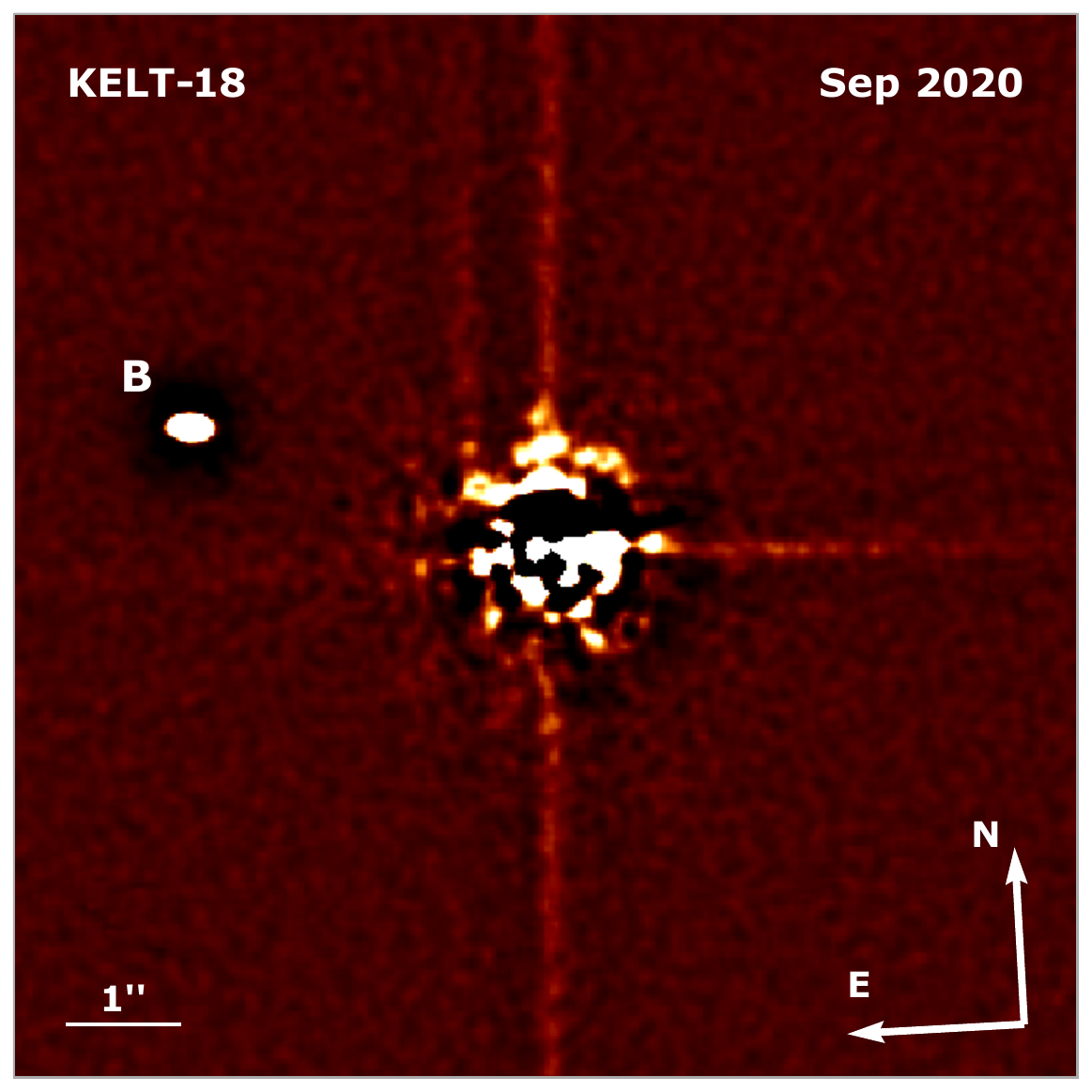}
\includegraphics[width=0.32\textwidth]{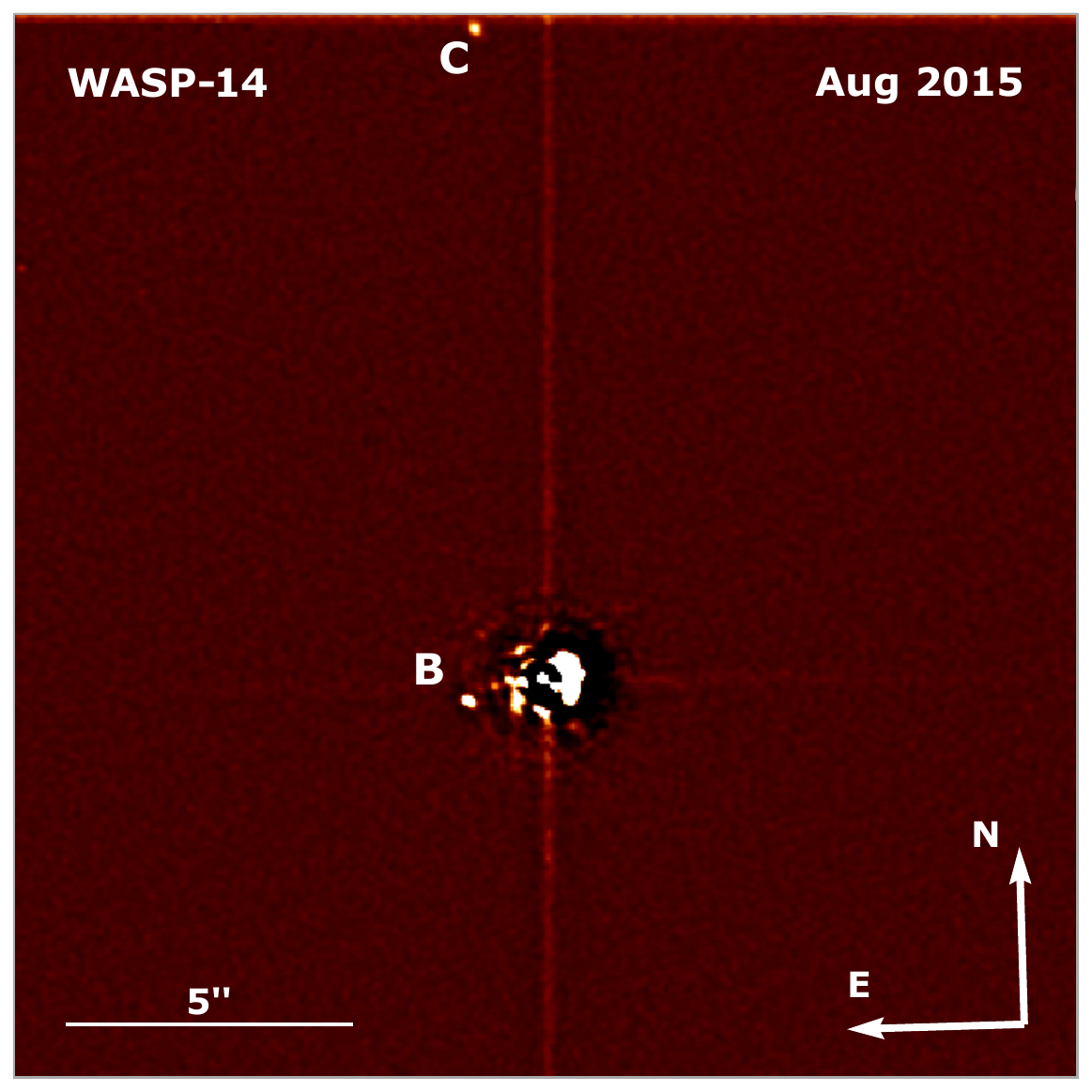}

\textbf{Figure\,\ref{fig_comps}} continued.
\end{figure*}
\begin{figure*}
\includegraphics[width=0.32\textwidth]{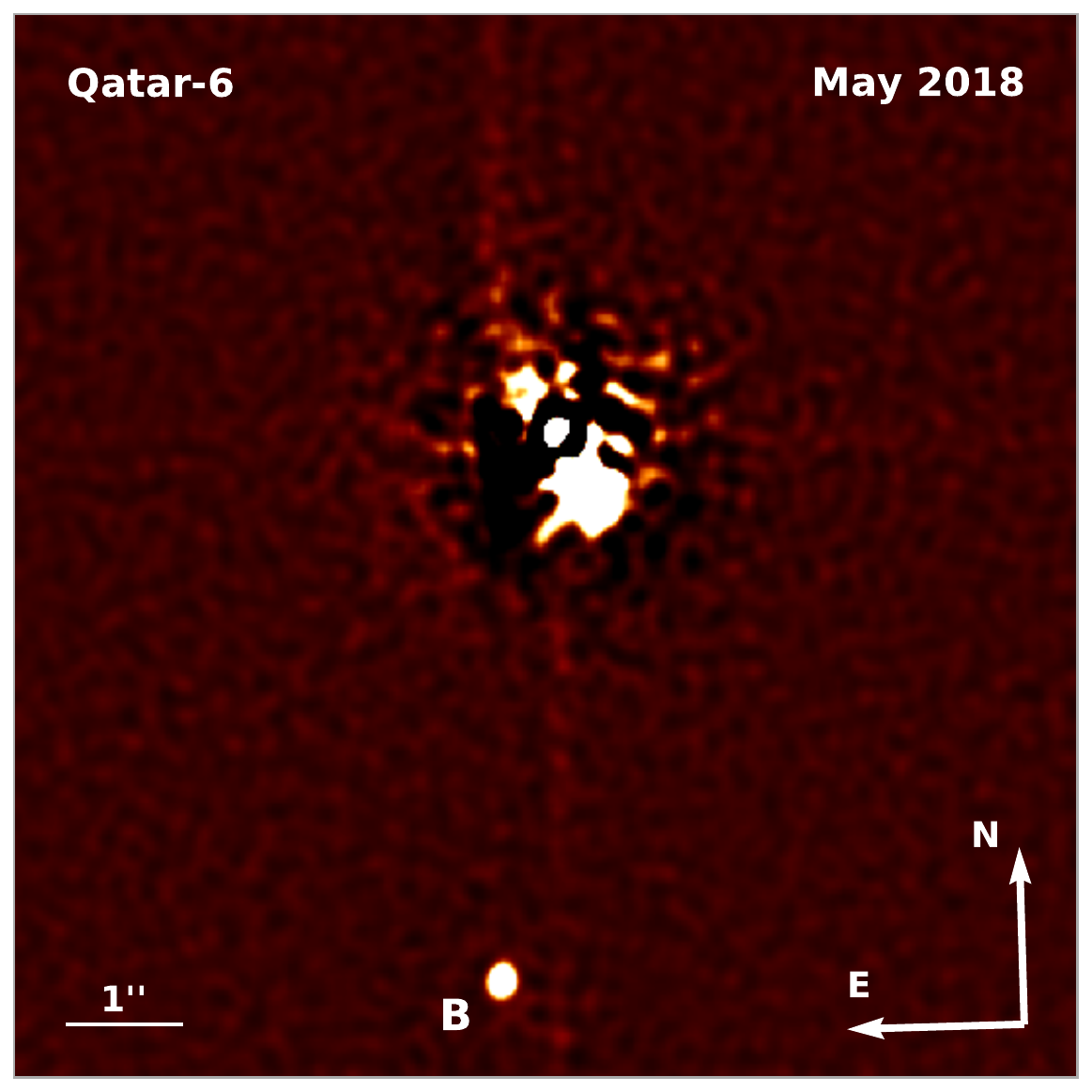}
\includegraphics[width=0.32\textwidth]{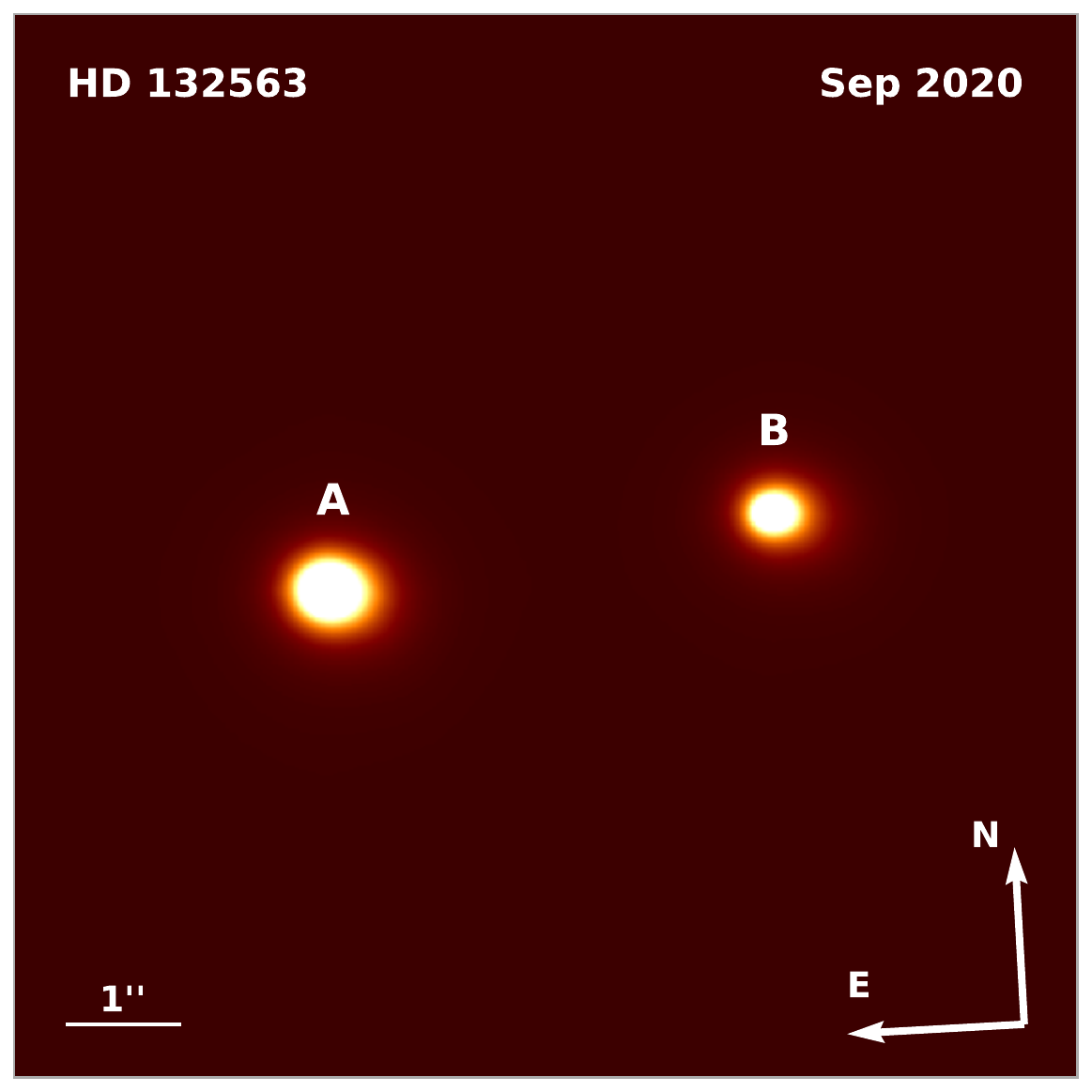}
\includegraphics[width=0.32\textwidth]{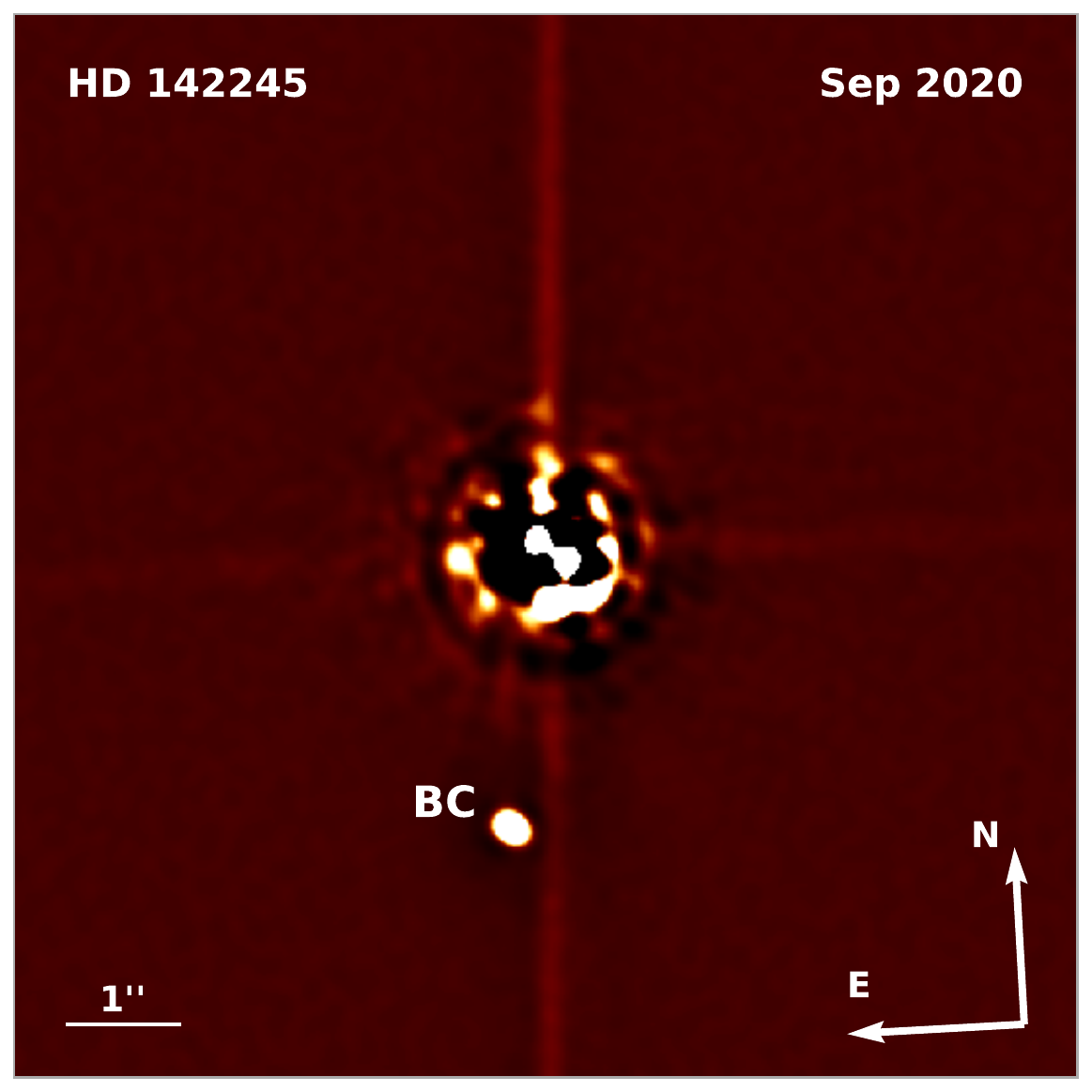}
\includegraphics[width=0.32\textwidth]{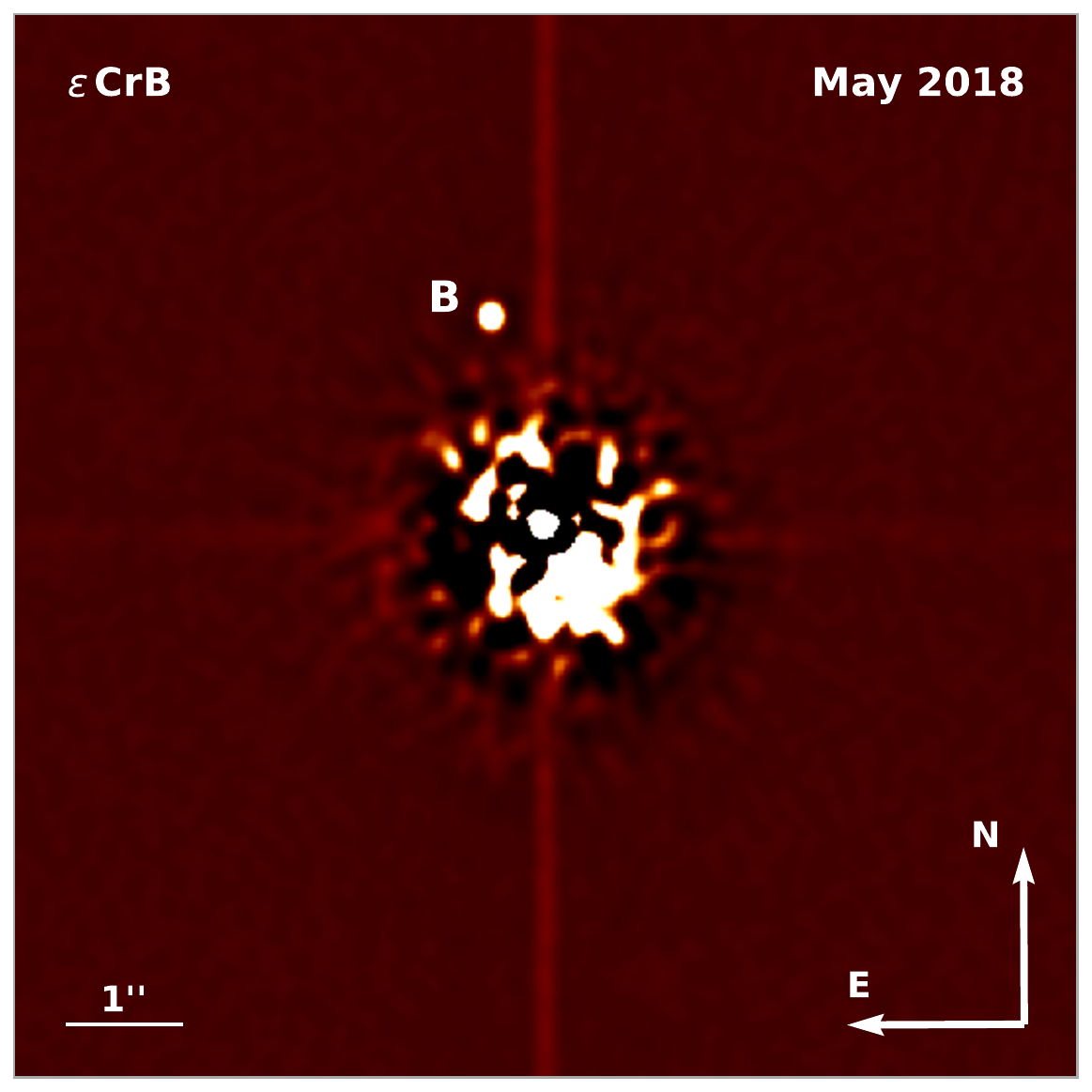}
\includegraphics[width=0.32\textwidth]{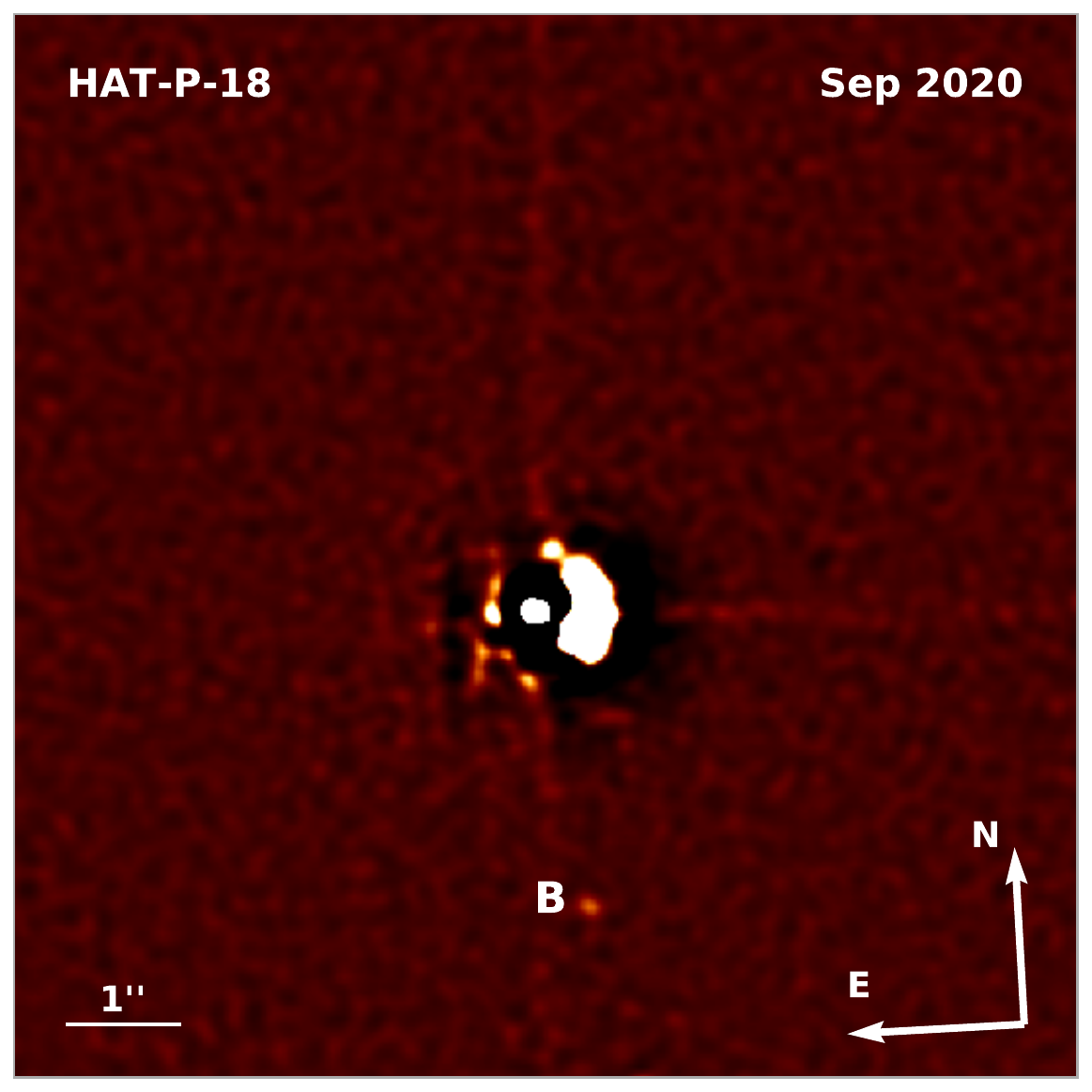}
\includegraphics[width=0.32\textwidth]{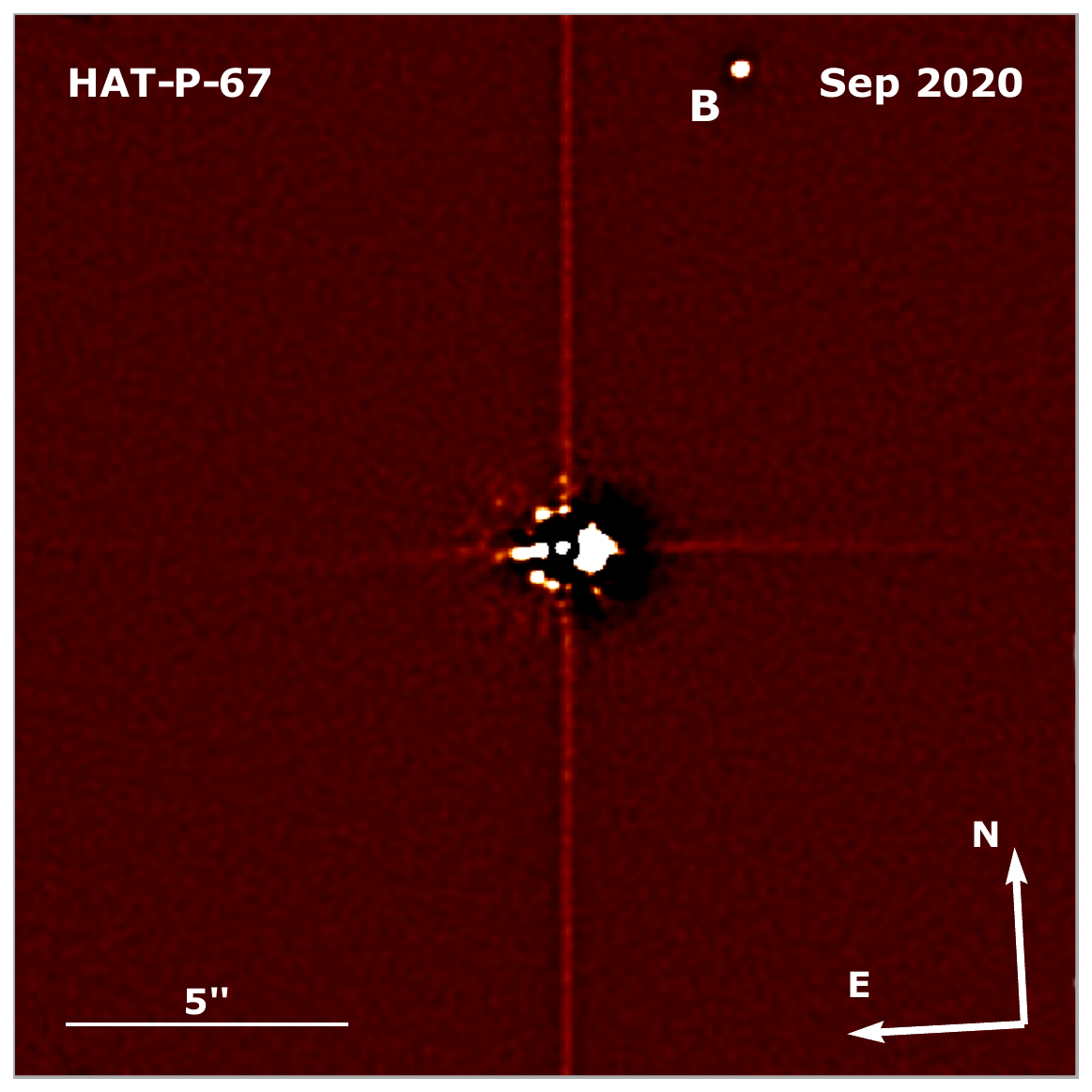}
\includegraphics[width=0.32\textwidth]{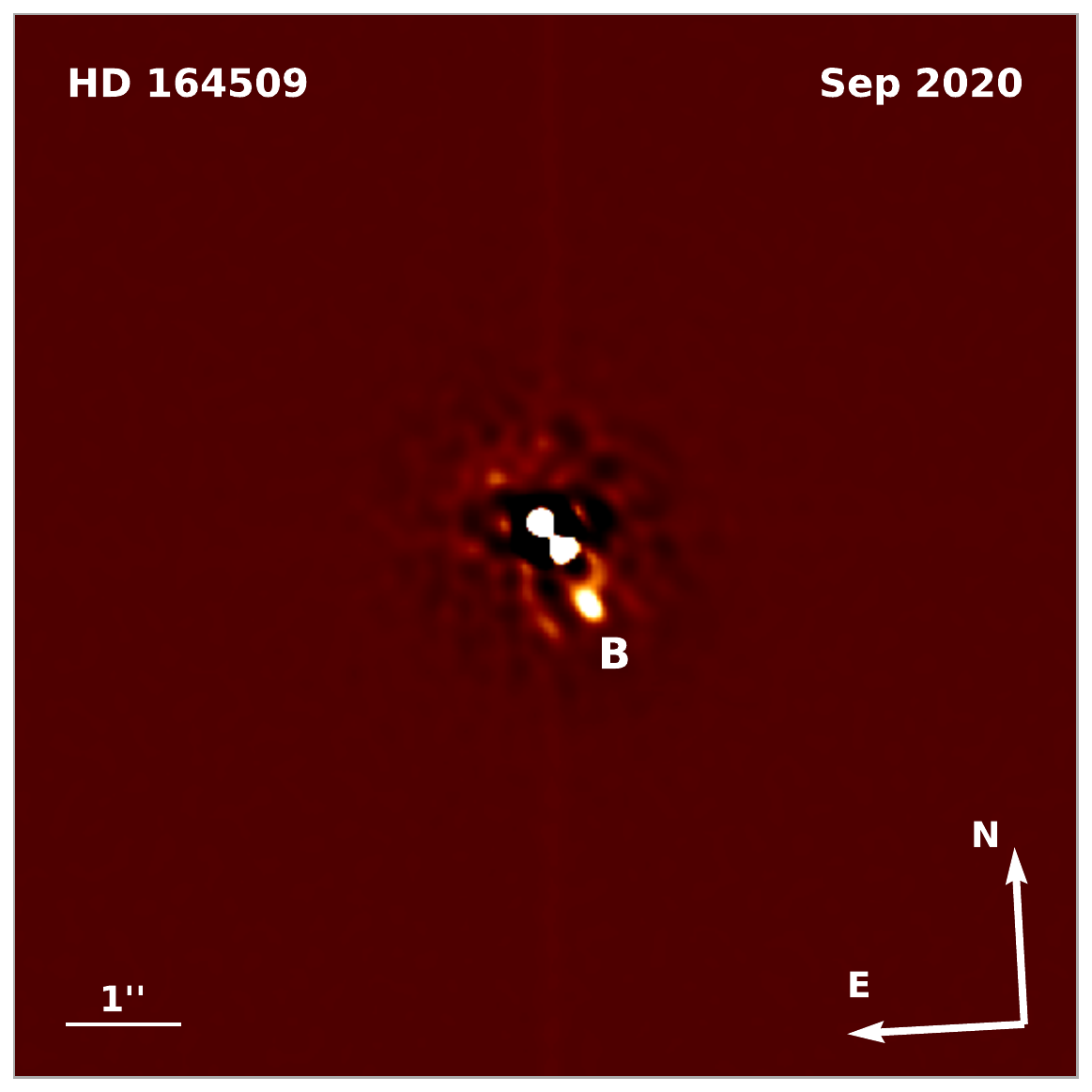}
\includegraphics[width=0.32\textwidth]{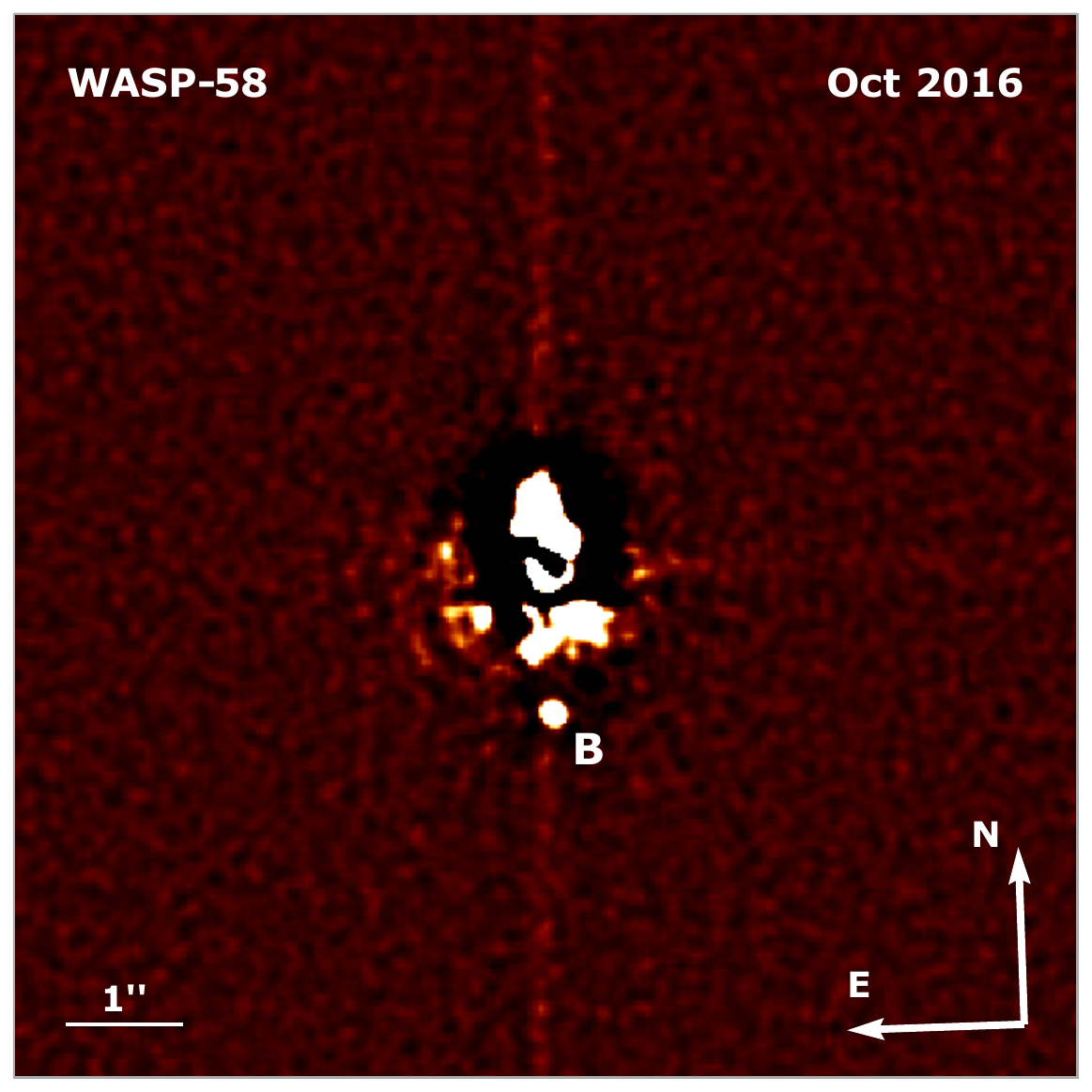}
\includegraphics[width=0.32\textwidth]{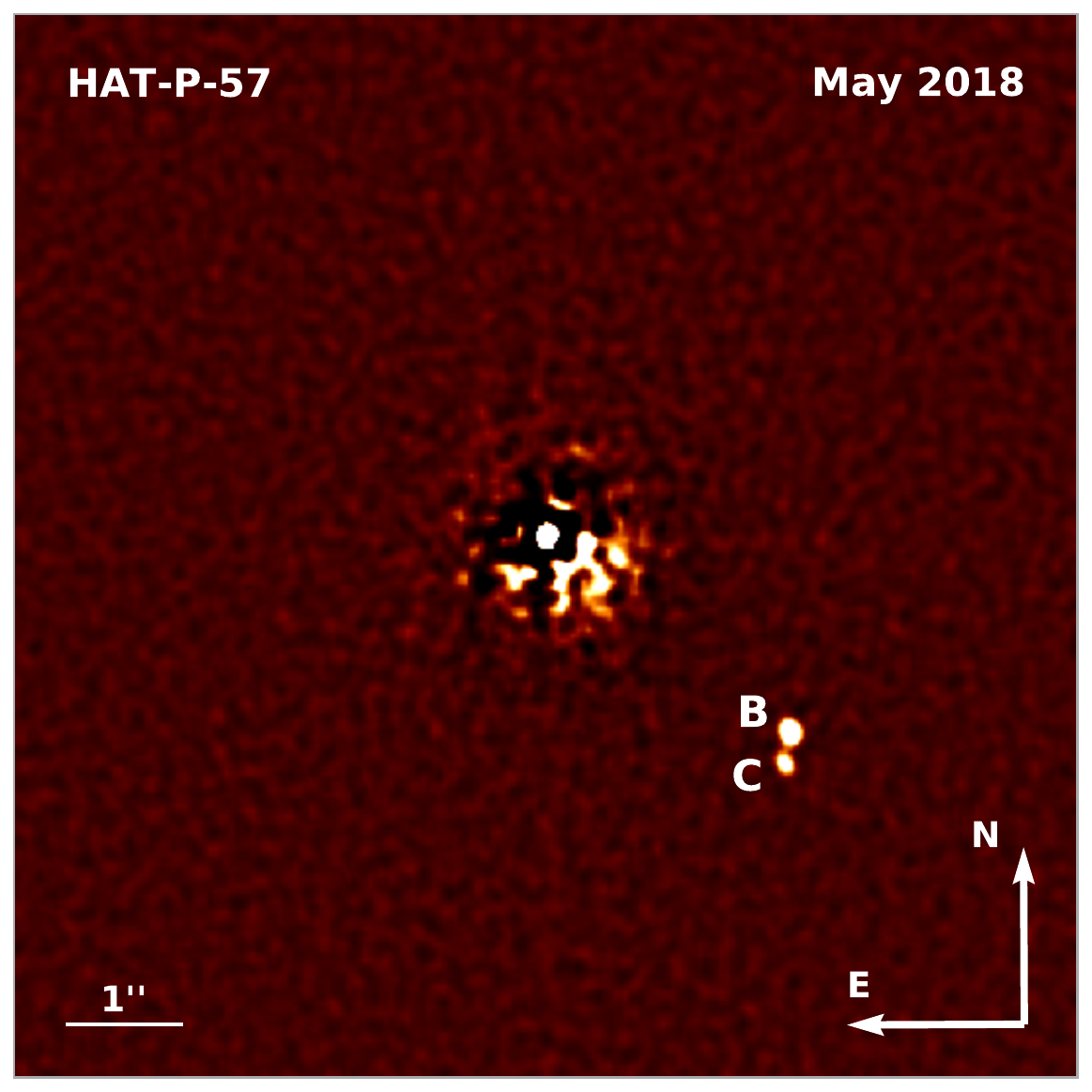}
\includegraphics[width=0.32\textwidth]{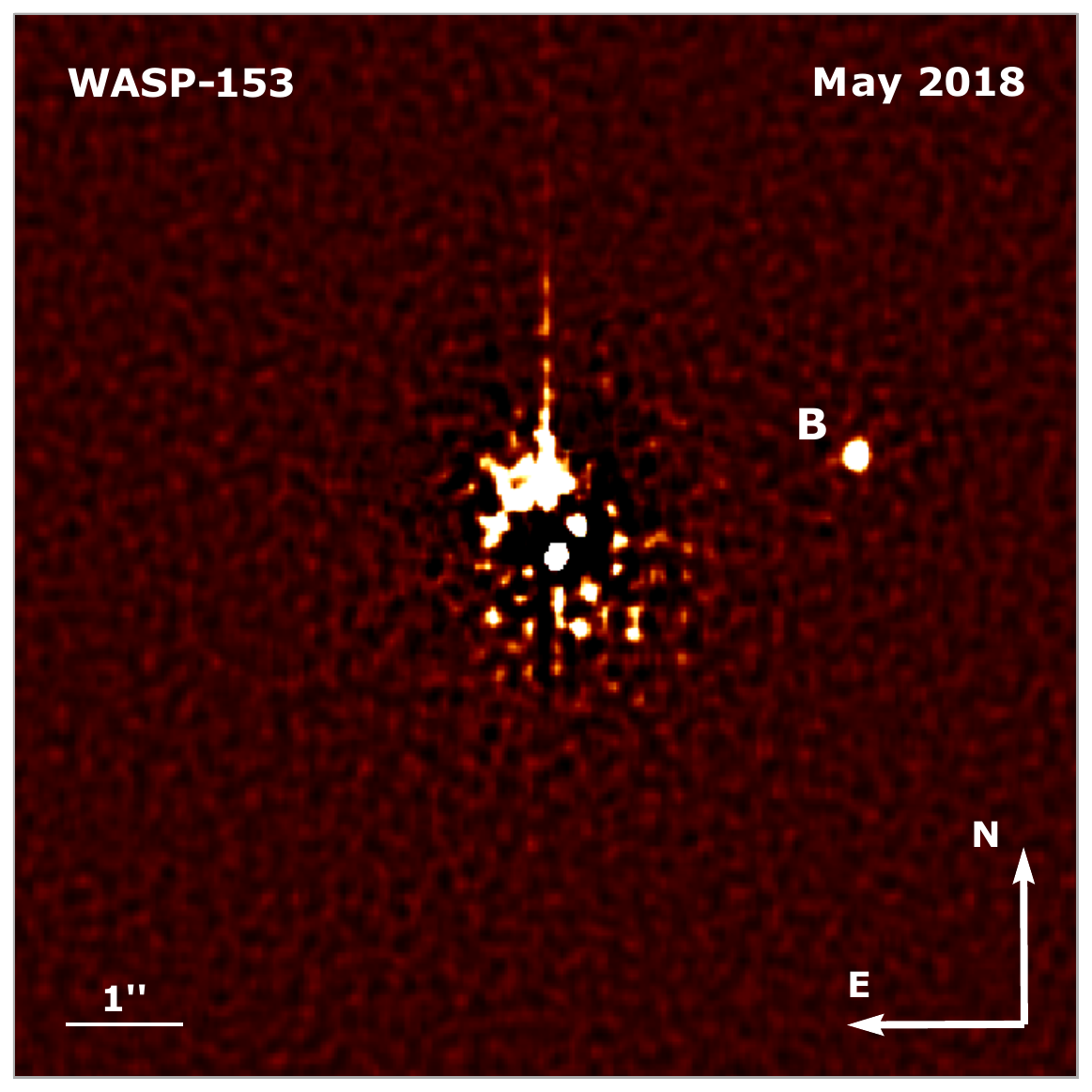}
\includegraphics[width=0.32\textwidth]{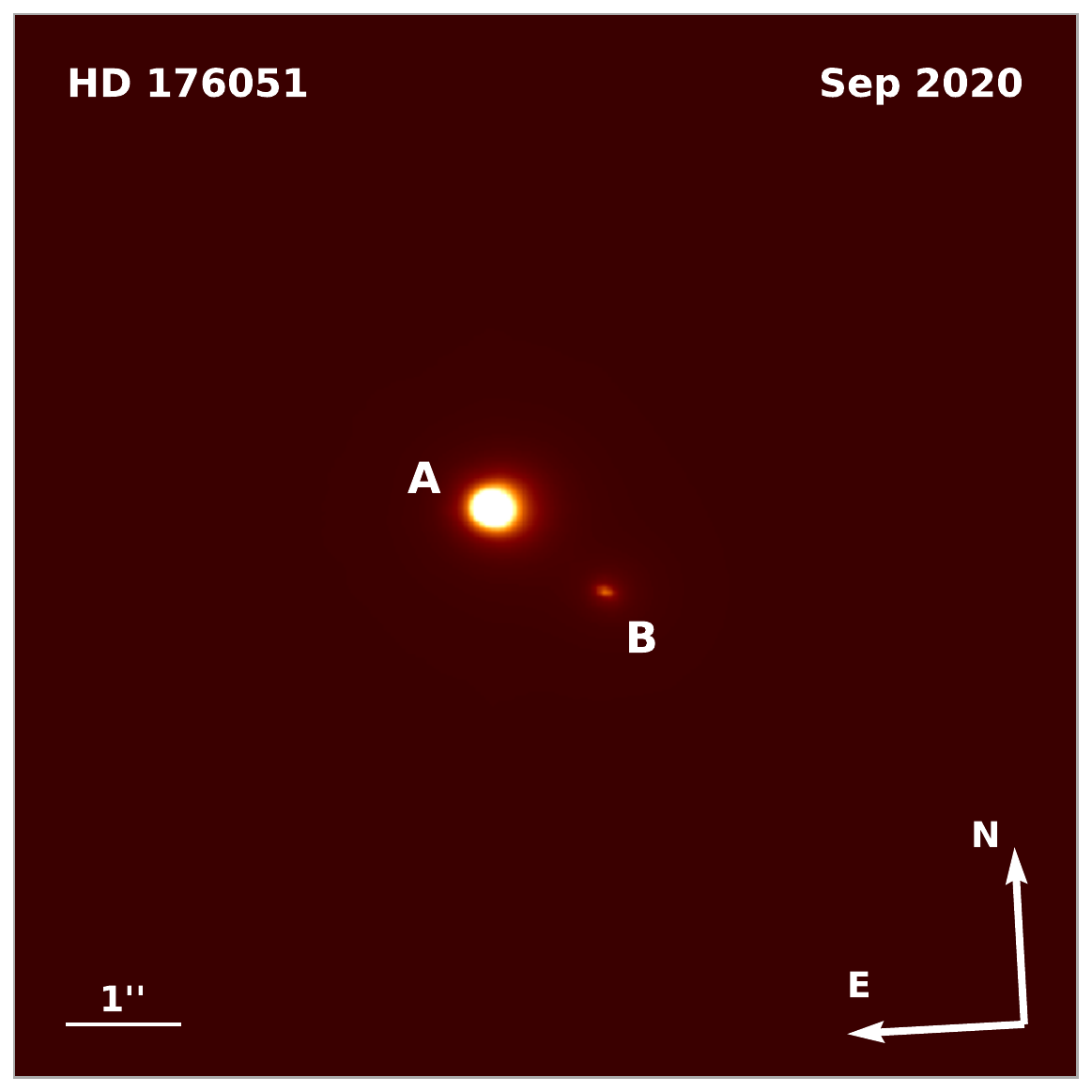}
\includegraphics[width=0.32\textwidth]{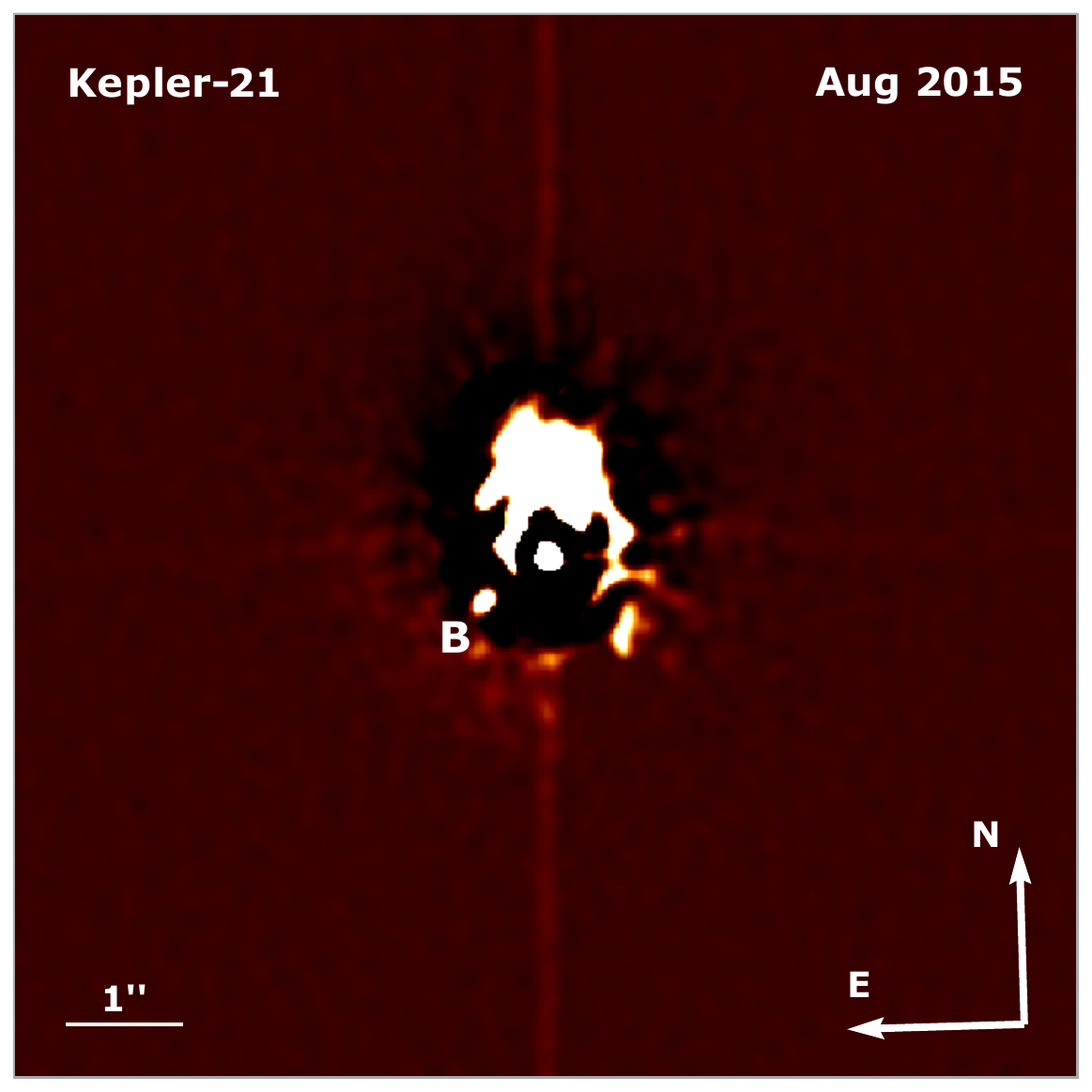}

\textbf{Figure\,\ref{fig_comps}} continued.
\end{figure*}
\begin{figure*}
\includegraphics[width=0.32\textwidth]{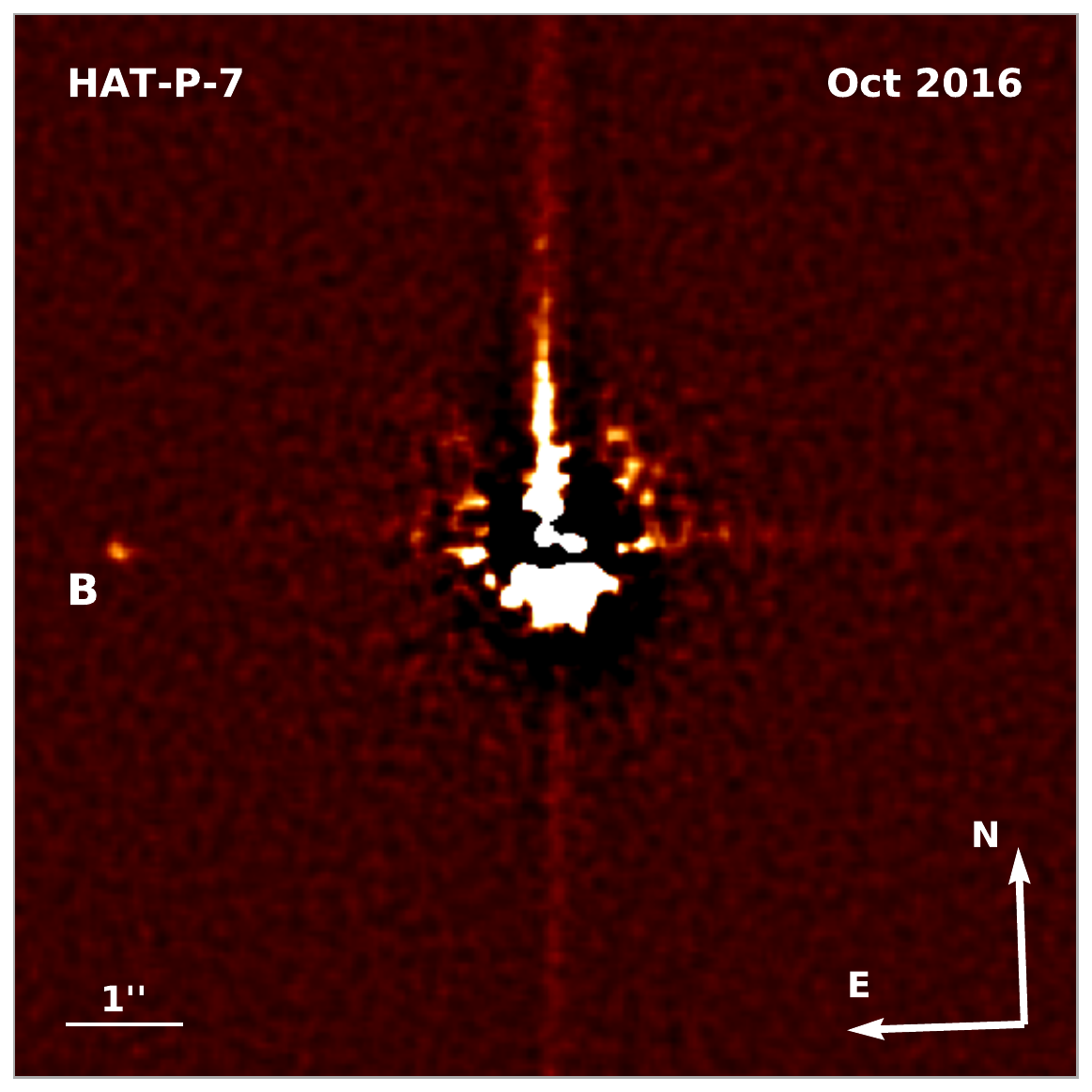}
\includegraphics[width=0.32\textwidth]{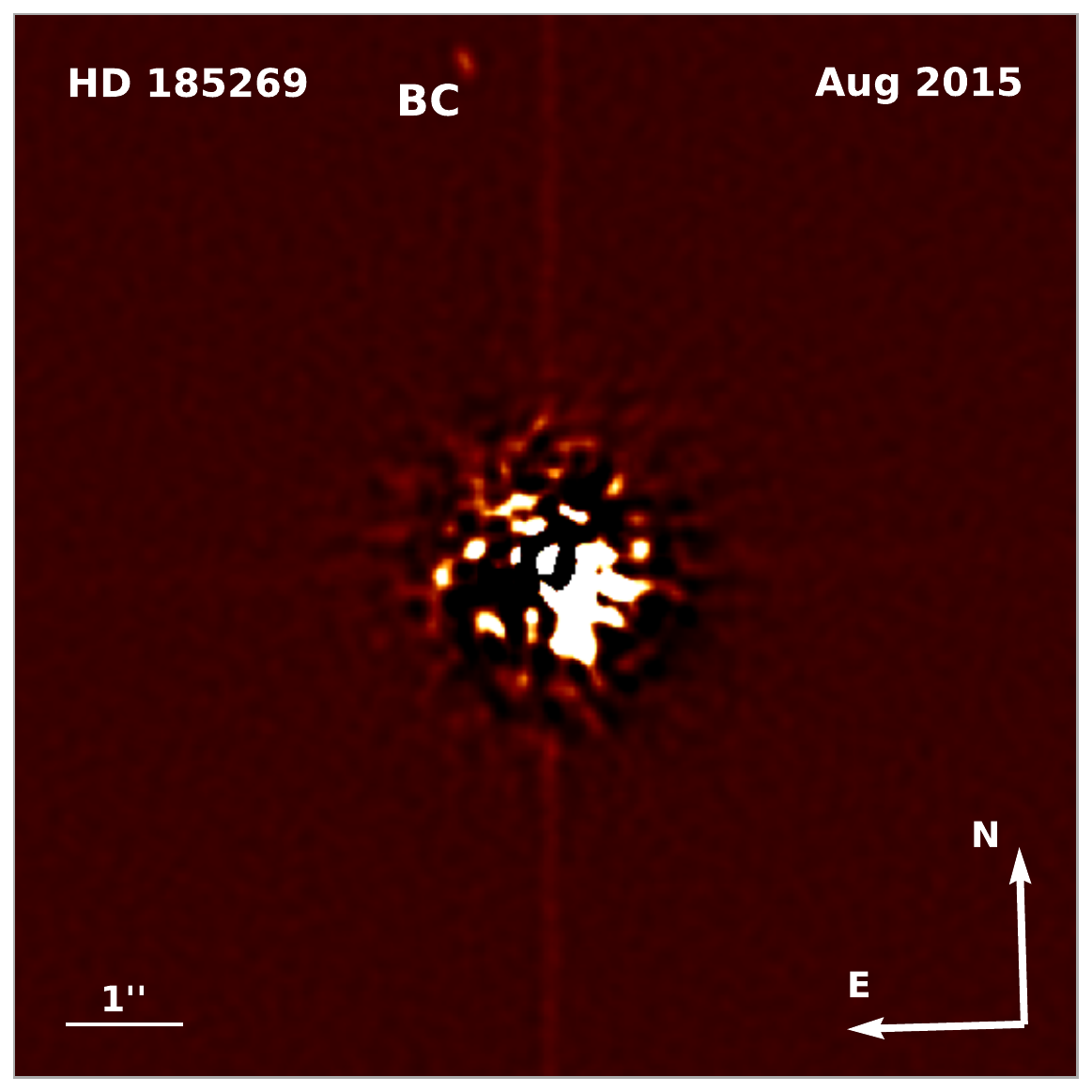}
\includegraphics[width=0.32\textwidth]{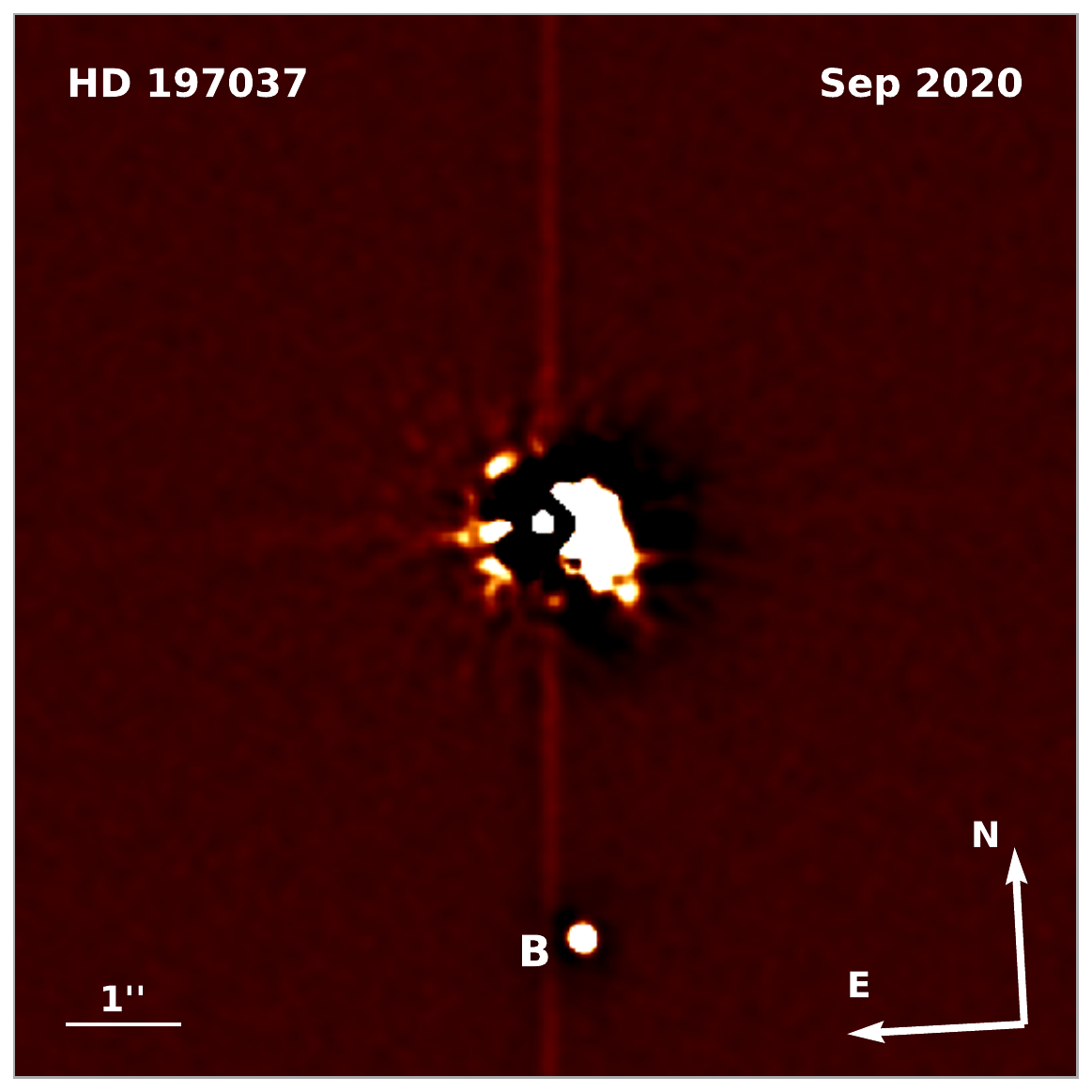}
\includegraphics[width=0.32\textwidth]{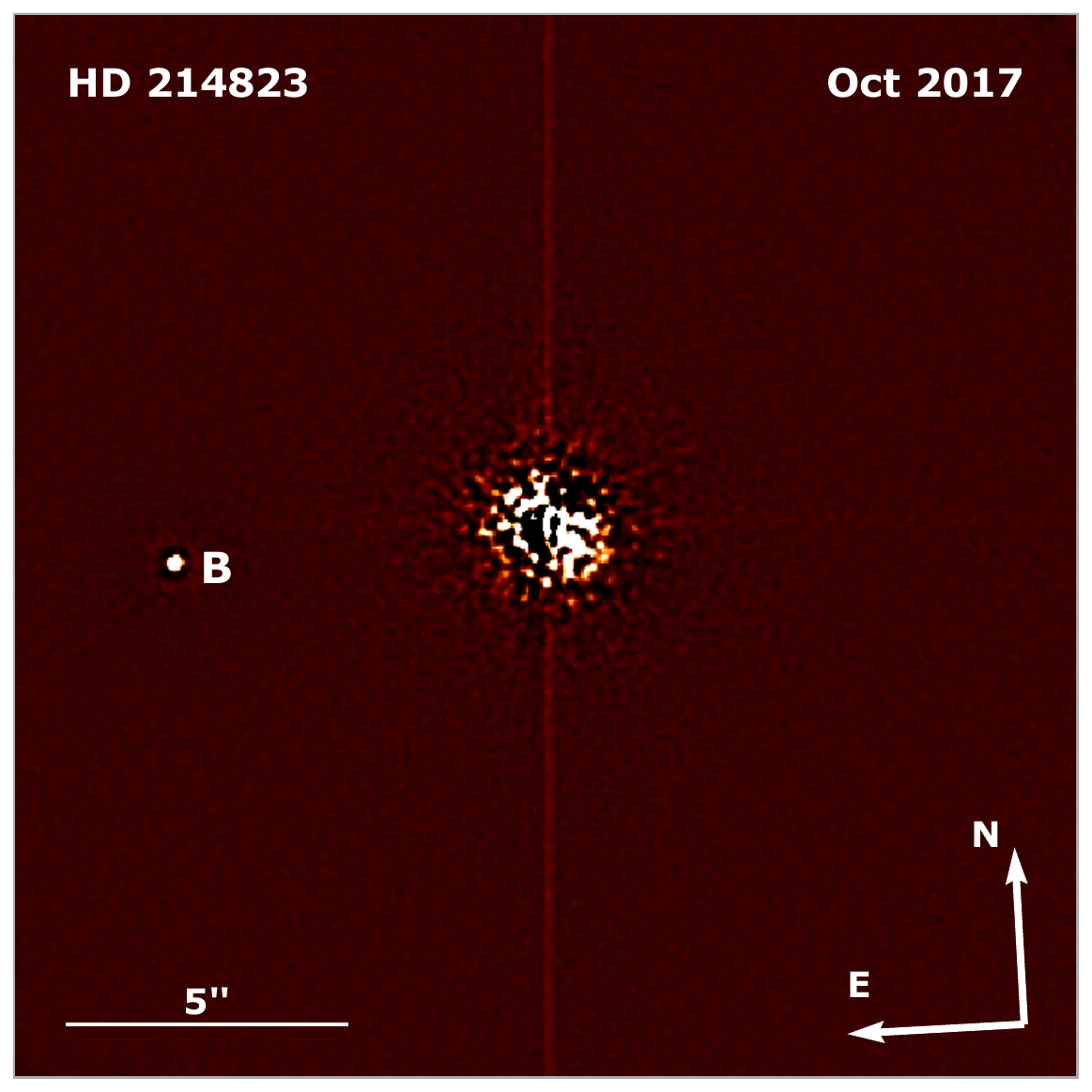}
\includegraphics[width=0.32\textwidth]{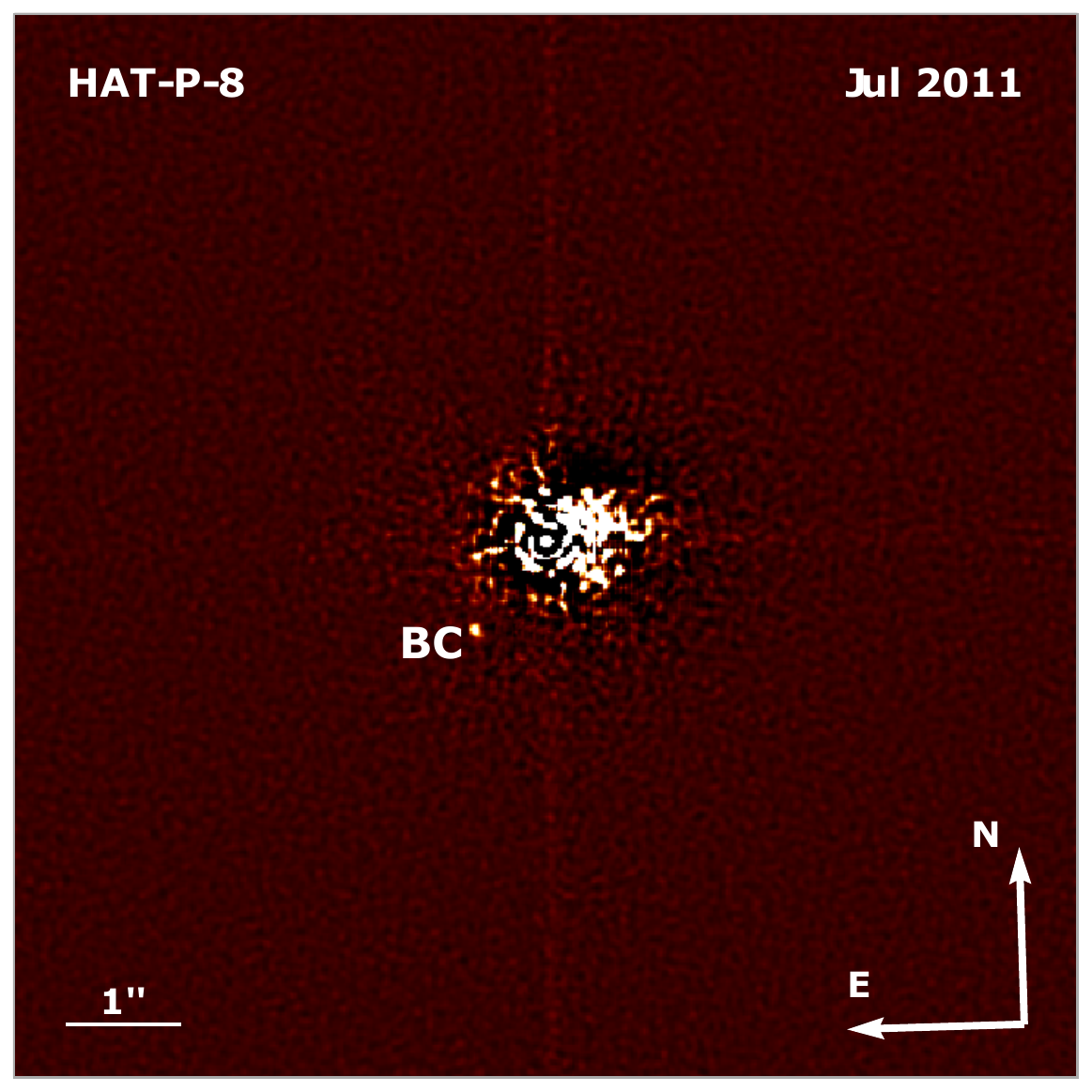}
\includegraphics[width=0.32\textwidth]{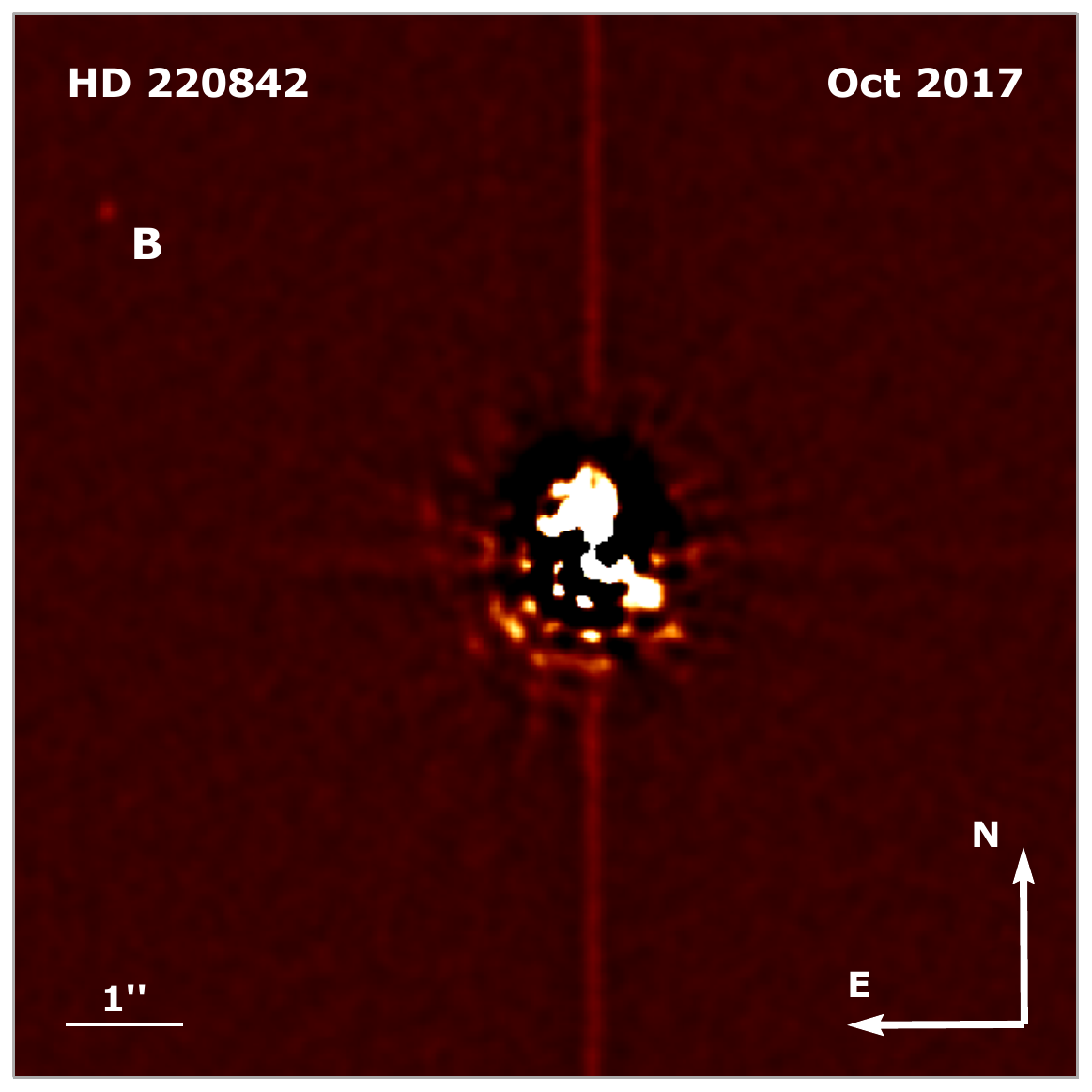}
\includegraphics[width=0.32\textwidth]{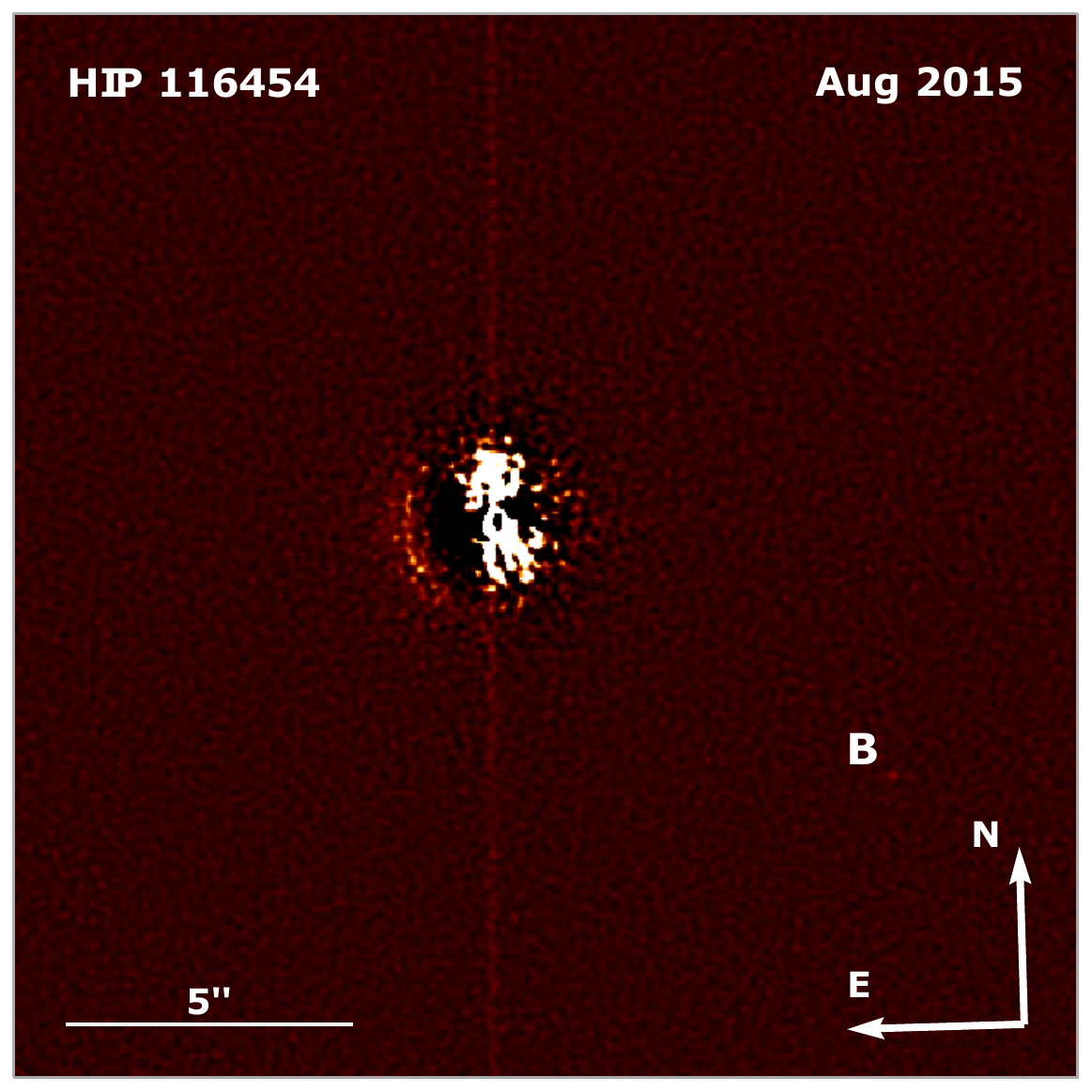}
\includegraphics[width=0.32\textwidth]{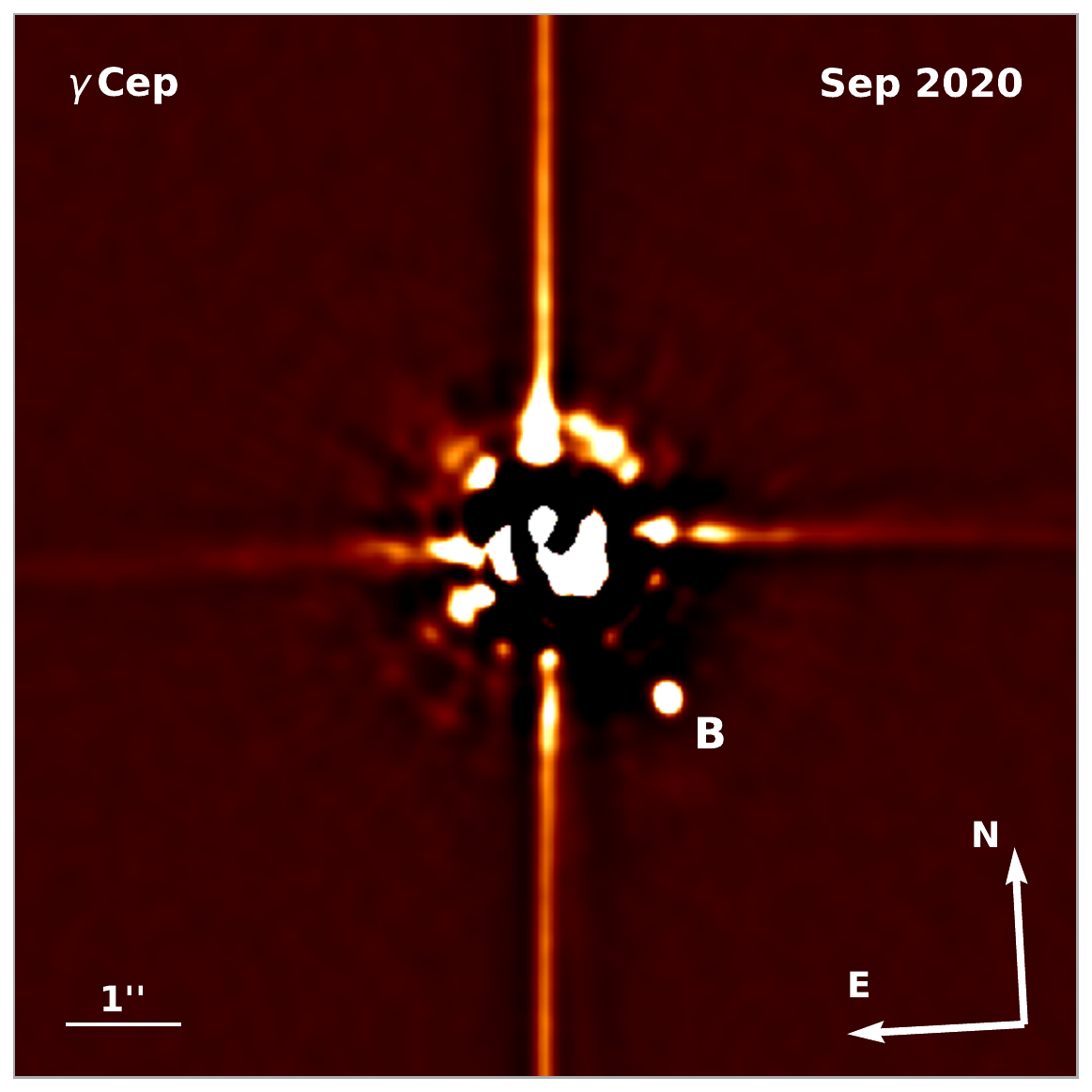}

\textbf{Figure\,\ref{fig_comps}} continued.
\end{figure*}

\section{Detection Limit}\label{sec_detlim}

For all targets the $S/N=3$ detection limit, as described in \cite{Ginski2012,Ginski2016}, is determined to explore the detection space of potential companions of the observed exoplanet host stars. Further companions than those reported above are not detected within the AstraLux field of view around the observed targets and can be excluded based on the detection limit achieved. The detection limits for all observed exoplanet host stars are shown in Figure\,\ref{fig_detlims}. The angular separation in arcsec (bottom axis) is converted to the projected separation in au (top axis) using the Gaia DR3 parallax of the targets. On the left-hand axis, the limiting apparent i$'$-band magnitude of a detectable companion is shown and converted to its mass (right-hand axis) using the given age of the targets and the stellar evolutionary models from \cite{Baraffe2015}.

\begin{figure}
\centering
\includegraphics[width=1\columnwidth]{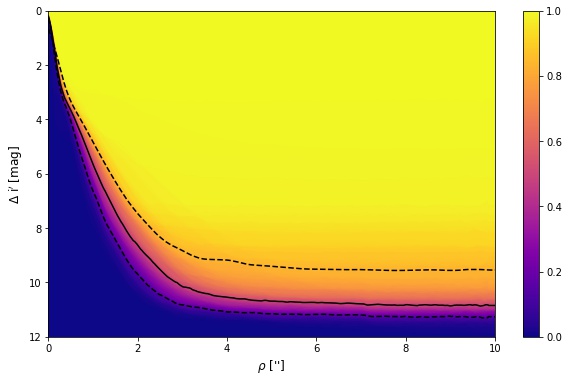}
\includegraphics[width=1\columnwidth]{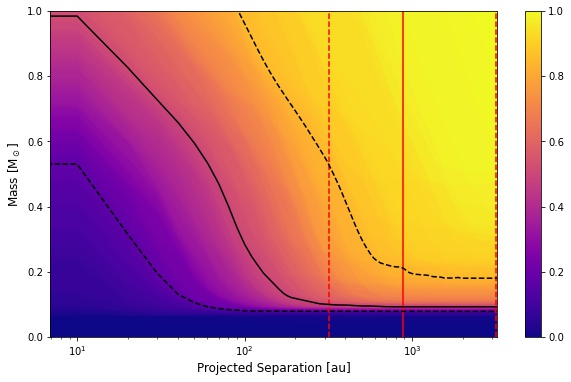}
\caption{\textbf{Top:} AstraLux detection limit for this survey. In the light yellow area, companions are detectable around all targets. In the dark blue area, companions are too faint and/or too close to be detectable with AstraLux around all exoplanet host stars. The solid black line represents the median of all detection limits.
\textbf{Bottom:} The detection space for close, low-mass stellar companions detectable in this survey, derived from the detection limits of all observed exoplanet host stars. The light yellow area indicates the detection space where companions can be imaged with AstraLux around all targets in this survey. The dark blue area shows the region where companions are not detectable. The solid black line is the median of the detection spaces of all targets, the dashed black lines are the 16\% and 84\% percentiles of their distribution. The red lines show the corresponding values of the maximum projected distance covered by the field of view of the AstraLux detector around all targets. This plot characterizes the overall detection space reached in this survey, the detection space of the individual stars is shown in Figure \ref{fig_detlims}.}
\label{fig_detlim}
\end{figure}

By comparing the AstraLux detection limits of all observed exoplanet host stars, the median detection limit of the entire survey can be determined. The relationship between the i$'$-band magnitude difference $\Delta \rm i'$ and the angular separation $\rho$ between detectable companions and the targets is shown in the top panel of Figure\,\ref{fig_detlim}. The solid black line represents the median of all detection limits and the dashed black lines indicate the 16\,\% and 84\,\% percentiles of their distribution. Within an angular separation of 1\,arcsec around the targets, companions with a magnitude difference of less than 5.7\,mag to the exoplanet host stars can be detected on median, while companions with i$'$-band magnitude differences of 8\,mag and 10\,mag are detectable beyond 1.7 and 2.9\,arcsec of angular separation, respectively. In the background limited region (beyond about 3.7\,arcsec) companions with a magnitude difference up to 10.5\,mag are detectable on median.

From the detection limit of each target, we derived the detection space for potential companions that can be imaged with AstraLux around all observed exoplanet host stars. The projected separation was derived from the angular separation and parallax of the targets. With parallax and extinction, the limiting apparent magnitude is converted into the limiting absolute magnitude of a detectable companion, which was converted into its mass using the age of the exoplanet host star and the stellar evolution models from \cite{Baraffe2015}. The derived detection space for low-mass stellar companions, i.e. minimum mass versus projected separation of a detectable companion, is shown in the bottom panel of Figure\,\ref{fig_detlim}. The solid red line represents the median of the maximum projected separations of companions that can be imaged with AstraLux in this survey around all targets. The dashed red lines represent the 16\,\% and 84\,\% percentiles of the corresponding distribution. On median, companions with a mass greater than 0.15\,$\rm M_\odot$ are detectable beyond about 160\,au of the projected distance. Companions with a mass of 0.5\,$\rm M_\odot$ (median mass of all detected companions) can be detected beyond 65\,au. The median of the maximum projected separation of the companions imaged in this survey is about 890\,au.

\section{Conclusion}

The multiplicity of exoplanet host stars is a key aspect of the diversity of exoplanets and helps to unravel planet formation and planetary system dynamics themselves, and also impacts the habitability of the planets. The Lucky Imaging technique has proven to be an excellent tool for finding stellar companions in these systems. Therefore, this study contributes to the understanding of the formation and evolution of planetary systems by investigating the multiplicity status of exoplanet host stars and by characterizing the properties of the detected stellar companions. The determined detection limit for each target further defines the detection space of potential companions of the observed exoplanet host stars. In total, 46 companions were found among 43 exoplanet host stars, which reside in 33 binary, and ten hierarchical triple star systems. One of the detected companions is the primary, 41 the secondary and four the tertiary component of their stellar systems. For all detected companions, we have determined their mass and projected separation. The companions have projected separations between 19\,au and 4731\,au and masses between 0.11\,$\rm M_\odot$ and 1.64\,$\rm M_\odot$.
With regards to planet formation, the closest binary systems, are the most interesting ones. The two planets HD\,2638\,B\,b and $\tau$\,Boo\,A\,b are both hot Jupiters with semi-major axis between 0.04\,au and 0.05\,au \citep{Paredes2021,Butler1997}. The planets around HD\,164509\,A, HD\,176051\,A and $\gamma$\,Cep\,A have minimum-masses of $0.48\,\rm M_{\rm{Jup}}$, $1.5\,\rm M_{\rm{Jup}}$, and 1.74\,$\rm M_{\rm{Jup}}$ with semi-major axes of 0.88\,au, 1.76\,au, and 2.15\,au, respectively \citep{Giguere2012,Muterspaugh2010,Huang2022}. Regarding the close stellar companions at projected separations of only 37\,au,  19\,au, and 22\,au, this indicates that these planets must have formed in a small protoplanetary disk, with only a few au, e.g. 4\,au in the case of $\gamma$\,Cep \citep{Jang-Condell2008}. This is especially remarkable in the $\gamma$\,Cep system where the true mass of the planet could already be determined via astrometry to be $9.4^{+0.7}_{-1.1}\,\rm M_{\rm{Jup}}$ \citep{Benedict2018}. The planetary system around HD\,87646\,A is an even more extreme case, as beside a massive planet ($m\sin(i) = 12.4 \pm 0.7\,\rm M_{\rm{Jup}}, a\sim0.1$\,au) also a brown dwarf companion ($m\sin(i) = 57.0 \pm 3.7\,\rm M_{\rm{Jup}}, a\sim1.6$\,au) has been detected close to the star, that both should have formed within that limited space around the star \citep{Ma2016}. Based on the determined detection space, additional companions with masses above 0.15\,$\rm M_\odot$ can be excluded outside a projected separation of 160\,au on median up to about 890\,au around the observed exoplanet host stars. According to our AstraLux observations, the sample of exoplanet host stars studied in this survey has a minimum multiplicity rate of about 20\,\%. The results of this survey, combined with those of other ongoing multiplicity studies of exoplanet host stars \citep[see e.g. ][]{Ginski2021}, will contribute to our understanding of planet formation and evolution in stellar systems.

\section*{Acknowledgements}

This work is based on observations made at the Centro Astron\'omico Hispano en Andaluc\'ia (CAHA) at Calar Alto, jointly operated by the Junta de Andaluc\'ia and the Consejo Superior de Investigaciones Cient\'ificas (IAA-CSIC). We would like to thank the members of the technical staff of the Calar Alto Observatory in Spain for all their help with the observations. In particular, we thank T. Eisenbeiss, A. Guijarro, C. Rodr\'iges L\'opez, and M. Seeliger for carrying out some of the observations.

MM would like to thank the German Science Foundation DFG for support in the programmes \mbox{MU2695/7-1}, \mbox{MU2695/12-1}, \mbox{MU2695/14-1}, \mbox{MU2695/20-1}, \mbox{MU2695/23-1}, \mbox{MU2695/24-1}, \mbox{MU2695/25-1}, \mbox{MU2695/28-1}, \mbox{MU2695/29-1} and \mbox{MU2695/30-1}.

MF acknowledges financial support from grants PID2019-109522GB-C52/AEI/10.13039/501100011033 of the Spanish Ministry of Science and Innovation (MICINN) and CEX2021-001131-S, funded by MCIN/AEI/ 10.13039/501100011033.

This research has used the SIMBAD and VizieR databases, both hosted at CDS in Strasbourg, France, and the Extrasolar Planets Encyclopaedia.

In addition, data were used from the European Space Agency (ESA) mission {\it Gaia} (\url{https://www.cosmos.esa.int/gaia}), processed by the {\it Gaia} Data Processing and Analysis Consortium (DPAC, \url{https://www.cosmos.esa.int web/gaia/dpac/consortium}). Funding for the DPAC has been provided by national institutions, in particular those participating in the {\it Gaia} Multilateral Agreement.

\section*{Data Availability}

Additional data can be found in the online appendices, namely:

\begin{itemize}

\item the observation log of this AstraLux survey
\item the summarized catalogue data of all targets
\item the astrometric and photometric observations
\item the AstraLux detection limit achieved for all targets

\end{itemize}

\bibliographystyle{mnras}
\bibliography{literature}

\begin{thebibliography}{}
\makeatletter
\relax
\def\mn@urlcharsother{\let\do\@makeother \do\$\do\&\do\#\do\^\do\_\do\%\do\~}
\def\mn@doi{\begingroup\mn@urlcharsother \@ifnextchar [ {\mn@doi@}
  {\mn@doi@[]}}
\def\mn@doi@[#1]#2{\def\@tempa{#1}\ifx\@tempa\@empty \href
  {http://dx.doi.org/#2} {doi:#2}\else \href {http://dx.doi.org/#2} {#1}\fi
  \endgroup}
\def\mn@eprint#1#2{\mn@eprint@#1:#2::\@nil}
\def\mn@eprint@arXiv#1{\href {http://arxiv.org/abs/#1} {{\tt arXiv:#1}}}
\def\mn@eprint@dblp#1{\href {http://dblp.uni-trier.de/rec/bibtex/#1.xml}
  {dblp:#1}}
\def\mn@eprint@#1:#2:#3:#4\@nil{\def\@tempa {#1}\def\@tempb {#2}\def\@tempc
  {#3}\ifx \@tempc \@empty \let \@tempc \@tempb \let \@tempb \@tempa \fi \ifx
  \@tempb \@empty \def\@tempb {arXiv}\fi \@ifundefined
  {mn@eprint@\@tempb}{\@tempb:\@tempc}{\expandafter \expandafter \csname
  mn@eprint@\@tempb\endcsname \expandafter{\@tempc}}}

\bibitem[\protect\citeauthoryear{{Anders} et~al.,}{{Anders}
  et~al.}{2019}]{Anders2019}
{Anders} F.,  et~al., 2019, \mn@doi [\aap] {10.1051/0004-6361/201935765}, \href
  {https://ui.adsabs.harvard.edu/abs/2019A&A...628A..94A} {628, A94}

\bibitem[\protect\citeauthoryear{{Anders} et~al.,}{{Anders}
  et~al.}{2022}]{Anders2022}
{Anders} F.,  et~al., 2022, \mn@doi [\aap] {10.1051/0004-6361/202142369}, \href
  {https://ui.adsabs.harvard.edu/abs/2022A&A...658A..91A} {658, A91}

\bibitem[\protect\citeauthoryear{{Baraffe}, {Homeier}, {Allard}  \&
  {Chabrier}}{{Baraffe} et~al.}{2015}]{Baraffe2015}
{Baraffe} I.,  {Homeier} D.,  {Allard} F.,   {Chabrier} G.,  2015, \mn@doi
  [\aap] {10.1051/0004-6361/201425481}, \href
  {https://ui.adsabs.harvard.edu/abs/2015A&A...577A..42B} {577, A42}

\bibitem[\protect\citeauthoryear{{Bechter} et~al.,}{{Bechter}
  et~al.}{2014}]{Bechter2014}
{Bechter} E.~B.,  et~al., 2014, \mn@doi [\apj] {10.1088/0004-637X/788/1/2},
  \href {https://ui.adsabs.harvard.edu/abs/2014ApJ...788....2B} {788, 2}

\bibitem[\protect\citeauthoryear{{Benedict}, {Harrison}, {Endl}  \&
  {Torres}}{{Benedict} et~al.}{2018}]{Benedict2018}
{Benedict} G.~F.,  {Harrison} T.~E.,  {Endl} M.,   {Torres} G.,  2018, \mn@doi
  [Research Notes of the American Astronomical Society]
  {10.3847/2515-5172/aabe7e}, \href
  {https://ui.adsabs.harvard.edu/abs/2018RNAAS...2....7B} {2, 7}

\bibitem[\protect\citeauthoryear{{Beuzit} et~al.,}{{Beuzit}
  et~al.}{2019}]{Beuzit2019}
{Beuzit} J.~L.,  et~al., 2019, \mn@doi [\aap] {10.1051/0004-6361/201935251},
  \href {https://ui.adsabs.harvard.edu/abs/2019A&A...631A.155B} {631, A155}

\bibitem[\protect\citeauthoryear{{Bohn}, {Southworth}, {Ginski}, {Kenworthy},
  {Maxted}  \& {Evans}}{{Bohn} et~al.}{2020}]{Bohn2020}
{Bohn} A.~J.,  {Southworth} J.,  {Ginski} C.,  {Kenworthy} M.~A.,  {Maxted}
  P.~F.~L.,   {Evans} D.~F.,  2020, \mn@doi [\aap]
  {10.1051/0004-6361/201937127}, \href
  {https://ui.adsabs.harvard.edu/abs/2020A&A...635A..73B} {635, A73}

\bibitem[\protect\citeauthoryear{{Butler}, {Marcy}, {Williams}, {Hauser}  \&
  {Shirts}}{{Butler} et~al.}{1997}]{Butler1997}
{Butler} R.~P.,  {Marcy} G.~W.,  {Williams} E.,  {Hauser} H.,   {Shirts} P.,
  1997, \mn@doi [\apjl] {10.1086/310444}, \href
  {https://ui.adsabs.harvard.edu/abs/1997ApJ...474L.115B} {474, L115}

\bibitem[\protect\citeauthoryear{{Cuntz}, {Luke}, {Millard}, {Boyle}  \&
  {Patel}}{{Cuntz} et~al.}{2022}]{Cuntz2022}
{Cuntz} M.,  {Luke} G.~E.,  {Millard} M.~J.,  {Boyle} L.,   {Patel} S.~D.,
  2022, \mn@doi [\apjs] {10.3847/1538-4365/ac9302}, \href
  {https://ui.adsabs.harvard.edu/abs/2022ApJS..263...33C} {263, 33}

\bibitem[\protect\citeauthoryear{{Desidera} et~al.,}{{Desidera}
  et~al.}{2011}]{Desidera2011}
{Desidera} S.,  et~al., 2011, \mn@doi [\aap] {10.1051/0004-6361/201117191},
  \href {https://ui.adsabs.harvard.edu/abs/2011A&A...533A..90D} {533, A90}

\bibitem[\protect\citeauthoryear{{Devillard}}{{Devillard}}{1997}]{Devillard1997}
{Devillard} N.,  1997, The Messenger, \href
  {https://ui.adsabs.harvard.edu/abs/1997Msngr..87...19D} {87, 19}

\bibitem[\protect\citeauthoryear{{Eastman} et~al.,}{{Eastman}
  et~al.}{2016}]{Eastman2016}
{Eastman} J.~D.,  et~al., 2016, \mn@doi [\aj] {10.3847/0004-6256/151/2/45},
  \href {https://ui.adsabs.harvard.edu/abs/2016AJ....151...45E} {151, 45}

\bibitem[\protect\citeauthoryear{{Fiorucci} \& {Munari}}{{Fiorucci} \&
  {Munari}}{2003}]{Fiorucci2003}
{Fiorucci} M.,  {Munari} U.,  2003, \mn@doi [\aap]
  {10.1051/0004-6361:20030075}, \href
  {https://ui.adsabs.harvard.edu/abs/2003A&A...401..781F} {401, 781}

\bibitem[\protect\citeauthoryear{{Fontanive} \& {Bardalez
  Gagliuffi}}{{Fontanive} \& {Bardalez Gagliuffi}}{2021}]{Fontanive2021}
{Fontanive} C.,  {Bardalez Gagliuffi} D.,  2021, \mn@doi [Frontiers in
  Astronomy and Space Sciences] {10.3389/fspas.2021.625250}, \href
  {https://ui.adsabs.harvard.edu/abs/2021FrASS...8...16F} {8, 16}

\bibitem[\protect\citeauthoryear{{Fukugita}, {Ichikawa}, {Gunn}, {Doi},
  {Shimasaku}  \& {Schneider}}{{Fukugita} et~al.}{1996}]{Fukugita1996}
{Fukugita} M.,  {Ichikawa} T.,  {Gunn} J.~E.,  {Doi} M.,  {Shimasaku} K.,
  {Schneider} D.~P.,  1996, \mn@doi [\aj] {10.1086/117915}, \href
  {https://ui.adsabs.harvard.edu/abs/1996AJ....111.1748F} {111, 1748}

\bibitem[\protect\citeauthoryear{{Gaia Collaboration} et~al.,}{{Gaia
  Collaboration} et~al.}{2016}]{GaiaCollaboration2016}
{Gaia Collaboration} et~al., 2016, \mn@doi [\aap]
  {10.1051/0004-6361/201629272}, \href
  {https://ui.adsabs.harvard.edu/abs/2016A&A...595A...1G} {595, A1}

\bibitem[\protect\citeauthoryear{{Gaia Collaboration} et~al.,}{{Gaia
  Collaboration} et~al.}{2018}]{GaiaCollaboration2018}
{Gaia Collaboration} et~al., 2018, \mn@doi [\aap]
  {10.1051/0004-6361/201833051}, \href
  {https://ui.adsabs.harvard.edu/abs/2018A&A...616A...1G} {616, A1}

\bibitem[\protect\citeauthoryear{{Gaia Collaboration} et~al.,}{{Gaia
  Collaboration} et~al.}{2023}]{GaiaCollaboration2023}
{Gaia Collaboration} et~al., 2023, \mn@doi [\aap]
  {10.1051/0004-6361/202243940}, \href
  {https://ui.adsabs.harvard.edu/abs/2023A&A...674A...1G} {674, A1}

\bibitem[\protect\citeauthoryear{{Giguere} et~al.,}{{Giguere}
  et~al.}{2012}]{Giguere2012}
{Giguere} M.~J.,  et~al., 2012, \mn@doi [\apj] {10.1088/0004-637X/744/1/4},
  \href {https://ui.adsabs.harvard.edu/abs/2012ApJ...744....4G} {744, 4}

\bibitem[\protect\citeauthoryear{{Ginski}, {Mugrauer}, {Seeliger}  \&
  {Eisenbeiss}}{{Ginski} et~al.}{2012}]{Ginski2012}
{Ginski} C.,  {Mugrauer} M.,  {Seeliger} M.,   {Eisenbeiss} T.,  2012, \mn@doi
  [\mnras] {10.1111/j.1365-2966.2012.20485.x}, \href
  {https://ui.adsabs.harvard.edu/abs/2012MNRAS.421.2498G} {421, 2498}

\bibitem[\protect\citeauthoryear{{Ginski}, {Schmidt}, {Mugrauer},
  {Neuh{\"a}user}, {Vogt}, {Errmann}  \& {Berndt}}{{Ginski}
  et~al.}{2014}]{Ginski2014}
{Ginski} C.,  {Schmidt} T.~O.~B.,  {Mugrauer} M.,  {Neuh{\"a}user} R.,  {Vogt}
  N.,  {Errmann} R.,   {Berndt} A.,  2014, \mn@doi [\mnras]
  {10.1093/mnras/stu1586}, \href
  {https://ui.adsabs.harvard.edu/abs/2014MNRAS.444.2280G} {444, 2280}

\bibitem[\protect\citeauthoryear{{Ginski} et~al.,}{{Ginski}
  et~al.}{2016}]{Ginski2016}
{Ginski} C.,  et~al., 2016, \mn@doi [\mnras] {10.1093/mnras/stw049}, \href
  {https://ui.adsabs.harvard.edu/abs/2016MNRAS.457.2173G} {457, 2173}

\bibitem[\protect\citeauthoryear{{Ginski}, {Mugrauer}, {Adam}, {Vogt}  \& {van
  Holstein}}{{Ginski} et~al.}{2021}]{Ginski2021}
{Ginski} C.,  {Mugrauer} M.,  {Adam} C.,  {Vogt} N.,   {van Holstein} R.~G.,
  2021, \mn@doi [\aap] {10.1051/0004-6361/202038964}, \href
  {https://ui.adsabs.harvard.edu/abs/2021A&A...649A.156G} {649, A156}

\bibitem[\protect\citeauthoryear{{Han}, {Lee}, {Kim}, {Mkrtichian}, {Hatzes}
  \& {Valyavin}}{{Han} et~al.}{2010}]{Han2010}
{Han} I.,  {Lee} B.~C.,  {Kim} K.~M.,  {Mkrtichian} D.~E.,  {Hatzes} A.~P.,
  {Valyavin} G.,  2010, \mn@doi [\aap] {10.1051/0004-6361/200912536}, \href
  {https://ui.adsabs.harvard.edu/abs/2010A&A...509A..24H} {509, A24}

\bibitem[\protect\citeauthoryear{{Harakawa} et~al.,}{{Harakawa}
  et~al.}{2015}]{Harakawa2015}
{Harakawa} H.,  et~al., 2015, \mn@doi [\apj] {10.1088/0004-637X/806/1/5}, \href
  {https://ui.adsabs.harvard.edu/abs/2015ApJ...806....5H} {806, 5}

\bibitem[\protect\citeauthoryear{{Henden}, {Levine}, {Terrell}  \&
  {Welch}}{{Henden} et~al.}{2015}]{Henden2015}
{Henden} A.~A.,  {Levine} S.,  {Terrell} D.,   {Welch} D.~L.,  2015, in
  American Astronomical Society Meeting Abstracts \#225. p. 336.16

\bibitem[\protect\citeauthoryear{{Henden}, {Levine}, {Terrell}, {Welch},
  {Munari}  \& {Kloppenborg}}{{Henden} et~al.}{2018}]{Henden2018}
{Henden} A.~A.,  {Levine} S.,  {Terrell} D.,  {Welch} D.~L.,  {Munari} U.,
  {Kloppenborg} B.~K.,  2018, in American Astronomical Society Meeting
  Abstracts \#232. p. 223.06

\bibitem[\protect\citeauthoryear{{Holman} \& {Wiegert}}{{Holman} \&
  {Wiegert}}{1999}]{Holman1999}
{Holman} M.~J.,  {Wiegert} P.~A.,  1999, \mn@doi [\aj] {10.1086/300695}, \href
  {https://ui.adsabs.harvard.edu/abs/1999AJ....117..621H} {117, 621}

\bibitem[\protect\citeauthoryear{{Hormuth}, {Hippler}, {Brandner}, {Wagner}  \&
  {Henning}}{{Hormuth} et~al.}{2008}]{Hormuth2008}
{Hormuth} F.,  {Hippler} S.,  {Brandner} W.,  {Wagner} K.,   {Henning} T.,
  2008, in {McLean} I.~S.,  {Casali} M.~M.,  eds,  Society of Photo-Optical
  Instrumentation Engineers (SPIE) Conference Series Vol. 7014, Ground-based
  and Airborne Instrumentation for Astronomy II. p. 701448 (\mn@eprint {arXiv}
  {0807.0497}), \mn@doi{10.1117/12.787384}

\bibitem[\protect\citeauthoryear{{Huang} \& {Ji}}{{Huang} \&
  {Ji}}{2022}]{Huang2022}
{Huang} X.,  {Ji} J.,  2022, \mn@doi [\aj] {10.3847/1538-3881/ac8f4c}, \href
  {https://ui.adsabs.harvard.edu/abs/2022AJ....164..177H} {164, 177}

\bibitem[\protect\citeauthoryear{{Ilic}, {Poppenhaeger}  \& {Hosseini}}{{Ilic}
  et~al.}{2022}]{Ilic2022}
{Ilic} N.,  {Poppenhaeger} K.,   {Hosseini} S.~M.,  2022, \mn@doi [\mnras]
  {10.1093/mnras/stac861}, \href
  {https://ui.adsabs.harvard.edu/abs/2022MNRAS.513.4380I} {513, 4380}

\bibitem[\protect\citeauthoryear{{Jang-Condell}, {Mugrauer}  \&
  {Schmidt}}{{Jang-Condell} et~al.}{2008}]{Jang-Condell2008}
{Jang-Condell} H.,  {Mugrauer} M.,   {Schmidt} T.,  2008, \mn@doi [\apjl]
  {10.1086/591791}, \href
  {https://ui.adsabs.harvard.edu/abs/2008ApJ...683L.191J} {683, L191}

\bibitem[\protect\citeauthoryear{{Johnson} et~al.,}{{Johnson}
  et~al.}{2011}]{Johnson2011}
{Johnson} J.~A.,  et~al., 2011, \mn@doi [\apjs] {10.1088/0067-0049/197/2/26},
  \href {https://ui.adsabs.harvard.edu/abs/2011ApJS..197...26J} {197, 26}

\bibitem[\protect\citeauthoryear{{Jordi}, {Grebel}  \& {Ammon}}{{Jordi}
  et~al.}{2006}]{Jordi2006}
{Jordi} K.,  {Grebel} E.~K.,   {Ammon} K.,  2006, \mn@doi [\aap]
  {10.1051/0004-6361:20066082}, \href
  {https://ui.adsabs.harvard.edu/abs/2006A&A...460..339J} {460, 339}

\bibitem[\protect\citeauthoryear{{Kaib}, {Raymond}  \& {Duncan}}{{Kaib}
  et~al.}{2013}]{Kaib2013}
{Kaib} N.~A.,  {Raymond} S.~N.,   {Duncan} M.,  2013, \mn@doi [\nat]
  {10.1038/nature11780}, \href
  {https://ui.adsabs.harvard.edu/abs/2013Natur.493..381K} {493, 381}

\bibitem[\protect\citeauthoryear{{Ma} et~al.,}{{Ma} et~al.}{2016}]{Ma2016}
{Ma} B.,  et~al., 2016, \mn@doi [\aj] {10.3847/0004-6256/152/5/112}, \href
  {https://ui.adsabs.harvard.edu/abs/2016AJ....152..112M} {152, 112}

\bibitem[\protect\citeauthoryear{{Mann}, {Feiden}, {Gaidos}, {Boyajian}  \&
  {von Braun}}{{Mann} et~al.}{2015}]{Mann2015}
{Mann} A.~W.,  {Feiden} G.~A.,  {Gaidos} E.,  {Boyajian} T.,   {von Braun} K.,
  2015, \mn@doi [\apj] {10.1088/0004-637X/804/1/64}, \href
  {https://ui.adsabs.harvard.edu/abs/2015ApJ...804...64M} {804, 64}

\bibitem[\protect\citeauthoryear{{Marzari} \& {Scholl}}{{Marzari} \&
  {Scholl}}{2000}]{Marzari2000}
{Marzari} F.,  {Scholl} H.,  2000, \mn@doi [\apj] {10.1086/317091}, \href
  {https://ui.adsabs.harvard.edu/abs/2000ApJ...543..328M} {543, 328}

\bibitem[\protect\citeauthoryear{{Mason}, {Wycoff}, {Hartkopf}, {Douglass}  \&
  {Worley}}{{Mason} et~al.}{2001}]{Mason2001}
{Mason} B.~D.,  {Wycoff} G.~L.,  {Hartkopf} W.~I.,  {Douglass} G.~G.,
  {Worley} C.~E.,  2001, \mn@doi [\aj] {10.1086/323920}, \href
  {https://ui.adsabs.harvard.edu/abs/2001AJ....122.3466M} {122, 3466}

\bibitem[\protect\citeauthoryear{{Mayer}, {Wadsley}, {Quinn}  \&
  {Stadel}}{{Mayer} et~al.}{2005}]{Mayer2005}
{Mayer} L.,  {Wadsley} J.,  {Quinn} T.,   {Stadel} J.,  2005, \mn@doi [\mnras]
  {10.1111/j.1365-2966.2005.09468.x}, \href
  {https://ui.adsabs.harvard.edu/abs/2005MNRAS.363..641M} {363, 641}

\bibitem[\protect\citeauthoryear{{Meshkat} et~al.,}{{Meshkat}
  et~al.}{2015}]{Meshkat2015}
{Meshkat} T.,  et~al., 2015, \mn@doi [\mnras] {10.1093/mnras/stv1732}, \href
  {https://ui.adsabs.harvard.edu/abs/2015MNRAS.453.2378M} {453, 2378}

\bibitem[\protect\citeauthoryear{{Michel} \& {Mugrauer}}{{Michel} \&
  {Mugrauer}}{2021}]{Michel2021}
{Michel} K.~U.,  {Mugrauer} M.,  2021, \mn@doi [Frontiers in Astronomy and
  Space Sciences] {10.3389/fspas.2021.624907}, \href
  {https://ui.adsabs.harvard.edu/abs/2021FrASS...8...14M} {8, 14}

\bibitem[\protect\citeauthoryear{{Michel} \& {Mugrauer}}{{Michel} \&
  {Mugrauer}}{2024}]{Michel2024}
{Michel} K.-U.,  {Mugrauer} M.,  2024, \mn@doi [\mnras]
  {10.1093/mnras/stad3196}, \href
  {https://ui.adsabs.harvard.edu/abs/2024MNRAS.527.3183M} {527, 3183}

\bibitem[\protect\citeauthoryear{{Mugrauer}}{{Mugrauer}}{2019}]{Mugrauer2019}
{Mugrauer} M.,  2019, \mn@doi [\mnras] {10.1093/mnras/stz2673}, \href
  {https://ui.adsabs.harvard.edu/abs/2019MNRAS.490.5088M} {490, 5088}

\bibitem[\protect\citeauthoryear{{Mugrauer} \& {Ginski}}{{Mugrauer} \&
  {Ginski}}{2015}]{Mugrauer2015}
{Mugrauer} M.,  {Ginski} C.,  2015, \mn@doi [\mnras] {10.1093/mnras/stv771},
  \href {https://ui.adsabs.harvard.edu/abs/2015MNRAS.450.3127M} {450, 3127}

\bibitem[\protect\citeauthoryear{{Mugrauer} \& {Michel}}{{Mugrauer} \&
  {Michel}}{2020}]{Mugrauer2020}
{Mugrauer} M.,  {Michel} K.-U.,  2020, \mn@doi [Astronomische Nachrichten]
  {10.1002/asna.202013825}, \href
  {https://ui.adsabs.harvard.edu/abs/2020AN....341..996M} {341, 996}

\bibitem[\protect\citeauthoryear{{Mugrauer} \& {Michel}}{{Mugrauer} \&
  {Michel}}{2021}]{Mugrauer2021}
{Mugrauer} M.,  {Michel} K.-U.,  2021, \mn@doi [Astronomische Nachrichten]
  {10.1002/asna.202113972}, \href
  {https://ui.adsabs.harvard.edu/abs/2021AN....342..840M} {342, 840}

\bibitem[\protect\citeauthoryear{{Mugrauer}, {Zander}  \& {Michel}}{{Mugrauer}
  et~al.}{2022a}]{Mugrauer2022b}
{Mugrauer} M.,  {Zander} J.,   {Michel} K.-U.,  2022a, \mn@doi [Astronomische
  Nachrichten] {10.1002/asna.20224017}, \href
  {https://ui.adsabs.harvard.edu/abs/2022AN....34324017M} {343, e24017}

\bibitem[\protect\citeauthoryear{{Mugrauer}, {Schlagenhauf}, {Buder}, {Ginski}
  \& {Fern{\'a}ndez}}{{Mugrauer} et~al.}{2022b}]{Mugrauer2022a}
{Mugrauer} M.,  {Schlagenhauf} S.,  {Buder} S.,  {Ginski} C.,   {Fern{\'a}ndez}
  M.,  2022b, \mn@doi [Astronomische Nachrichten] {10.1002/asna.20224014},
  \href {https://ui.adsabs.harvard.edu/abs/2022AN....34324014M} {343, e24014}

\bibitem[\protect\citeauthoryear{{Mugrauer}, {R{\"u}ck}  \&
  {Michel}}{{Mugrauer} et~al.}{2023}]{Mugrauer2023}
{Mugrauer} M.,  {R{\"u}ck} J.,   {Michel} K.~U.,  2023, \mn@doi [Astronomische
  Nachrichten] {10.1002/asna.20230055}, \href
  {https://ui.adsabs.harvard.edu/abs/2023AN....34430055M} {344, e20230055}

\bibitem[\protect\citeauthoryear{{Muterspaugh} et~al.,}{{Muterspaugh}
  et~al.}{2010}]{Muterspaugh2010}
{Muterspaugh} M.~W.,  et~al., 2010, \mn@doi [\aj]
  {10.1088/0004-6256/140/6/1657}, \href
  {https://ui.adsabs.harvard.edu/abs/2010AJ....140.1657M} {140, 1657}

\bibitem[\protect\citeauthoryear{{Ngo} et~al.,}{{Ngo} et~al.}{2017}]{Ngo2017}
{Ngo} H.,  et~al., 2017, \mn@doi [\aj] {10.3847/1538-3881/aa6cac}, \href
  {https://ui.adsabs.harvard.edu/abs/2017AJ....153..242N} {153, 242}

\bibitem[\protect\citeauthoryear{{Ofek}}{{Ofek}}{2008}]{Ofek2008}
{Ofek} E.~O.,  2008, \mn@doi [\pasp] {10.1086/592456}, \href
  {https://ui.adsabs.harvard.edu/abs/2008PASP..120.1128O} {120, 1128}

\bibitem[\protect\citeauthoryear{{Paredes}, {Henry}, {Quinn}, {Gies},
  {Hinojosa-Go{\~n}i}, {James}, {Jao}  \& {White}}{{Paredes}
  et~al.}{2021}]{Paredes2021}
{Paredes} L.~A.,  {Henry} T.~J.,  {Quinn} S.~N.,  {Gies} D.~R.,
  {Hinojosa-Go{\~n}i} R.,  {James} H.-S.,  {Jao} W.-C.,   {White} R.~J.,  2021,
  \mn@doi [\aj] {10.3847/1538-3881/ac082a}, \href
  {https://ui.adsabs.harvard.edu/abs/2021AJ....162..176P} {162, 176}

\bibitem[\protect\citeauthoryear{{Pecaut} \& {Mamajek}}{{Pecaut} \&
  {Mamajek}}{2013}]{Pecaut2013}
{Pecaut} M.~J.,  {Mamajek} E.~E.,  2013, \mn@doi [\apjs]
  {10.1088/0067-0049/208/1/9}, \href
  {https://ui.adsabs.harvard.edu/abs/2013ApJS..208....9P} {208, 9}

\bibitem[\protect\citeauthoryear{{Ricker} et~al.,}{{Ricker}
  et~al.}{2015}]{Ricker2015}
{Ricker} G.~R.,  et~al., 2015, \mn@doi [Journal of Astronomical Telescopes,
  Instruments, and Systems] {10.1117/1.JATIS.1.1.014003}, \href
  {https://ui.adsabs.harvard.edu/abs/2015JATIS...1a4003R} {1, 014003}

\bibitem[\protect\citeauthoryear{{Romanenko} \& {Kiselev}}{{Romanenko} \&
  {Kiselev}}{2014}]{Romanenko2014}
{Romanenko} L.~G.,  {Kiselev} A.~A.,  2014, \mn@doi [Astronomy Reports]
  {10.1134/S1063772914010053}, \href
  {https://ui.adsabs.harvard.edu/abs/2014ARep...58...30R} {58, 30}

\bibitem[\protect\citeauthoryear{{Thebault} \& {Haghighipour}}{{Thebault} \&
  {Haghighipour}}{2015}]{Thebault2015}
{Thebault} P.,  {Haghighipour} N.,  2015, in , Planetary Exploration and
  Science: Recent Results and Advances.
Springer Geophysics, pp 309--340, \mn@doi{10.1007/978-3-662-45052-9_13}

\bibitem[\protect\citeauthoryear{{Warmels}}{{Warmels}}{1992}]{Warmels1992}
{Warmels} R.~H.,  1992, in {Worrall} D.~M.,  {Biemesderfer} C.,   {Barnes} J.,
  eds,  Astronomical Society of the Pacific Conference Series Vol. 25,
  Astronomical Data Analysis Software and Systems I. p.~115

\bibitem[\protect\citeauthoryear{{W{\"o}llert} \& {Brandner}}{{W{\"o}llert} \&
  {Brandner}}{2015}]{Woellert2015}
{W{\"o}llert} M.,  {Brandner} W.,  2015, \mn@doi [\aap]
  {10.1051/0004-6361/201526525}, \href
  {https://ui.adsabs.harvard.edu/abs/2015A&A...579A.129W} {579, A129}

\bibitem[\protect\citeauthoryear{{Wu}, {Murray}  \& {Ramsahai}}{{Wu}
  et~al.}{2007}]{Wu2007}
{Wu} Y.,  {Murray} N.~W.,   {Ramsahai} J.~M.,  2007, \mn@doi [\apj]
  {10.1086/521996}, \href
  {https://ui.adsabs.harvard.edu/abs/2007ApJ...670..820W} {670, 820}

\bibitem[\protect\citeauthoryear{{Yee}, {Petigura}  \& {von Braun}}{{Yee}
  et~al.}{2017}]{Yee2017}
{Yee} S.~W.,  {Petigura} E.~A.,   {von Braun} K.,  2017, \mn@doi [\apj]
  {10.3847/1538-4357/836/1/77}, \href
  {https://ui.adsabs.harvard.edu/abs/2017ApJ...836...77Y} {836, 77}

\bibitem[\protect\citeauthoryear{{Zacharias}, {Finch}, {Girard}, {Henden},
  {Bartlett}, {Monet}  \& {Zacharias}}{{Zacharias}
  et~al.}{2013}]{Zacharias2013}
{Zacharias} N.,  {Finch} C.~T.,  {Girard} T.~M.,  {Henden} A.,  {Bartlett}
  J.~L.,  {Monet} D.~G.,   {Zacharias} M.~I.,  2013, \mn@doi [\aj]
  {10.1088/0004-6256/145/2/44}, \href
  {https://ui.adsabs.harvard.edu/abs/2013AJ....145...44Z} {145, 44}

\bibitem[\protect\citeauthoryear{{Zacharias} et~al.,}{{Zacharias}
  et~al.}{2015}]{Zacharias2015}
{Zacharias} N.,  et~al., 2015, \mn@doi [\aj] {10.1088/0004-6256/150/4/101},
  \href {https://ui.adsabs.harvard.edu/abs/2015AJ....150..101Z} {150, 101}

\bibitem[\protect\citeauthoryear{{da Silva}, {Milone}  \& {Rocha-Pinto}}{{da
  Silva} et~al.}{2015}]{daSilva2015}
{da Silva} R.,  {Milone} A. d.~C.,   {Rocha-Pinto} H.~J.,  2015, \mn@doi [\aap]
  {10.1051/0004-6361/201525770}, \href
  {https://ui.adsabs.harvard.edu/abs/2015A&A...580A..24D} {580, A24}

\bibitem[\protect\citeauthoryear{{van Leeuwen}}{{van
  Leeuwen}}{2007}]{vanLeeuwen2007}
{van Leeuwen} F.,  2007, \mn@doi [\aap] {10.1051/0004-6361:20078357}, \href
  {https://ui.adsabs.harvard.edu/abs/2007A&A...474..653V} {474, 653}

\makeatother
\end{thebibliography}

\onecolumn
\appendix
\renewcommand{\thesubsection}{\Alph{section}.\arabic{subsection}}

\section{AstraLux Observation Log}\label{app_obslog}

\tablefirsthead{\hline Target & Epoch & $N_\text{Frame}$ & $N_\text{sel}$ & DIT [ms] & TIT [s] & EM Gain\\ \hline}
\tablehead{\multicolumn{2}{l}{Table \ref{tab_oblsog} continued.}\\
\hline Target & Epoch & $N_\text{Frame}$ & $N_\text{sel}$ & DIT [ms] & TIT [s] & EM Gain\\ \hline}
\tabletail{\hline}
\tablelasttail{\hline}
\tablecaption{Observation log: For each target the individual observation epochs, the total number of frames ($N_\text{Frame}$), the number of frames selected for the Lucky Imaging data processing ($N_\text{sel}$), the used detector integration time (DIT), the resulting total integration time (TIT) of the fully reduced AstraLux images, and the utilized EM gain of the AstraLux detector are listed. The full table can be found in the online appendix.}
{\centering
\begin{supertabular}[center]{lcccccc}
HD\,984        & 2017-10-29 & 50000 & 5000 & 29.54 & 147.70 & 140 \\
\hline
GJ\,15         & 2017-10-27 & 50000 & 5000 & 29.54 & 147.70 & 195 \\
\hline
HD\,1502       & 2015-08-25 & 50000 & 5000 & 29.54 & 147.70 & 170 \\
\hline
HD\,1605       & 2015-08-26 & 50000 & 5000 & 29.54 & 147.70 & 160 \\
\hline
HD\,1666       & 2015-08-26 & 50000 & 5000 & 29.54 & 147.70 & 200 \\
               & 2016-10-26 & 50000 & 5000 & 29.54 & 147.70 & 205 \\
\end{supertabular} \label{tab_oblsog}
\par}

\section{Catalogue Data}\label{app_catalogues}

\tablefirsthead{\hline Target & RA (J2000) & Dec (J2000) & \multicolumn{2}{c}{$\varpi$ [mas]} & \multicolumn{2}{c}{$\mu_{\rm RA}$ [mas/yr]} & \multicolumn{2}{c}{$\mu_{\rm Dec}$ [mas/yr]}\\ \hline}
\tablehead{\multicolumn{2}{l}{Table \ref{tab_cat_gaia} continued.}\\
\hline Target & RA(J2000) & Dec(J2000) & \multicolumn{2}{c}{$\varpi$ [mas]} & \multicolumn{2}{c}{$\mu_{\rm RA}$ [mas/yr]} & \multicolumn{2}{c}{$\mu_{\rm Dec}$ [mas/yr]}\\ \hline}
\tabletail{\hline}
\tablelasttail{\hline}
\tablecaption{\label{tab_cat_gaia} The equatorial coordinates (RA, Dec) for all targets, together with their parallax $\varpi$, as well as the proper motion in right ascension $\mu_\text{RA}$ and declination $\mu_\text{Dec}$, are taken from the Gaia DR3 catalogue \citep{GaiaCollaboration2023}, unless otherwise stated. The full table can be found in the online appendix.}
{\centering
\begin{mpsupertabular}[center]{lcc r@{$\pm$}l r@{$\pm$}l r@{$\pm$}l}
HD\,984                      & 00 14 10.3 & $-$07 11 57 &  21.8770 & 0.0249 &  104.775 & 0.036 & -68.016 & 0.022 \\
GJ\,15                       & 00 18 22.9 & $+$44 01 23 & 280.7068 & 0.0203 & 2891.518 & 0.015 & 411.832 & 0.013 \\
HD\,1502                     & 00 19 17.1 & $+$14 03 17 &   5.2018 & 0.0341 &   74.472 & 0.034 & -17.069 & 0.027 \\
HD\,1605                     & 00 20 31.5 & $+$30 58 29 &  11.3555 & 0.0210 &   84.836 & 0.023 &  59.208 & 0.020 \\
HD\,1666                     & 00 20 52.3 & $-$19 55 52 &   8.4536 & 0.0226 &   -8.528 & 0.026 & -59.727 & 0.020 \\
\end{mpsupertabular}
\par}
\vspace{5mm}
\tablefirsthead{\hline Target & $ T_{\rm eff}$ [K] & $log(g[{\rm cm^2/s}])$ & $A_{\rm V}$ [mag]\\ \hline}
\tablehead{\multicolumn{2}{l}{Table \ref{tab_cat_starhorse} continued.}\\
\hline
Target & $T_{\rm eff}$ [K] & $log(g[{\rm cm^2/s}])$ & $A_{\rm V}$ [mag]\\
\hline}
\tabletail{\hline}
\tablelasttail{\hline \\}
\tablecaption{Effective temperature $T_{\rm eff}$, surface gravity $log(g)$ and V-band extinction estimate $A_{\rm V}$ for all targets listed in the SHC. The full table is available in the online appendix}
\nopagebreak
{\centering
\begin{supertabular}[center]{lccc}
HD\,984          & 6402 $^{+ 372}_{-328}$ & 4.30 $^{+0.06}_{-0.08}$ & 0.088 $^{+0.216}_{-0.088}$\\
GJ\,15           & 3443 $^{+ 55~}_{-47~}$ & 4.81 $^{+0.01}_{-0.02}$ & 0.000 $^{+0.020}_{-0.000}$\\
HD\,1502         & 5102 $^{+ 170}_{-173}$ & 2.97 $^{+0.08}_{-0.09}$ & 0.211 $^{+0.130}_{-0.168}$\\
HD\,1605         & 4936 $^{+ 179}_{-161}$ & 3.31 $^{+0.07}_{-0.06}$ & 0.019 $^{+0.188}_{-0.019}$\\
HD\,1666         & 6714 $^{+ 684}_{-476}$ & 3.97 $^{+0.08}_{-0.08}$ & 0.255 $^{+0.372}_{-0.255}$\\
\end{supertabular} \label{tab_cat_starhorse}
\par}
\vspace{0mm}
\tablefirsthead{\hline Target & i$'$ [mag] & Reference & Age [Gyr] & Reference\\ \hline}
\tablehead{\multicolumn{2}{l}{Table \ref{tab_cat_imag_age} continued.}\\
\hline
Target & i$'$ [mag] & Reference & Age [Gyr] & Reference\\
\hline}
\tabletail{\hline}
\tablelasttail{\hline}
\tablecaption{Apparent i$'$-band magnitude and age of all targets from the literature. Where more than one reference is given, the mean and standard deviation of all values are given. The full table is available in the online appendix.}

\begin{mpsupertabular*}{\textwidth}[center]{l c p{0.4\textwidth} c p{0.4\textwidth}}
HD\,984  & $7.46 \pm 0.33$  & \tiny{\cite{Zacharias2013}, \cite{Henden2018}}                                       & $0.08^{+0.12}_{-0.05}$   & \tiny{\cite{Meshkat2015}}\\
GJ\,15   & $7.28 \pm 0.01$  & \tiny{\cite{Zacharias2013}}                                                          & $3.8 \pm 1.1$            & \tiny{\cite{Yee2017}, \cite{Mann2015}}\\
HD\,1502 & $7.90 \pm 0.16$  & \tiny{\cite{Zacharias2013},\cite{Zacharias2015},\cite{Henden2015},\cite{Henden2018}} & $2.4 \pm 0.5$            & \tiny{\cite{Johnson2011}}\\
HD\,1605 & $7.06 \pm 0.02$  & \tiny{\cite{Ofek2008}}                                                               & $4.59 \pm 1.37$          & \tiny{\cite{Harakawa2015}}\\
HD\,1666 & $7.75 \pm 0.47$  & \tiny{\cite{Zacharias2013}, \cite{Henden2015}, \cite{Henden2018}}                    & $1.76 \pm 0.2$           & \tiny{\cite{Harakawa2015}}\\
\end{mpsupertabular*} \label{tab_cat_imag_age}

\newpage

\section{Measurements}\label{app_measurements}

\subsection*{Astrometric Measurements}

\tablefirsthead{\hline
Companion & Epoch & $\rho$ [arcsec] & $PA$ [$^\circ$] & $\sigma (\Delta\varpi)$ & cpm & $\mu_{\rm rel}$ [mas/yr] & $\mu_{\rm esc}$ [mas/yr] & orbit?\\
 \hline}
\tablehead{\hline
Companion & Epoch & $\rho$ [arcsec] & $PA$ [$^\circ$] & $\sigma (\Delta\varpi)$ & cpm & $\mu_{\rm rel}$ [mas/yr] & $\mu_{\rm esc}$ [mas/yr] & orbit?\\
\hline}
\tabletail{\hline}
\tablelasttail{\hline \vspace{-5mm}\footnotetext[0]{$^1$Relative astrometry of the companion is based on Gaia DR3 astrometry.} \\}
\tablecaption{The angular separation ($\rho$) and position angle ($PA$) of all companions of exoplanet host stars detected with AstraLux in this survey, whose parallax and proper motion are listed in the Gaia DR3 catalogue \citep{GaiaCollaboration2023}. The significance of the parallax difference between the detected companions and the associated exoplanet host stars $\sigma (\Delta\varpi)$, their cpm-index, the differential proper motion $\mu_{\rm rel}$ and the derived estimate of the escape velocity $\mu_{\rm esc}$ are given. Furthermore, the last column indicates whether the orbital motion of the system (orbit?) is significantly detected. The full table is available in the online appendix.}
{\centering
\nopagebreak
\begin{mpsupertabular}[center]{lccccccccc}
K2-267\,B     & 2016\footnotemark[1]& $~5.899 \pm 0.001$ & $223.50 \pm 0.01$ & 1.1 & 30.8 & $0.84 \pm 0.26$ & $0.33 \pm 0.04$ & yes\\
              & 2020-09-06          & $~5.898 \pm 0.011$ & $223.51 \pm 0.10$ &     &      &                 &                 &    \\
\hline
KELT-2\,B     & 2016\footnotemark[1]& $~2.381 \pm 0.001$ & $332.38 \pm 0.01$ & 0.7 & 19.2 & $1.83 \pm 0.04$ & $4.22 \pm 0.02$ & yes\\
              & 2017-10-29          & $~2.372 \pm 0.009$ & $332.59 \pm 0.23$ &     &      &                 &                 &    \\
              & 2020-09-05          & $~2.361 \pm 0.007$ & $332.80 \pm 0.14$ &     &      &                 &                 &    \\
\hline
HAT-P-20\,B   & 2016\footnotemark[1]& $~6.989 \pm 0.001$ & $320.76 \pm 0.01$ & 0.7 & 45.8 & $4.16 \pm 0.64$ & $4.84 \pm 0.08$ & yes\\
              & 2017-10-27          & $~6.919 \pm 0.009$ & $321.09 \pm 0.07$ &     &      &                 &                 &    \\
              & 2017-10-29          & $~6.917 \pm 0.013$ & $320.95 \pm 0.10$ &     &      &                 &                 &    \\
              & 2018-05-04          & $~6.908 \pm 0.016$ & $320.78 \pm 0.10$ &     &      &                 &                 &    \\
\hline
HAT-P-35\,C   & 2016\footnotemark[1]& $~9.018 \pm 0.001$ & $213.96 \pm 0.01$ & 0.8 & 77.4 & $0.31 \pm 0.16$ & $0.27 \pm 0.01$ &    \\
              & 2016-04-02          & $~8.974 \pm 0.023$ & $213.96 \pm 0.15$ &     &      &                 &                 &    \\
              & 2016-10-26          & $~9.011 \pm 0.012$ & $214.01 \pm 0.09$ &     &      &                 &                 &    \\
              & 2017-10-29          & $~9.007 \pm 0.014$ & $213.97 \pm 0.09$ &     &      &                 &                 &    \\
\hline
HAT-P-22\,B   & 2016\footnotemark[1]& $~9.163 \pm 0.001$ & $~23.47 \pm 0.01$ & 0.3 & 106  & $1.66 \pm 0.03$ & $3.82 \pm 0.03$ & yes\\
              & 2017-10-29          & $~9.176 \pm 0.015$ & $~23.70 \pm 0.09$ &     &      &                 &                 &    \\
              & 2018-05-05          & $~9.151 \pm 0.017$ & $~23.46 \pm 0.06$ &     &      &                 &                 &    \\
\end{mpsupertabular} \label{tab_astr1}
\vspace{5mm}
\tablefirsthead{\hline Companion & Epoch & $\rho$ [arcsec] & PA [$^\circ$] & $\sigma_{\rho;{\rm co}}$ & $\sigma_{PA; {\rm co}}$ & $\sigma_{\rho; {\rm bg}} $ & $\sigma_{PA;{\rm bg}}$ & cpm? & orbit?\\ \hline}
\tablehead{\multicolumn{2}{l}{Table \ref{tab_astr2} continued.}\\
\hline
Companion & Epoch & $\rho$ [arcsec] & PA [$^\circ$] & $\sigma_{\rho;{\rm co}}$ & $\sigma_{PA; {\rm co}}$ & $\sigma_{\rho; {\rm bg}} $ & $\sigma_{PA;{\rm bg}}$ & cpm? & orbit?\\
\hline}
\tabletail{\hline}
\tablelasttail{\hline \vspace{-5mm} \footnotetext[0]{$^1$Relative astrometry of the companion is based on Gaia DR3.}  \footnotetext[0]{$^2$Relative astrometry of the companion is taken from \cite{Ginski2016}.}\\}
\tablecaption{Angular separation ($\rho$) and position angle ($PA$) of all companions of exoplanet host stars detected with AstraLux in this survey, whose parallax and proper motion are not listed in the Gaia DR3. The last observation epoch is used as the astrometric reference to test the co-moving and background hypotheses. For all other epochs, the significance to reject the co-moving hypothesis is given in angular separation $\sigma_{\rho;co}$ and position angle $\sigma_{\text{PA};co}$, and the significance to reject the background hypothesis is given in angular separation $\sigma_{\rho;bg}$ and position angle $\sigma_{\text{PA};bg}$. In addition, the two right columns indicate whether a common proper motion (cpm?) of the exoplanet host star and the companion, or the orbital motion of the system (orbit?) is significantly detected. The full table is available in the online appendix.}
{\centering
\begin{mpsupertabular}[center]{lccccccccc}
HD\,1666\,B         & 2015-08-26                 & $~2.092 \pm 0.008$ & $~64.35 \pm 0.14$ & 0.4 & 1.7 & 3.5 & 5.4 &     &    \\
                    & 2016\footnotemark[1]       & $~2.092 \pm 0.001$ & $~63.19 \pm 0.01$ & 0.6 & 4.3 & 2.1 & 11  &     &    \\
                    & 2016-10-26                 & $~2.097 \pm 0.009$ & $~63.97 \pm 0.18$ & --- & --- & --- & --- & yes &    \\
\hline
HD\,2638\,C         & 2014-08-20\footnotemark[2] & $~0.520 \pm 0.004$ & $168.00 \pm 0.35$ & 1.2 & 3.5 & 117 & 16  &     &    \\
                    & 2017-10-29                 & $~0.512 \pm 0.005$ & $173.41 \pm 1.56$ & --- & --- & --- & --- & yes & yes\\
\hline
WASP-76\,B          & 2014-08-20\footnotemark[2] & $~0.444 \pm 0.005$ & $214.92 \pm 0.56$ & 1.5 & 1.2 & 4.4 & 12  &     &    \\
                    & 2015-08-24                 & $~0.492 \pm 0.010$ & $211.51 \pm 0.97$ & 1.6 & 1.1 & 0.4 & 3.9 &     &    \\
                    & 2016-10-26                 & $~0.465 \pm 0.013$ & $213.28 \pm 1.30$ & --- & --- & --- & --- & yes &    \\
\hline
HD\,15779\,B        & 2016\footnotemark[1]       & $11.525 \pm 0.000$ & $115.46 \pm 0.00$ & 2.8 & 6.5 & 5.1 & 27  &     &    \\
                    & 2020-09-04                 & $11.503 \pm 0.008$ & $115.72 \pm 0.04$ & --- & --- & --- & --- & yes & yes\\
\hline
HD\,43691\,BC        & 2015-03-10\footnotemark[2] & $~4.435 \pm 0.016$ & $~40.77 \pm 0.24$ & 2.6 & 0.3 & 1.4 & 4.6 &     &    \\
                    & 2016\footnotemark[1]       & $~4.474 \pm 0.001$ & $~40.89 \pm 0.01$ & 1.1 & 0.1 & 0.1 & 3.4 &     &    \\
                    & 2016-04-02                 & $~4.468 \pm 0.009$ & $~40.91 \pm 0.12$ & 1.3 & 0.2 & 1.4 & 2.4 &     &    \\
                    & 2016-10-26                 & $~4.490 \pm 0.014$ & $~40.87 \pm 0.19$ & --- & --- & --- & --- & yes &    \\
\end{mpsupertabular} \label{tab_astr2}
\par}

\newpage

\tablefirsthead{\hline Companion & Epoch & $\rho$ [arcsec]& $PA$ [$^\circ$] & Survey Reference\\ \hline}
\tablehead{\multicolumn{2}{l}{Table \ref{tab_astr_extra} continued.}\\
\hline
Companion & Epoch & $\rho$ [arcsec] & $PA$ [$^\circ$] & Comment\\
\hline}
\tabletail{\hline}
\tablelasttail{\hline \\}
\tablecaption{Angular separation ($\rho$) and positional angle ($PA$) of all companions of exoplanet host stars whose parallax and proper motion are not listed in the Gaia DR3 and which are detected in this survey in only one observing epoch with AstraLux, but which could be observed earlier in other surveys.}
{\centering
\begin{mpsupertabular}[center]{lcccccccc}
HAT-P-35\,C      & 2016-10-26 & $0.914 \pm 0.012$ & $139.07 \pm 0.79$ & \cite{Woellert2015}\\
$\gamma\,$Leo\,B & 2018-05-04 & $4.725 \pm 0.016$ & $126.66 \pm 0.17$ & \cite{Romanenko2014}\\
HAT-P-57\,B      & 2018-05-06 & $2.692 \pm 0.007$ & $231.41 \pm 0.14$ & \cite{Bohn2020}\\
HAT-P-57\,C      & 2018-05-06 & $2.790 \pm 0.010$ & $227.25 \pm 0.17$ & \cite{Bohn2020}\\
\end{mpsupertabular} \label{tab_astr_extra}
\par}\vspace{5mm}

\tablefirsthead{\hline Candidate & Epoch & $\rho$ [arcsec] & $PA$ [$^\circ$] & i$'$ [mag] & $\sigma (\Delta\varpi)$ & cpm-index\\ \hline}
\tablehead{\multicolumn{2}{l}{Table \ref{tab_astr3} continued.}\\
\hline
Candidate & Epoch & $\rho$ [arcsec] & $PA$ [$^\circ$] & i$'$ [mag] & $\sigma (\Delta\varpi)$ & cpm-index\\
\hline}
\tabletail{\hline}
\tablelasttail{\hline\\}
\tablecaption{The angular separation ($\rho$) and position angle ($PA$) for the Gaia reference epoch of all companion-candidates detected with AstraLux in this survey, whose companionship with the associated exoplanet host stars is clearly excluded by their accurate Gaia DR3 astrometry. For each candidate we give its i$'$-band photometry as measured in our AstraLux images. In addition, for all candidates we list the significance of the difference between their parallax and that of the associated exoplanet host star $\sigma (\Delta\varpi)$, as well as their cpm-index. The full table can be found in the online appendix.}
{\centering
\begin{supertabular}[center]{lccccccc}
HD\,10697\,CC1     & 2016 &  8.862 & 287.64 & 13.2 & 98.1 & 1.0 \\
BD\,+49\,828\,CC1  & 2016 &  6.045 & 131.79 & 14.0 & 12.5 & 2.1 \\
BD\,+49\,828\,CC2  & 2016 &  5.194 & 265.61 & 17.4 &  6.1 & 0.9 \\
GJ\,180\,CC1       & 2016 & 10.999 & 248.69 & 20.2 &  605 & 1.0 \\
BD\,-06\,1339\,CC1 & 2016 &  8.479 & 252.02 & 15.5 &  115 & 1.0 \\
\end{supertabular} \label{tab_astr3}
\par}
\vspace{5mm}
\tablefirsthead{\hline Candidate & Epoch & $\rho$ [arcsec] & $PA$ [$^\circ$] & i$'$ [mag] & $\sigma_{\rho;{\rm co}}$ & $\sigma_{PA;{\rm co}}$ & $\sigma_{\rho;{\rm bg}}$ & $\sigma_{PA;{\rm bg}}$\\ \hline}
\tablehead{\multicolumn{2}{l}{Table \ref{tab_astr4} continued.}\\
\hline
Candidate & Epoch & $\rho$ [arcsec] & $PA$ [$^\circ$] & i$'$ [mag] & $\sigma_{\rho;{\rm co}}$ & $\sigma_{PA;{\rm co}}$ & $\sigma_{\rho;{\rm bg}}$ & $\sigma_{PA;{\rm bg}}$\\
\hline}
\tabletail{\hline}
\tablelasttail{\hline \vspace{-5mm} \footnotetext[0]{$^1$Relative astrometry of the companion is based on Gaia DR3 astrometry.}\\}
\tablecaption{Companion-candidates detected with AstraLux in this survey, without Gaia DR3 parallaxes and proper motions, whose companionship is ruled out based on their AstraLux astrometry. We list the angular separation ($\rho$) and the positional angle ($PA$) of all these candidates for the individual observation epochs, as well as their AstraLux i$'$-band photometry. The last observation epoch is used as an astrometric reference epoch to test the co-moving and background hypotheses. For all other epochs, the significance to reject the co-moving hypothesis in separation $\sigma_{\rho;co}$ and position angle $\sigma_{\text{PA};co}$, and the significance to reject the background hypothesis in separation $\sigma_{\rho;bg}$ and position angle $\sigma_{\text{PA};bg}$ are given.}
{\centering
\begin{mpsupertabular}[center]{lcccccccc}
BD\,-06\,1339\,CC2   & 2016\footnotemark[1] & $8.868 \pm 0.002$ & $257.19 \pm 0.01$ & 15.5 & 36  & 262 & 8.5 & 9.0\\
                     & 2016-10-27           & $8.829 \pm 0.011$ & $259.18 \pm 0.07$ &      & 13  & 109 & 1.2 & 2.5\\
                     & 2017-10-27           & $8.780 \pm 0.010$ & $261.48 \pm 0.04$ &      & 9.6 & 115 & 2.1 & 2.4\\
                     & 2020-09-05           & $8.673 \pm 0.005$ & $267.98 \pm 0.04$ &      & --- & --- & --- & ---\\
\hline
HD\,44219\,CC1       & 2016\footnotemark[1] & $8.951 \pm 0.000$ & $232.70 \pm 0.00$ & 16.9 & 8.4 & 15  & 0.3 & 4.7\\
                     & 2017-10-29           & $9.060 \pm 0.013$ & $233.73 \pm 0.07$ &      & --- & --- & --- & ---\\
\end{mpsupertabular} \label{tab_astr4}
\par}

\begin{flushleft}
\subsection*{Photometric Measurements}
\end{flushleft}
\tablefirsthead{\hline Companion & i$'$ [mag] & $M_{\rm i'}$ [mag] & Mass [$\rm M_\odot$] & Mass Reference\\ \hline}
\tablehead{}
\tabletail{\hline}
\tablelasttail{\hline}
\tablecaption{Photometry of all co-moving companions of exoplanet host stars detected with AstraLux in this survey. For each companion, we list its apparent i$'$-band magnitude as measured in our AstraLux images, its absolute i$'$-band magnitude, and its mass, which is derived from the absolute magnitude of the companion and stellar evolution models, or otherwise taken from the literature.}
{\centering
\begin{mpsupertabular}[center]{lcccc}
\nopagebreak
HD\,1666\,B  & $14.71 \pm 0.48$ & $9.19 \pm 0.53$ & $0.38 \pm 0.08$ & \\
HD\,2638\,C  & $12.53 \pm 0.02$ & $8.52 \pm 0.18$ & $0.47 \pm 0.03$ & \\
WASP-76\,B   & $12.20 \pm 0.14$ & $5.62 \pm 0.31$ & $0.83 \pm 0.05$ & \\
HD\,15779\,B & $12.09 \pm 0.09$ & $8.33 \pm 0.09$ & $0.50 \pm 0.02$ & \\
K2-267       & $17.10 \pm 0.09$ & $8.71 \pm 0.23$ & $0.45 \pm 0.04$ & \\
\end{mpsupertabular} \label{tab_photo1} \newline\newline
\par}\vspace{5mm}}
\markright{AstraLux Detection Limit for All Targets}\label{app_detlim}

\renewcommand{\thefigure}{D\arabic{figure}}
\begin{figure}
\textbf{APPENDIX D:~ASTRALUX DETECTION LIMIT FOR ALL TARGETS}
\\
\\
\includegraphics[width=0.5\textwidth]{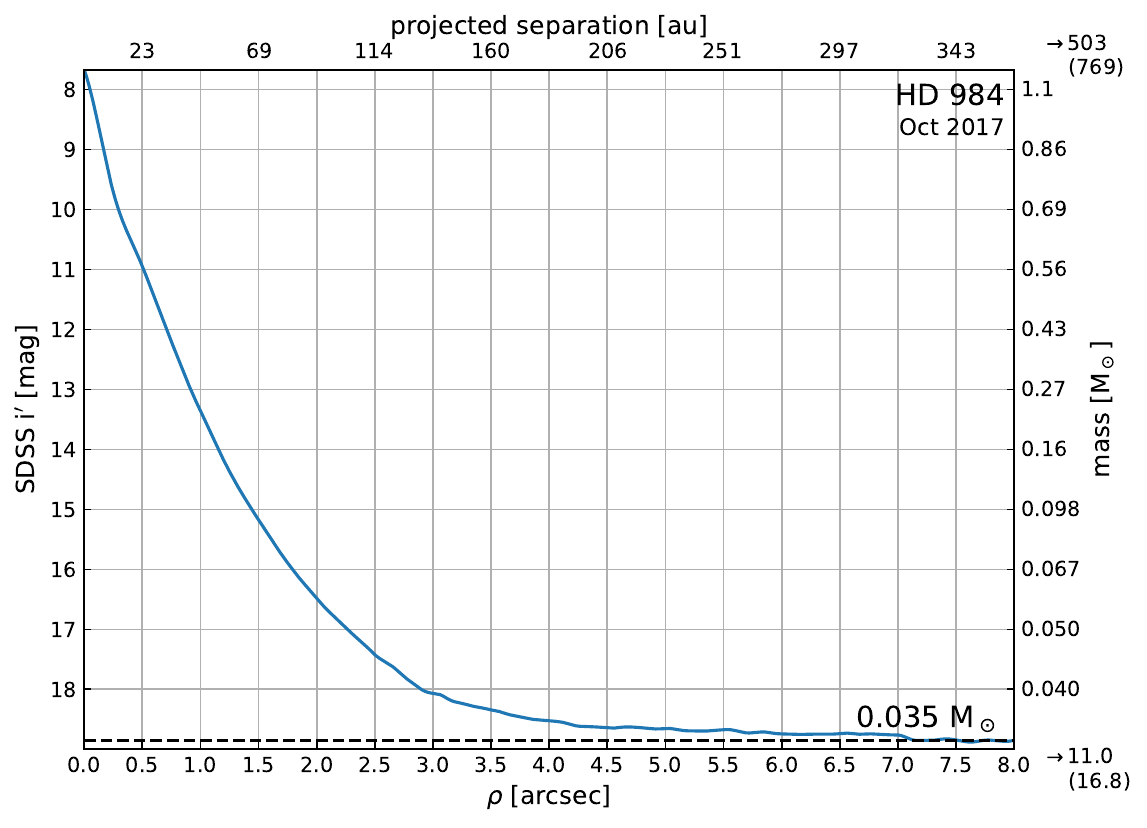} \includegraphics[width=0.5\textwidth]{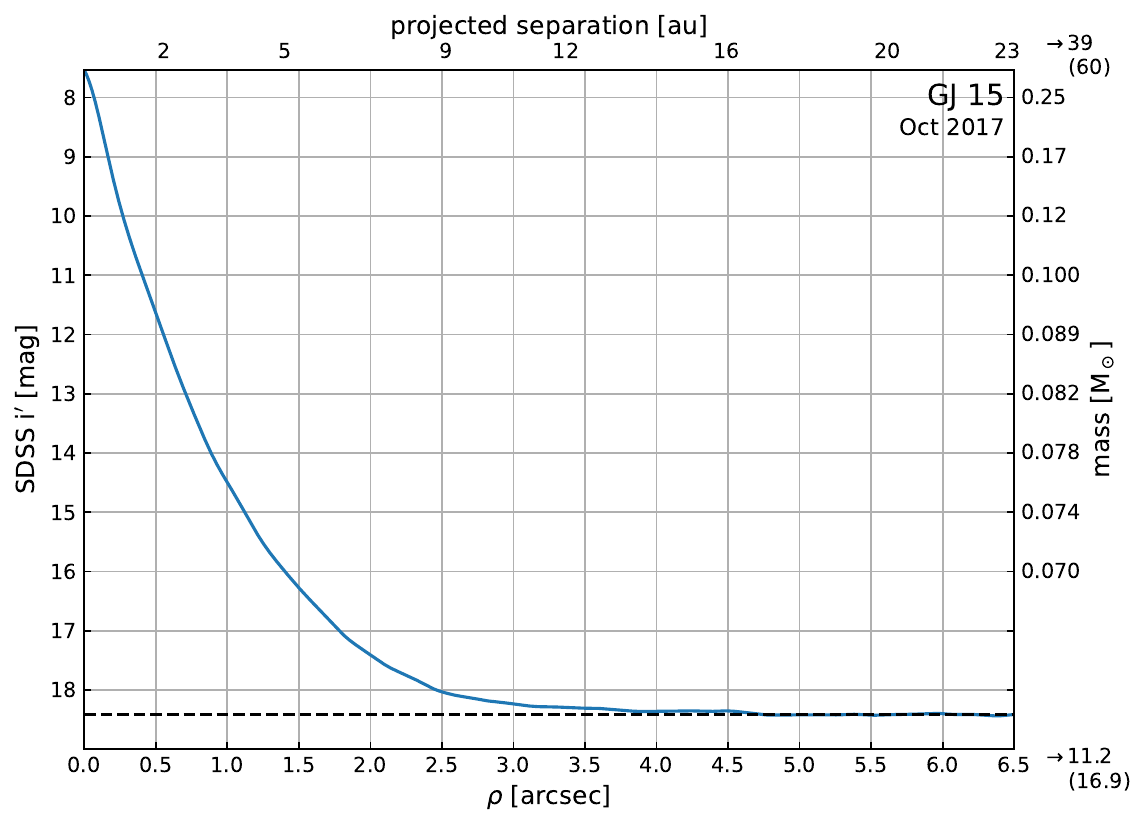}
\caption{The AstraLux detection limit ($S/N=3$) of all targets observed in this survey. The dashed black line indicates the detection limit reached in the background limited region around the bright exoplanet host stars. On the right, the maximum angular and projected radius of the radial field of view around the targets, fully covered by the AstraLux detector, are given in arcsec and au respectively. The values in brackets give the maximum angular and projected separation of companions detectable within the square field of view of the AstraLux detector. The masses of the (sub)stellar objects detectable in the AstraLux images are given on the right axis. The individual detection limits of all observed targets are given in the online appendix.}\label{fig_detlims}
\end{figure}

\bsp
\label{lastpage}
\end{document}